\documentclass
[rmp,aps,twocolumn,showpacs,preprintnumbers,amsmath,amssymb]{revtex4}

\usepackage{graphicx}  
\usepackage{bm}        
\usepackage{paralist}  
\usepackage{amsthm}    

\theoremstyle{plain}
\newtheorem{thm}{Theorem}

\newtheorem*{Darmois}{Darmois junction conditions (DJC)}
\newtheorem*{SSJC}{Spherically symmetric junction conditions (SSJC)}

\theoremstyle{definition}
\newtheorem{defi}[thm]{Definition}

\theoremstyle{remark}

\newcommand{\ul}[1]{\underline{#1}}

\newcommand{\sss}{\scriptscriptstyle}
\newcommand{\ind}[1]{{\sss \text{#1}}}        
\DeclareMathAlphabet{\bi}{OML}{cmm}{b}{it}    
\newcommand{\br}[1]{\mathbf{#1}}              

\newcommand{\Plane}{\mathit{\Pi}}             
\newcommand{\tr}{\mathrm{tr}}                 
\newcommand{\diag}{\mathrm{diag}}             
\newcommand{\abs}[1]{\lvert#1\rvert}          
\newcommand{\norm}[1]{\lVert#1\rVert}         
\newcommand{\sprod}[2]{\langle#1,#2\rangle}   
\newcommand{\vecp}[1]{{\vec #1}\;\!'}         
\newcommand{\vecpp}[1]{{\vec #1}\;\!''}       
\newcommand{\Ord}{\mathcal{O}}                

\DeclareMathOperator{\LC}{\bm{\nabla}}        
\DeclareMathOperator{\Grad}{\br{grad}}        
\DeclareMathOperator{\Hess}{\br{Hess}}        
\DeclareMathOperator{\Div}{\br{div}}          
\DeclareMathOperator{\Riem}{\br{Riem}}        
\DeclareMathOperator{\R}{\bi{R}}              
\DeclareMathOperator{\Weyl}{\br{Weyl}}        
\DeclareMathOperator{\Ric}{\br{Ric}}          
\DeclareMathOperator{\Ricci}{\br{Ricci}}      
\DeclareMathOperator{\Scal}{\mathrm{Scal}}    
\DeclareMathOperator{\Ein}{\br{Ein}}          
\DeclareMathOperator{\Laplacian}{\Delta}      
\newcommand{\bpartial}{\boldsymbol{\partial}} 
\newcommand{\bnabla}{\boldsymbol{\nabla}}     
\newcommand{\bdot}{\boldsymbol{\cdot}}
\newcommand{\Lie}{\mathbf{L}}                 
\newcommand{\ed}{\bi{d}}                      

\newcommand{\M}{\mathcal{M}}                      
\newcommand{\B}{\mathcal{B}}                      
\newcommand{\Stwo}{{S^2}}                         
\newcommand{\sB}{{\sss \B}}                       
\newcommand{\sStwo}{{\sss \Stwo}}                 
\newcommand{\gB}{\g_\sB}                          
\newcommand{\gG}{\g_\varGamma}                    
\newcommand{\gStwo}{\g_\sStwo}                    
\newcommand{\ScalB}{\sideset{^\sB}{}\Scal}        
\DeclareMathOperator{\LCB}{\sideset{^\sB}{}{\LC}} 
\DeclareMathOperator{\LCStwo}{\sideset{^\Stwo}{}\LC}  
\DeclareMathOperator{\LCG}{\sideset{^\varGamma}{}{\LC}} 
\DeclareMathOperator{\GradB}{\sideset{^\sB}{}\Grad} 
\DeclareMathOperator{\HessB}{\sideset{^\sB}{}\Hess} 
\DeclareMathOperator{\LaplacianB}{\sideset{^\sB\!}{}\Laplacian} 

\newcommand{\Id}{\mathbf{id}}                   
\newcommand{\g}{\bg}                            
\newcommand{\gammadot}{{\dot{\bgamma}}}         

\newcommand{\ba}{\bi{a}}
\newcommand{\bb}{\bi{b}}
\newcommand{\bd}{\bi{d}}
\newcommand{\be}{\bi{e}}

\newcommand{\bg}{\bi{g}}
\newcommand{\bh}{\bi{h}}
\newcommand{\bj}{\bi{j}}
\newcommand{\bk}{\bi{k}}
\newcommand{\bl}{\bi{l}}

\newcommand{\bn}{\bi{n}}

\newcommand{\bs}{\bi{s}}

\newcommand{\bu}{\bi{u}}
\newcommand{\bv}{\bi{v}}
\newcommand{\bz}{\bi{z}}

\newcommand{\bB}{\bi{B}}
\newcommand{\bC}{\bi{C}}

\newcommand{\bK}{\bi{K}}

\newcommand{\bP}{\bi{P}}
\newcommand{\bQ}{\bi{Q}}

\newcommand{\bS}{\bi{S}}
\newcommand{\bT}{\bi{T}}
\newcommand{\bU}{\bi{U}}
\newcommand{\bV}{\bi{V}}
\newcommand{\bW}{\bi{W}}
\newcommand{\bX}{\bi{X}}
\newcommand{\bY}{\bi{Y}}
\newcommand{\bZ}{\bi{Z}}

\newcommand{\balpha}{\boldsymbol{\alpha}}
\newcommand{\bbeta}{{\boldsymbol{\beta}}}
\newcommand{\bgamma}{\boldsymbol{\gamma}}
\newcommand{\bdelta}{\boldsymbol{\delta}}

\newcommand{\bfeta}{\boldsymbol{\eta}}
\newcommand{\btheta}{\boldsymbol{\theta}}

\newcommand{\bmu}{\boldsymbol{\mu}}

\newcommand{\bsigma}{\boldsymbol{\sigma}}
\newcommand{\btau}{\boldsymbol{\tau}}

\newcommand{\bomega}{\boldsymbol{\omega}}

\newcommand{\Lemaitre}{Lema\^{\i}tre}
\newcommand{\Schuecking}{Sch\"ucking}

\begin{document}



\title{Influence of global cosmological expansion 
       on local dynamics and kinematics}

\author{Matteo Carrera}
\email{matteo.carrera@physik.uni-freiburg.de}
\affiliation{Institute of Physics, University of Freiburg, 
Hermann-Herder-Strasse~3, D-79104 Freiburg, Germany}
\author{Domenico Giulini}
\email{domenico.giulini@itp.uni-hannover.de}
\affiliation{University of Hannover,
Appelstrasse~2, D-30167 Hannover, Germany}
\altaffiliation[Also at: ]{ZARM, University of Bremen, Am Fallturm\\ 
D-28359 Bremen, Germany}

\date{\today}

\begin{abstract}
We review attempts to estimate the influence of global cosmological
expansion on local systems. Here `local' is taken to mean that 
the sizes of the considered systems are much smaller than cosmologically 
relevant scales. For example, such influences can affect orbital 
motions as well as configurations of compact objects, like 
black holes. We also discuss how measurements based on the exchange of 
electromagnetic signals of distances, velocities, etc. of moving 
objects are influenced. As an application we compare orders of 
magnitudes of such effects with the scale set by the apparently 
anomalous acceleration of the Pioneer 10 and 11 spacecrafts, 
which is $10^{-9}{\rm m/s^2}$. We find no reason to believe that 
the latter is of cosmological origin. However, the general problem 
of gaining a qualitative and quantitative understanding of how 
the cosmological dynamics influences local systems remains 
challenging, with only partial clues being so far provided by 
exact solutions to the field equations of General Relativity. 
\end{abstract}

\pacs{95.30.Sf, 04.20.Cv, 04.20.Jb, 98.80.Jk}


\maketitle

\tableofcontents


\section{Introduction}
\label{sec:Introduction}
There is by now ample evidence that our Universe is expanding on average. 
This means that on the largest scales one observes redshifts from 
structures that are interpreted as recessional motion, also called the 
Hubble flow. To first approximation, the relative velocity between two 
structures grows linearly with their mutual distance. The constant of 
proportionality is the so-called Hubble constant, $H_0$, whose value is 
now fairly accurately measured as being close to 
$70\,\mathrm{Km\cdot s^{-1}\cdot Mpc^{-1}}$, see e.g.~\cite{Komatsu.etal:2008}. 
This means that for any additional mega-parsec 
($\mathrm{Mpc}=3.262\times10^6\,\mathrm{lightyears}=
3.086\times 10^{19}\,\mathrm{km}$) the recessional velocity picks up an 
extra $70$ kilometers per second. Clearly, typical peculiar velocities 
superimpose on the global Hubble flow. For galaxies they can be up to 
$1000$ kilometers per second, so that the Hubble flow definitely dominates 
at distances above $200\,\mathrm{Mpc}$, i.e.~above supercluster scale. In 
this respect it is remarkable that Hubble's classic paper \cite{Hubble:1929} 
of 1929 plots the velocity-distance relation of extra-galactic nebulae only 
up to $2\,\mathrm{Mpc}$, though it has to be added that in those days 
distances where generally underestimated, sometimes up to a factor of 10. 

For pedagogical purposes the global expansion is sometimes represented 
by the two-dimensional balloon model, in which three-dimensional space 
corresponds to the two-dimensional surface of an inflating rubber balloon; 
see e.g.~\S\,27.5 in \cite{Misner.Thorne.Wheeler:Gravitation}. 
At each point attached to the rubber material an observer sees other points 
attached in a state of radial recessional motion, the faster the further 
they are away. This picture is used to stress that each point is locally 
(i.e.~with respect to the local rubber material) at rest but receding from 
all other points because space in-between is itself expanding. However, this 
global expansion does not affect all structures: Local overdensities in the 
matter distribution may inhibit space from expanding. In the balloon model 
of \cite{Misner.Thorne.Wheeler:Gravitation} this is represented by little 
pennies being glued onto the balloon. The rubber material underneath the 
coins does not expand due to the stiff glue which holds it in place. The 
question arises what, in reality, are the structures corresponding to the 
coin and what dynamical mechanism provides the glue? It is often heard that 
`bound systems' do not participate in the global expansion, or that 
systems below the scale of galaxy clusters `break away' from the Hubble 
flow. But what does `bound' and `break away' really mean?%
\footnote{For a recent discussion on the meaning of `joining the 
Hubble flow' see~\cite{Barnes.etal:2006}.} 
For example, is it obvious that the Astronomical Unit is not affected by 
global expansion (compare \cite{Krasinsky.Brumberg:2004,Standish:2004}) 
or can it even be, as e.g.~suggested in \cite{Fahr.Siewert:2008}, that our 
Universe is contracting on small scales while it expands in the large? 
If so, what precisely would rule the relation between contracting and 
expanding scales? 

The purpose of this paper is to review and discuss attempts that aim to 
make precise and answer some of these fundamental questions, taking due 
account of the dynamical laws and the kinematical framework of General 
Relativity. We will emphasize the changes in kinematical relations within 
time-dependent spacetime geometries, which seem to be widely neglected 
in related discussions. 

Next to being a question of fundamental interest, the raised issue also needs 
to be clarified quantitatively in connection with more practical aims, like, 
e.g., the modeling of celestial reference frames~\cite{Klioner.Soffel:2005}. 
The specific question of whether the global expansion has any influence on 
the local dynamics and kinematics within the Solar System has recently also 
attracted increasing attention in connection with the so-called 
`Pioneer-Anomaly'~\cite{Anderson.etal:1998,Anderson.etal:2002a,%
Markwardt:2002,Nieto.Turyshev:2004,Turyshev.etal:2005a,Turyshev.etal:2005b}, 
henceforth abbreviated by PA. Here frequency-measurements in Doppler tracking 
are translated into standard kinematical quantities, like velocity and 
acceleration. The result shows an anomalous acceleration of the Pioneer 
satellites directed towards the center of the Solar System. 
In \cite{Markwardt:2002} the magnitude of this acceleration is reported to 
be $a=8.6\pm 1.34\times 10^{-10}\,\mathrm{m\cdot s^{-2}}$. Note that such 
an apparently small acceleration amounts to variations in spatial 
localization of nearly 500 kilometers after 10 years. It so happens that the 
magnitude of this acceleration is very close to the product of the current 
value of the Hubble constant, $H_0$, and the velocity of light in vacuum: 
\begin{equation}\label{eq:HubbleTimesC}
\begin{split}
  H_0c&\approx 
  \bigl(70\,\mathrm{km\cdot s^{-1}\cdot Mpc^{-1}}\bigr)
  \bigl(3\times 10^5\,\mathrm{km\cdot s^{-1}}\bigr)\\
  &=7\times 10^{-10}\,\mathrm{m\cdot s^{-2}}\,.\\
\end{split}
\end{equation}
Whether this `almost coincidence' of numbers does indeed have any deeper 
significance can and should only be decided on the basis of reliable 
estimates within the dynamical framework of General Relativity. There already 
exist various speculations and claims in the literature that try to attribute 
the PA to either simple kinematical (e.g.~\cite{Rosales.Sanchez-Gomez:1998}) 
or dynamical (e.g.~\cite{Fahr.Siewert:2008}) effects of a time varying 
background geometry, though none of them does justice to the requirements 
posed by General Relativity.%
\footnote{The reader will soon find out that we disagree with all such 
claims.} 
This is clearly a very difficult task: There is very little analytical 
knowledge of how to model in terms of exact solutions, or at least in terms 
of controlled approximations to exact solutions, the hierarchy of mutually 
embedded systems: 
Solar System $\rightarrow$ Galaxy $\rightarrow$ Local Group $\rightarrow$ 
Cluster $\rightarrow$ Supercluster $\rightarrow$ Standard-Cosmological 
Solution. Usually we expect each such system to define a typical length 
scale beyond which we may consider it as quasi isolated~\cite{Cox:2007}. 
But, clearly, whether this is a valid assumption or not can only be 
decided on the basis of a self-consistent dynamical consideration.   
In our context all this suggests to first study the influence 
of cosmic expansion on the most simple systems immersed in an 
otherwise homogeneous cosmological background. We will see that 
this already poses a number of non-trivial analytical as well as 
conceptual problems. 

In this article we will derive upper bounds for various effects of global 
expansion on local systems in the context of such simple models. The idea 
here is that the upper bounds so derived will \emph{a fortiori} be upper 
bounds in more realistic models, since a further embedding of the system 
we consider into a higher structure of local overdensities will further 
suppress the influence of cosmological expansion. This is evidently true 
in situations in which the spherically symmetric Einstein--Straus model 
applies, but can also be argued for as a result of taking into account 
small-scale anisotropies in the matter distribution, as has been done 
from first-order perturbations of the Newtonian 
equations~\cite{Dominguez.Gaite:2001}. We conclude form this that if we 
find the relevant upper bounds to be outside current experimental reach, 
this will maintain to be the case in more realistic contexts.

\section{Strategic outline and results}
\label{sec:Outline}

\subsection{Improved Newtonian equations}
\label{sec:ImprovedNewtonian}
The strategies that so far have been followed are twofold: Either one 
studies \emph{modified} Newtonian or special relativistic equations of 
motions for two point-particles with a force of mutual attraction 
(gravitational or electromagnetic). The modifications are derived from 
putting the system into a fixed standard-cosmological background (usually 
spatially flat) without back-reactions being taken into account. We shall 
discuss this approach in Sections \ref{sec:NewtonianApproach} and 
\ref{sec:DickePeebles}. Our discussion, based on \cite{Carrera.Giulini:2005}, 
complements the perturbative analysis in \cite{Cooperstock.etal:1998} which 
misses all orbits which are unstable under cosmological expansion (which do 
exist). In this respect we follow a very similar strategy as, e.g., in the 
more recent papers by~\cite{Price:2005} (the basic idea of which goes back at 
least to~\cite{Pachner:1963,Pachner:1964}) and also~\cite{Adkins.etal:2007}, 
though we think that there are also useful differences. We also supply 
quantitative estimates and clarify that the improved Newtonian equations 
of motion are written in terms of the right coordinates (non-rotating and 
metrically normalized). The purpose of this model is to develop a good 
physical intuition for the qualitative as well as quantitative features 
of any \emph{dynamical} effects involved. 

Eventually the Newtonian model just mentioned has to be understood as a 
limiting case of a genuinely relativistic treatment. For the gravitational 
case this is done in Section~\ref{sec:McVittie} (an alternative and more 
geometric derivation is given in Section~\ref{sec:DopplerTracking}), where 
we employ the McVittie metric to model a spherically symmetric mass embedded 
in a spatially flat Friedmann--\Lemaitre--Robertson--Walker (FLRW) universe. 
The geodesic equation is then, in a suitable limit, shown to lead to the 
improved Newtonian model discussed above (see also 
\cite{Carrera.Giulini:2005}). The same holds for the electromagnetic case, 
as we show in Section~\ref{sec:DickePeebles}. 
There we take a slight detour to also reconsider a classic argument by 
Dicke \& Peebles~\cite{Dicke.Peebles:1964}, which allegedly shows the 
absence of any relevant dynamical effect of global expansion. Its original 
form only involved the dynamical action principle together with some simple 
scaling argument. Since this reference is one of the most frequently cited 
in this field, and since the simplicity of the argument (which hardly 
involves any real analysis) is definitely deceptive, we give an independent 
treatment that makes no use of any hypothetical scaling rules for physical 
quantities other than spatial lengths and times. Our treatment, which 
follows \cite{Carrera.Giulini:2005}, also reveals that the original argument 
by Dicke \& Peebles is insufficient to discuss leading order effects of 
cosmological expansion. It is therefore also ineffective in its attempt 
to contradict~\cite{Pachner:1963,Pachner:1964}.

\subsection{Exact solutions}
\label{sec:ExactSolutions}
The other approach consists of finding exact solutions to Einstein's field 
equations for an inhomogeneous situation that, in the most simple case, 
models a single, quasi-localized, non-rotating, electrically neutral 
inhomogeneity within a FLRW universe. Using this inhomogeneous solution as 
background one can then study the motion of test particles (following 
geodesics in the background geometry) and, in particular, the influence of 
expansion on this motion. 

This approach can be subdivided into two strategies. The first tries 
to literally construct a new exact solution out of two known ones, so 
that the new solution contains a connected piece from each of the two 
old ones as isometric submanifolds. These we refer to as 
\emph{matched solutions}. This is relaxed in the second, more 
general strategy, where the new solution is merely required to 
somehow approximate the relevant part of each of the two old 
solutions in some region. These we refer to as \emph{melted solutions}.
Needless to say that melted solutions offer a much greater 
variety for construction than matched ones. However, it is 
also true that often not much is known about the proper physical 
interpretation of the former. In this respect the matching 
solutions usually provide a much clearer picture. 

According to the above requirements, in both cases we shall restrict 
attention to spherically symmetric spacetimes which, loosely 
speaking, approximate a FLRW solution of standard cosmology for 
`large radii' and a non-charged, non-rotating compact object 
characterized by the exterior Schwarzschild solution for `small radii'. 
(Clearly there must be some characteristic radius in terms of which 
`large' and `small' radii are defined.) 
Also, one often restricts attention to the spatially flat 
FLRW models for simplicity, which also seems justified in view 
of current cosmological data which are compatible with spatial 
flatness.

\subsubsection{Matched solutions}
\label{sec:MatchedSolutions}
A first approach to the matching idea was initiated by Einstein and Straus 
\cite{Einstein.Straus:1945,Einstein.Straus:1946} in 1945 and later worked 
out in more analytical detail by \Schuecking~\cite{Schuecking:1954}. Here 
the matched solution is really such that for radii smaller than a certain 
matching radius, $R_v$ (henceforth called the \emph{vacuole} or 
\emph{\Schuecking\ radius}), it is \emph{exactly} given by the Schwarzschild 
solution (exterior for a black hole, exterior plus interior for a star) and 
for radii above this radius it is \emph{exactly} given by a FLRW universe 
for dust matter without cosmological constant (this can be generalized, see 
below). The radius  $R_v$ is a function of the central gravitational mass 
$M$ and the cosmological mass-density $\varrho$, through the latter of which 
it also depends on the cosmological time~$t$. It is determined by 
\begin{equation}\label{eq:MatchingRadius}
  \frac{4\pi}{3}\,R_v^3\cdot\varrho=M\,.
\end{equation} 
This formula holds for flat as well as curved FLRW models if `radius' 
is taken to mean `areal radius', the definition of which is that a 
two-sphere of areal radius~$R$ has a proper surface area of $4\pi R^2$. 
In flat space the areal radius coincides with the proper radius (the 
geodesic distance between the center and any point on the sphere), so 
that $\frac{4\pi}{3}\,R_v^3$ is just the proper volume inside the sphere 
of radius~$R_v$ (cf.~Section~\ref{sec:spher-symm-match}). 
However, in backgrounds of positive (negative) curvature this expression 
is smaller (larger) than the proper volume (the proper volume grows 
faster (slower) with areal radius) and hence, for given $\varrho$, the 
left-hand side of (\ref{eq:MatchingRadius}) is also smaller (larger) 
than the proper mass of the dust contained within a sphere of areal 
radius~$R_v$. 

Here we recall that the gravitational mass of a lump of matter 
is not just proportional to the amount of matter (baryons) in that 
region. For example, the kinetic energy as well as the gravitational 
binding energy also contribute to the gravitational mass. This is 
expressed in formula (\ref{eq:MSE-int-expr}) of 
Appendix~\ref{sec:ss-perfect-fluids}, where further explanations will 
be provided. As is well known, the mathematical characterization of 
appropriate notions of quasi-local gravitational mass that would 
apply to general spacetimes is a notoriously difficult problem to
which various attempts for solutions exist; see 
\cite{SzabadosLivingReviews:2004} for the current status. However, 
in the spherically-symmetric case, to which we restrict attention, 
the so-called Misner--Sharp energy gives a satisfying and convenient 
concept of active gravitational mass. Its definition will be given 
in Section~\ref{sec:spher-symm-match} and more details, including its 
equality in value to the Hawking mass, are discussed in the 
Appendices~\ref{sec:MS-energy} and~\ref{sec:ss-perfect-fluids}.

The original construction by Einstein and Straus and its analytical 
completion by \Schuecking\ were quite complicated. We will give a 
much simpler and conceptually clearer description in 
Section~\ref{sec:ESS-vacuole}, using a suitable reformulation of the 
condition for the matching of solutions.  However, it is not hard to 
gain some intuitive understanding for the matching construction and 
the value of $R_v$ as defined by~(\ref{eq:MatchingRadius}). 
Let us for the moment restrict to the spatially flat case and consider 
the homogeneous and isotropic dust-filled universe at some moment of 
time~$t$. The dust within a 3-ball of proper radius $R_v$ represents an 
amount of matter of total mass $M$ as given by~(\ref{eq:MatchingRadius}). 
Now compress this amount of matter in a spherically symmetric fashion 
until it becomes a compact star or a black hole. In Newtonian gravity 
the gravitational field outside a spherically symmetric mass 
distribution only depends on the total mass and not on its radial 
density distribution. This is also true in General Relativity, 
which is essentially the content of Birkhoff's theorem.%
\footnote{An elegant proof of Birkhoff's theorem will appear as a 
by-product from our considerations in Appendix~\ref{sec:MS-energy}.} 
Hence the above compression preserves equilibrium (albeit an unstable 
one, see below) for the dust particles just outside the boundary-sphere of 
radius $R_v$. For radii smaller than $R_v$ we have the Schwarzschild 
solution (which is the unique non-trivial spherically symmetric 
vacuum solution according to Birkhoff's theorem) which therefore 
matches to the FLRW solution for $R \geq R_v$ at the boundary $R=R_v$ 
where the matter density is discontinuous. The spatial two-sphere 
$R=R_v$ is comoving with the Hubble flow, meaning that its proper 
surface area grows in case of expansion. Finally, in case of constant 
positive (negative) spatial curvature, (\ref{eq:MatchingRadius}) tells us 
that the matched Schwarzschild solution has a smaller (larger) mass than 
the mass that the amount of dust represents within the ball of areal 
radius $R_v$ within the FLRW universe. 

The Einstein--Straus model can be generalized in several ways. 
Instead of cutting out one ball, one can cut several non-overlapping ones 
and fill in the interiors with Schwarzschild geometries of appropriate 
masses. For obvious reasons these are sometimes referred to as 
`Swiss-Cheese models'. These, in turn, can be generalized to the cases of 
non-vanishing cosmological constant~\cite{Balbinot.etal:1988} or non-vanishing 
pressure~\cite{Bona.Stela:1987}. Finally, the Einstein--Straus model can be 
generalized to spherically symmetric but inhomogeneous 
\Lemaitre--Tolman--Bondi (LTB) cosmological backgrounds~\cite{Bonnor:2000b}. 

Since for the Einstein--Straus model the geometry within $R \leq R_v$ is 
exactly Schwarzschild (for vanishing cosmological constant) or 
Schwarzschild--de\,Sitter spacetime (for non-vanishing cosmological 
constant), it is clear that any dynamical system situated in this 
background geometry (no back reaction) only detects that part of 
the cosmic expansion that is due to a non-vanishing cosmological 
constant. In particular, for vanishing cosmological constant, the 
cosmic expansion that goes on outside the expanding vacuole $R=R_v$ is
not felt from within. Hence global expansion due to ordinary 
(localizable) matter can, in principle, be completely inhibited 
by local inhomogeneities. 

There are, however, several severe problems concerning the Einstein--Straus 
approach. First of all, it cannot provide a realistic model for the 
environment of small structures in our Universe, `small' meaning below the 
scales of galaxy clusters or superclusters. To see this, apply 
(\ref{eq:MatchingRadius}) to a spatially flat universe whose background 
matter density $\varrho$ is given by the critical density 
\begin{equation}\label{eq:DefCriticalDensity}
  \varrho_{\mathrm{crit}} := \frac{3H_0^2}{8\pi G}\,,
\end{equation}
where $G$ is Newton's constant. Then (\ref{eq:MatchingRadius}) gives 
\begin{equation}\label{eq:MatchingRadius-1}
  R_v = \bigl(R_S\,R^2_H\bigr)^{1/3}  
     \approx \, \left( \frac{M}{M_\odot} \right)^{1/3} 400\,\mathrm{ly} \,,
\end{equation}
where 
\begin{alignat}{2}
\label{eq:DefSchwarzschildRadius}
& R_S&&\,:=\,\frac{2GM}{c^2}
     \,\approx\,\frac{M}{M_{\odot}}\ 3\,\mathrm{km}\,,\\
\label{eq:DefHubbleRadius}
&R_H&&\,:=\,\frac{c}{H_0}
   \,\approx\,4\,\mathrm{Gpc}\approx 1.3\times 10^{23}\,\mathrm{km}\,,
\end{alignat}
are the \emph{Schwarzschild radius} for the mass $M$ and the 
\emph{Hubble radius}, respectively. 
$M_{\odot}=2\times 10^{30}\,\mathrm{kg}$ is the solar mass. 

For a single solar mass this gives a vacuole radius of almost 
400 lightyears, which is almost two orders of magnitude larger than the 
average distance of stars in our Galaxy. Therefore, the 
Swiss-Cheese model cannot apply at the scale of stars in galaxies.    
This changes as one goes to larger scales. For example, the Virgo 
cluster is estimated to have a mass of approximately $10^{15}$ solar 
masses~\cite{Fouque.etal}\footnote{Their considerations are based on 
a LTB model for the cluster.}, which makes its vacuole radius $10^5$ 
times larger than that for a single solar mass, so that it is approximately 
given by $10\,\mathrm{Mpc}$. 
This is just a little smaller than the average distance of groups 
and clusters of galaxies within the Virgo supercluster. Hence the 
Einstein--Straus approach might well give viable models above cluster 
scales. Similar conclusion can be drawn for the vacuole construction 
in LTB spacetimes~\cite{Bonnor:2000b}: There it is argued that the 
vacuole might be as big as the Local Group.

The Einstein--Straus solution (as well as its generalization for LTB 
spacetimes given by Bonnor) may also be criticized on theoretical grounds. 
An obvious one is its dynamical instability: slight perturbations of the 
matching radius to larger radii will let it increase without bound, slight 
perturbations to smaller radii will let it collapse. This can be proven 
formally (e.g.~\cite{Krasinski:1998}, Ch.\,3 and~\cite{Bonnor:2000b}) but 
it is also rather obvious, since $R_v$ is defined by the equal and opposite 
gravitational pull of the central mass on one side and the cosmological 
masses on the other. Both pulls increase as one moves towards their side, 
so that the equilibrium position must correspond to a local maximum of 
the gravitational potential. Another criticism of the Einstein--Straus 
solution concerns the severe restrictions under which it may be 
generalized to non spherically-symmetric situations; see 
e.g.~\cite{Senovilla.Vera:1997,Mena.etal:2002,Mena.etal:2003,Mena.etal:2005}.

\subsubsection{Melted solutions}
\label{sec:MeltedSolutions}
The above discussion shows that the Einstein--Straus approach does not 
give us useful information regarding the dynamical impact of cosmic 
expansion on structures well below the scales of galaxy clusters.
For this reason other exact solutions are sought. In this respect 
we wish to remind the reader on the following general aspect: In physics 
we are hardly 
ever in the position to mathematically rigorously model physically 
realistic scenarios. Usually we are at best either able to provide 
approximate solutions for realistic models or exact solutions for 
approximate models, and in most cases approximations are made on 
both sides. The art of physics then precisely consists in finding the 
right mixture in each given case. However, in this process our 
intuition usually strongly rests on the existence of at least some 
`nearby' exact solutions. Accordingly, one seeks exact solutions in 
General Relativity that, with some degree of physical approximation, 
model a spherically symmetric body immersed in an expanding 
universe. However, it is not as easy as one might think at first to 
characterize `body' and `immersed'.%
\footnote{In a linear theory, the `simultaneous presence' of two  
structures, like a local inhomogeneity in an `otherwise' homogeneous
background, naturally corresponds to the mathematical operation of
addition of the corresponding individual solutions. In a non-linear 
theory, however, no such simple recipe exists.} 
Clearly it is associated with some inhomogeneity 
in form of a spatial region with an overdense matter distribution, as 
compared to that of the approximately homogeneous distribution far out. 
But a body should also be quasi-isolated in order to be distinguishable 
form a mere local density fluctuation with smooth transition. 
Typical exact solutions that models the latter are the LTB solutions, 
in which matter is represented by pressureless dust that freely falls 
into the local overdense inhomogeneity. In some sense, these form 
the other extreme to the Einstein--Straus solutions in that they 
make the transition as smooth and mild as one wishes. Here we shall 
be interested in models that somewhat lie in-between these extremes.

An attempt to combine an interior Schwarzschild solution (representing a 
star) and a flat FLRW universe was made by Gautreau~\cite{Gautreau:1984b}.
Here the matter model consists of two components, a perfect fluid 
with pressure and equation of state $p=p(\varrho)$ outside the star, 
and the superposition of this with the star's dust-matter inside 
the star. However, Gautreau also made the assumption that the matter 
outside the star moves on radially infalling geodesics, which is only 
consistent if the pressure outside is spatially constant. Thus one 
is reduced to exact FLRW outside the star~\cite{vdBergh.Wils:1984} 
or the LTB model. (Further remarks may be found in~\cite{Krasinski:1998}, 
e.g.~p.\,113 and 165.) Other solutions, modeling a black hole in a 
cosmological spacetime, have been given in the literature. However, these 
solutions model objects which are either 
rotating~\cite{Vaidya:1977,Vaidya:1984,Ramachandra.etal:2003}, 
charged~\cite{Gao.Zhang:2004}, or both~\cite{Patel.Trivedi:1982}. 
Surveys on the subject of cosmological black holes 
are~\cite{Vishveshwara:2000} and~\cite{McClure:2006}. 
Further interesting solutions are given in~\cite{RajeshNayak.etal:2001} 
and in~\cite{Sultana.Dyer:2005,Faraoni.Jacques:2007}. The solutions 
proposed in the latter two works can be seen as generalizations of 
McVittie's model~\cite{McVittie:1933}, which we extensively discuss in 
Section~\ref{sec:McVittieModel}. A crucial feature of these solutions is, 
however, that the strength of the inhomogeneity%
\footnote{In Section~\ref{sec:McVittieInterpretation} we will identify the 
strength of the inhomogeneity with the Weyl part of the Misner--Sharp energy.}
varies in time, whereas for the McVittie model it remains constant. 
These solutions are of interest in their own right (for a detailed analysis 
see~\cite{Carrera.Giulini:2009a}), but our goal here is to focus on the 
effects due to cosmological expansion and not on the effects due to a 
changing strength of the central inhomogeneity. 
The solution proposed in the former work~\cite{RajeshNayak.etal:2001} is 
the melting of a Schwarzschild spacetime in an Einstein's static universe. 
This is a purely static solution whose properties and geodesics where studied 
in~\cite{Ramachandra.Vishveshwara:2002}. For our purposes, however, this 
spacetime is not interesting since it is asymptotically an Einstein universe, 
and hence not in agreement with the present picture of our Universe at large 
scales.


For these reasons in Section~\ref{sec:McVittieModel} we shall pay special 
attention to the McVittie model. This contains a distinguished central 
object in the sense that the mass within a sphere centered at the 
inhomogeneity splits into a piece that comes from the continuously 
distributed cosmological fluid (with pressure) and a constant piece 
that does not depend on the radius of the enclosing sphere; see our 
Eq.~(\ref{eq:McV-MSE-wEeq}). Moreover, the latter piece is also 
constant in time, meaning that the strength of the central inhomogeneity 
remains constant. By the way, McVittie's solutions contain the 
Schwarzschild--de\,Sitter one as a special case, which was recently used 
in the literature to estimate the effects of cosmological expansion on 
local systems~\cite{Kagramanova.etal:2006,Hackmann:Laemmerzahl:2008c}. 
In Section~\ref{sec:MotionInMcVittie} we show that in a suitable weak-field 
and slow-motion approximation the geodesic equation in McVittie spacetime 
reduces to the improved Newtonian equations discussed earlier. An alternative 
and more geometric derivation of the improved Newtonian equations for the 
McVittie case is presented in Section~\ref{sec:DopplerTracking} (see 
Eq.~(\ref{eq:McV-Newton-eq})).

\subsection{Kinematical effects}
\label{sec:Kinematical Effects}
\subsubsection{Timing and distances}
\label{sec:TimingAndDistances}
Neither the improved Newtonian model nor other general \emph{dynamical} 
arguments make any statement about possible \emph{kinematical} 
effects, i.e.~effects in connection with measurements of 
\emph{spatial distances} and \emph{time durations} in a cosmological 
environment whose geometry changes with time. This is an important 
issue if one wants to perform the tracking of a spacecraft, that is 
a `mapping out' of its trajectory, which basically means to determine 
its simultaneous spatial distance to the observer at given observer times. 
But we know from General Relativity that the concepts of `simultaneity' 
and `spatial distance' are not uniquely defined. This fact needs to be 
taken due care of when analytical expressions for trajectories, 
e.g.~solutions to the equations of motion in some arbitrarily chosen 
coordinate system, are compared with experimental findings. In those 
situations it is likely that different kinematical notions of simultaneity 
and distance are involved which need to be properly transformed into each 
other before being compared. For example, these transformations can 
result in additional acceleration terms involving the product 
(\ref{eq:HubbleTimesC}). Accordingly, there were claims in the 
literature that these kinematical effects could account for the PA; 
see e.g.~\cite{Palle:2005,Nottale:2003,Rosales:2002,%
Rosales.Sanchez-Gomez:1998,Ranada:2005,Nieto.etal:2005} and also 
statements to the contrary \cite{Laemmerzahl.etal:2006}. 
In Section~\ref{sec:SimultaneityComparison}, following 
\cite{Carrera.Giulini:2005}, we will confirm the existence of kinematical 
acceleration terms proportional to $H_0c$, but they are suppressed with 
additional powers of $\beta=v/c$, which renders them irrelevant as far as 
the PA is concerned.

\subsubsection{Doppler Tracking}
\label{sec:DopplerTrackingIntro}
The discussion in Section~\ref{sec:DopplerTracking} is based on 
\cite{Carrera.Giulini:2006b}. We explain in some detail the geometric 
theory for setting up the kinematical framework in which Doppler 
tracking should be discussed in order to properly speak of relative 
velocities and accelerations. This is a non-trivial issue which is, 
in our opinion, not properly appreciated in the literature on this 
subject (related general discussions are \cite{Bini.etal:1995,Bolos:2007}). 
Using this setting, we show how to derive an exact Doppler-tracking 
formula for a flat FLRW universe. This we use to give reliable 
upper bounds for kinematical effects caused by cosmic expansion. 
We also discuss generalizations to McVittie spacetime. Even though 
such effects exist, they again turn out to be irrelevant for the PA.

\section{Newtonian approach}
\label{sec:NewtonianApproach}
In order to gain intuition we consider a simple bounded system, say 
an atom or a planetary system, immersed in an expanding cosmos. 
We ask for the effects of this expansion on our local system. 
Does our system expand with the cosmos? Does it expand only partially? 
Or does it not expand at all? Here we shall not be concerned with the 
far more complex problem of how stable large-scale structures may 
emerge from unstable local gravitational dynamics in an expanding 
universe. This has been discussed in~\cite{Buchert.Dominguez:2005}
and references therein.

\subsection{Restricted two-body problem in an expanding universe}
We consider the dynamical problem of two bodies attracting 
each other via a force with $1/R^2$ fall-off. For simplicity 
we may think of one mass as being much smaller than the other 
one, though this is really inessential. One may think of 
two galaxies, a star and a planet, a planet and a spacecraft, 
or a (classical) atom given by an electron orbiting around a 
proton. The system is placed into an isotropically expanding 
ambient universe. We wish to know the leading order influence 
of the ambient expansion onto the relative two-body dynamics. 

To leading order, the global expansion is described by the 
simple linear \emph{Hubble law}, $\dot R=HR$, which states that 
the relative radial velocity of two comoving objects at a 
mutual distance $R$ grows proportional to that distance. 
More precisely, the term `distance' is here understood as 
the geodesic distance in the spacetime hypersurface of 
constant cosmological time~$t$ between its two intersection 
points with the two worldlines of the objects considered. 
$H$ denotes the 
\emph{Hubble parameter}, which generally depends on $t$ but not 
on space. It is given in terms of the \emph{scale parameter}, 
$a(t)$, via $H=\dot a/a$.

Taking into account $\dot H=(\ddot a/a)-H^2$, the 
acceleration that results from the Hubble law is simply given by 
\begin{equation}\label{eq:cosmological-acc}
  \ddot{R}|_{\mathrm{cosm. acc.}} 
  = \dot HR + H\dot R
  = \frac{\ddot{a}}{a}R
  = -q\,H^2\,R \,,
\end{equation}
where 
\begin{equation}\label{eq:def:DecelerationParameter}
  q:=-\,\frac{\ddot aa}{{\dot a}^2}
    =-\,\frac{\ddot a}{a}\,H^{-2}
\end{equation}
is the dimensionless \emph{deceleration parameter}. To get a feeling 
for the magnitude, we remark that for the current best-estimates 
for the parameters $H$ and $q$, 
$H_0\approx 70\,\mathrm{Km\cdot s^{-1}\cdot Mpc^{-1}}$ and 
$q_0\approx -0.6$ respectively, we get 
$\ddot a/a\approx 3\times 10^{-36}\,\mathrm{s}^{-2}$, which even 
at Pluto's distance of $40\,\mathrm{AU}$ merely amounts to a 
tiny outward pointing acceleration of 
$2\times 10^{-23}\,\mathrm{m\cdot s^{-1}}$. 

Now note that, in the sense of General Relativity, a body that is 
comoving with the cosmological expansion is moving on an inertial 
trajectory, i.e.~it is force free. On the other hand, according 
to Newton, a dynamical force is, by definition, the cause for 
\emph{deviations} from  inertial motion. In the 
present context this would mean that dynamical forces are the 
causes for deviations from the motions described by 
(\ref{eq:cosmological-acc}), which suggests that in Newton's 
law, $m\ddot{\vec x}=\vec F$, we should make the replacement 
\begin{equation}\label{eq:ReplaceAcceleration}
  \ddot R \mapsto \ddot R-(\ddot a/a)\,R
\end{equation}
in order to apply to the (sufficiently slow) motion of interacting 
point masses in an expanding universe. Note that this also applies 
to gravitational interactions in a Newtonian approximation in which 
gravity is considered to be a force in the above sense. 

As we will see, the replacement (\ref{eq:ReplaceAcceleration}) can be 
justified rigorously in a variety of contexts, like for gravitationally 
bound systems, using the equation of geodesic deviation in General 
Relativity. Whenever we attempt to justify the replacement 
(\ref{eq:ReplaceAcceleration}) we must not forget that the Newtonian 
equations of motion (without Coriolis and centrifugal type `forces' in 
them) necessarily refer to preferred systems of coordinates which are 
1)~locally non rotating,
2)~whose origin is freely falling, and 
3)~in which the coordinate values directly refer to (local) inertial 
time (time coordinate) and spatial geodesic distance (space coordinates), 
as measured by comoving clocks and rods.
This is achieved by using so-called Fermi normal coordinates (see, e.g., 
\S\,13.6 of \cite{Misner.Thorne.Wheeler:Gravitation}) in a neighborhood 
of a geodesic worldline---e.g.~that of the Sun or the proton. This is also 
the approach followed in~\cite{Cooperstock.etal:1998}. Note that a 
Fermi system of coordinates can be defined for worldlines of arbitrary 
acceleration and correspond to locally non-rotating frames, which may 
physically be realized by a system of at least two non collinear gyros 
in torque-free suspensions taken along the worldline. Along a geodesic, 
that is a worldline of zero acceleration, this system corresponds to 
a local inertial observer and is called Fermi \emph{normal}. 
The equation of geodesic deviation in these coordinates now gives the 
variation of the spatial geodesic distance to a neighboring geodesically 
moving object, e.g.~a planet or spacecraft. It reads%
\footnote{By construction of the coordinates, the Christoffel symbols 
$\Gamma^\mu_{\alpha\beta}$ vanish along the worldline of the first observer. 
Since this worldline is geodesic, Fermi--Walker transportation just reduces 
to parallel transportation.} 
\begin{equation}\label{eq:GeodDev}
  \frac{d^2x^k}{d\tau^2} + R^k_{\ 0l0 }x^l = 0\,.
\end{equation}
Here the $x^k$ are the spatial non-rotating normal coordinates whose 
values directly refer to the proper spatial distance. In these 
coordinates we further have~\cite{Cooperstock.etal:1998}
\begin{equation}\label{eq:GeodDevCurv}
  R^k_{\ 0l0 } = - \delta^k_l \, \ddot a/a
\end{equation}
on the worldline of the first observer, where the overdot refers to 
differentiation with respect to the cosmological time, which reduces 
to the proper time along the observer's worldline.    

Equations (\ref{eq:GeodDev},\ref{eq:GeodDevCurv}) simply state that 
in Fermi normal coordinates around one inertial observer another 
nearby inertial observer is radially accelerating away at a magnitude 
$(\ddot a/a)R$, just as envisaged before. In linear approximation, 
this acceleration has to be added to that resulting from the 
other metric perturbation that is caused by the mass at the position 
of the first observer. As a result, in the case of purely gravitational 
interaction, we obtain the equation of motion of a test particle 
(whose metric perturbation we neglect) in the gravitational field 
of a heavier object whose metric perturbation away from the FLRW 
cosmological background we approximate to linear order. 
In the case of charged objects, we neglect the metric 
perturbations caused by the masses of both charges as well as 
their electromagnetic field, and simply take into account their 
mutual electromagnetic interaction. 

Neglecting large velocity effects (i.e.~terms quadratic or higher 
order in $v/c$) we can now write down the equation of motion 
for the familiar two-body problem. After specification of a scale 
function $a(t)$, we get two ODEs for the variables $(R,\varphi)$, which 
describe the position\footnote{Recall that `position' refers to 
Fermi normal coordinates, i.e.~$R$ is the radial geodesic distance 
to the observer at $R=0$.} of the orbiting body with respect to the 
central one:
\begin{subequations}\label{eq:impr-N-eqs}
\begin{align}
  \label{eq:r-eq}
  &\ddot{R} = \frac{L^2}{R^3} - \frac{C}{R^2} + \frac{\ddot{a}}{a}R \\
  \label{eq:phi-eq}
  &R^2\dot{\varphi}=L \,.
\end{align}
\end{subequations}
These are the $(\ddot a/a)$--improved Newtonian equations of motion
for the two-body problem, where $L$ represents the (conserved) angular 
momentum of the planet (or electron) per unit mass and $C$ the 
strength of the attractive force. In the gravitational case $C=GM$, 
where $M$ is the mass of the central body, and in the electromagnetic 
case, for the electron-proton system, $C=e^2/m$ (Gaussian unit),
where $e$ and $m$ are the electron's charge and mass, respectively.
In Sections~\ref{sec:McVittie} and~\ref{sec:DickePeebles} we will show how 
to obtain (\ref{eq:impr-N-eqs}) in appropriate limits from the full general 
relativistic treatments. 

We now wish to study the effect the $\ddot a$ term has on the unperturbed 
Kepler orbits. We start with the obvious remark that this term results 
from the \emph{acceleration} and not just the expansion of the universe. 

Next we point out that in the concrete physical cases 
of interest, the time dependence of this term is negligible to a very 
good approximation. Indeed, putting $f:=\ddot{a}/a$, the relative time 
variation of the coefficient of $R$ in~(\ref{eq:cosmological-acc}) is 
$\dot{f}/f$. For an exponential scale function $a(t)\propto\exp(\lambda t)$ 
(vacuum-energy-dominated universe) this vanishes, and for a power law
$a(t)\propto t^\lambda$ (for example matter-, or 
radiation-dominated universes) this is $-2H/\lambda$, and hence of 
the order of the inverse age of the universe. If we consider a planet 
in the Solar System, the relevant time scale of the problem is the 
period of its orbit around the Sun. The relative error in the 
disturbance, when treating the factor $\ddot{a}/a$ as constant during 
an orbit, is hence smaller than $10^{-9}$. For atoms it is much smaller, 
of course. In principle, 
a time varying $\ddot{a}/a$ causes changes in the semi-major 
axis and eccentricity of Kepler orbits~\cite{Sereno.Jetzer:2007}.
But here we shall neglect the time-dependence of~(\ref{eq:cosmological-acc})
and set $\ddot{a}/a$ equal to a constant $A$. Because 
of~(\ref{eq:def:DecelerationParameter}) we have $A:=-q_0H_0^2$.
Then (\ref{eq:r-eq}) can be immediately integrated: 
\begin{equation}\label{eq:r-eq-first-order}
\frac{1}{2}\dot{R}^2 + U(R) = E\,,
\end{equation}
where the effective potential is 
\begin{equation}\label{eq:potential}
U(R)=\frac{L^2}{2R^2}-\frac{C}{R}-\frac{A}{2}R^2 \,.
\end{equation}
We will see below that the three parameters $(L,C,A)$ can 
be effectively reduced to two.

\subsection{Specifying the initial-value problem}
\label{sec:SpecifyingIVP}
Solutions of (\ref{eq:r-eq-first-order}) and (\ref{eq:phi-eq}) 
are specified by initial conditions 
$(R,\dot R,\varphi,\dot \varphi)(t_0)=(R_0,V_0,\varphi_0,\omega_0)$ 
at the initial time $t_0$. The discussion of the dynamical behavior of $R$ 
is most effectively done in terms of the effective potential. Moreover, 
since perturbations are best discussed in terms of dimensionless parameters, 
we also introduce a length scale and a time scale that appropriately 
characterize the dynamical perturbation and the solution to be perturbed. 

The length scale is defined as the radius at which the acceleration due to 
the cosmological expansion has the same magnitude as the two-body attraction. 
This happens precisely at the critical radius
\begin{equation}\label{eq:r-star}
  R_c := \left( \frac{C}{|A|} \right)^{1/3} \,.
\end{equation}
For $R<R_c$ the two-body attraction dominates, whereas for $R>R_c$ 
the effect of the cosmological expansion is the dominant one. 

In order to gain an understanding of the length scales of the critical radius
it is instructive to express it in terms of the physical parameters. In the 
case of gravitational interaction we have $C/\abs{A}=GM/(\abs{q_0}H_0^2)$ and 
thus
\begin{equation}\label{eq:r-star-grav}
  R_c = \left( \frac{R_S R_H^2}{2\abs{q_0}} \right)^{1/3} \,.
\end{equation}
Inserting the approximate value $q_0=-1/2$ of the present epoch, this 
reduces to the \Schuecking\ radius (\ref{eq:MatchingRadius-1}). 

In the electromagnetic case, e.g.~for an electron-proton system, we have 
$C/\abs{A}=(e^2/m)/(\abs{q_0}H_0^2)$. Defining, in analogy 
with~(\ref{eq:DefSchwarzschildRadius}), the length scale 
\begin{equation}\label{eq:electron-proton-length}
  R_e := \frac{2e^2}{mc^2} \approx 5.64 \cdot 10^{-15}\,\textrm{m} \,,
\end{equation}
the critical radius~(\ref{eq:r-star}) becomes 
\begin{equation}\label{eq:r-star-elm}
  R_c = \left( \frac{R_e R_H^2}{2\abs{q_0}} \right)^{1/3}
  \approx \, 30 \, \mathrm{AU}\,,
\end{equation}
where in the last step we inserted $q_0=-1/2$. This 
is about as big as the Neptune orbit! 

From~(\ref{eq:r-star-grav}) and~(\ref{eq:r-star-elm}) one sees that, 
in both cases, a larger (smaller) $\abs{q_0}$ implies a smaller (larger) 
critical radius, according to expectations. 

So much for the length scale. The time scale is defined to be 
the period of the unperturbed Kepler orbit (a solution to the 
above problem for $A=0$) of semi-major axis $R_0$. By Kepler's 
third law it is given by 
\begin{equation}\label{eq:T-Kepler}
  T_K := 2\pi \left( \frac{R_0^3}{C} \right)^{1/2} \,.
\end{equation}
It is convenient to introduce two dimensionless parameters which 
essentially encode the initial conditions $R_0$ and $\omega_0$.
\begin{alignat}{3}
\label{eq:def-lambda}
&\lambda &&\,:=\,\quad\left( \frac{\omega_0}{2\pi/T_K} \right)^2
         &&\,=\,\frac{L^2}{C R_0}\,, \\
\label{eq:def-alpha}
&\alpha  &&\,:=\,\mathrm{sign}(A)\left( \frac{R_0}{R_c} \right)^3
         &&\,=\, A\frac{R_0^3}{C} \, .
\end{alignat}
For close to Keplerian orbits $\lambda$ is close to one. For reasonably 
sized orbits $\alpha$ is close to zero. For example, in the Solar 
System, where $R_0 < 100$ AU, one has $|\alpha| < 10^{-16}$. For an atom 
whose radius is smaller than $10^4$ Bohr-radii we have $|\alpha|<10^{-57}$.

Now, defining 
\begin{equation}\label{eq:def-x}
  x(t):=R(t)/R_0\,,
\end{equation}
equations~(\ref{eq:r-eq-first-order}) and~(\ref{eq:phi-eq}) can be 
written as
\begin{align}
  \label{eq:x-eq}
  &\frac{1}{2}\dot{x}^2 + (2\pi/T_K)^2 \, u_{\lambda,\alpha}(x) = e \\
  \label{eq:phi-x-eq}
  &x^2\dot{\varphi} = \omega_0\,,
\end{align}
where $e:=E/r_0^2$ now plays the role of the energy-constant and 
where the reduced two-parameter effective potential $u_{\lambda,\alpha}$ 
is given by  
\begin{equation}\label{eq:potential-x}
u_{\lambda,\alpha}(x) := \frac{\lambda}{2x^2}
                        -\frac{1}{x}
                        -\frac{\alpha}{2} x^2\,.
\end{equation}
The initial conditions now read
\begin{equation}\label{eq:InitCond} 
  (x,\dot{x},\varphi,\dot{\varphi})(t_0) = (1,V_0/R_0,\varphi_0,\omega_0)\,.
\end{equation}
The point of introducing the dimensionless variables is that the 
three initial parameters $(L,C,A)$ of the effective potential 
could be reduced to two: $\lambda$ and $\alpha$. This will be 
convenient in the discussion of the potential.

\subsection{Discussion of the reduced effective potential}
\label{sec:DiscEffPot}
%
\begin{figure}[ht]
\centering
\includegraphics[width=\linewidth]{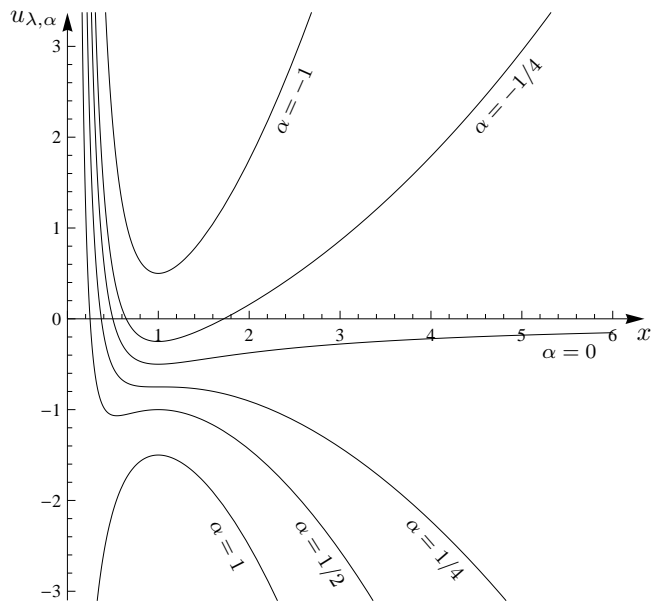}
\caption{The figure shows the effective potential $u_{\lambda,\alpha}$
for circular orbits (for which $\lambda=1-\alpha$) for some values of $\alpha$. 
The initial conditions are $x=1$ and $\dot x=0$ (see~(\ref{eq:def-x})).
At $x=1$ the potential has an extremum, which for $\alpha<1/4$ is a 
local minimum corresponding to stable circular orbits. For 
$1/4 \leq \alpha<1$ these become unstable. The value $\alpha=0$ corresponds 
to the Newtonian case.}
\label{fig:potential}
\end{figure}

Circular orbits correspond to extrema of the effective 
potential~(\ref{eq:potential}). Expressed in terms of the 
dimensionless variables this is equivalent to 
$u'_{\lambda,\alpha}(1) = -\lambda + 1 - \alpha = 0$.
By its very definition~(\ref{eq:def-lambda}), $\lambda$ is always 
non-negative, implying $\alpha \leq 1$. For negative $\alpha$ (decelerating 
case) this is always satisfied. On the contrary, for positive $\alpha$ 
(accelerating case), this implies, in view of~(\ref{eq:def-alpha}), the 
existence of a critical radius, given by $R_c$, beyond which no circular 
orbit exists.
These orbits are stable if the considered extremum is a true minimum, 
i.e.~if the second derivative of the potential evaluated at the 
critical value is positive. Now, 
$u''_{\lambda,\alpha}(1)=3\lambda-2-\alpha=1-4\alpha$, showing stability 
for $\alpha < 1/4$ and instability for $\alpha \geq 1/4$. For the 
accelerating case, in view of~(\ref{eq:def-alpha}), this implies that 
the circular orbits are stable iff $R_0$ is smaller than the critical value 
\begin{equation}\label{eq:def-r-sco}
  R_{\rm sco} := (1/4)^{1/3} R_c \approx 0.63 \,R_c \,,
\end{equation}
where `sco' stands for `stable circular orbits'.

Summarizing, we have the following situation: in the decelerating case 
(i.e.~for negative $\alpha$ or, equivalently, for negative $A$) stable 
circular orbits exist for every radius $R_0$; one just has to increase 
the angular velocity by some amount stated below in 
(\ref{eq:omega-circle}). On the contrary, in the  accelerating case 
(i.e.~for positive $\alpha$, or, equivalently, for positive $A$), 
we have three regions:
\begin{itemize}
\item $R_0 < R_{\rm sco}$, 
      where circular orbits exist and are stable, 
\item $R_{\rm sco} \leq R_0 \leq R_c$, 
      where circular orbits exist but are unstable, and 
\item $R_0 > R_c$, 
      where no circular orbits exist.
\end{itemize}
Generally, there exist no bounded orbits that extend beyond the 
critical radius $R_c$, the reason being simply that there is no 
$R>R_c$ where $U'(R)>0$. Bigger systems will just be slowly 
pulled apart by the cosmological acceleration and approximately 
move with the Hubble flow at later times.\footnote{This genuine 
non-perturbative behavior was not seen in the perturbation analysis 
performed in \cite{Cooperstock.etal:1998}.} Modifications of this 
strict qualitative distinction implied by time dependencies of $A$ 
in (\ref{eq:potential}) were discussed in~\cite{Faraoni.Jacques:2007}.

Turning back to the case of circular orbits, we now express the 
condition for an extrema derived above, $\lambda=1-\alpha$, in 
terms of the physical quantities, which leads to 
\begin{equation}\label{eq:omega-circle}
  \omega_0 = (2\pi/T_K)\sqrt{1-\mathrm{sign}(A)(R_0/R_c)^3} \,.
\end{equation}
This equation says that, in order to get a circular orbit, our planet, 
or electron, must have a smaller or bigger angular velocity according 
to the universe expanding in an accelerating or decelerating fashion, 
respectively. This is just what one would expect, since the effect of 
a cosmological `pulling apart' or `pushing together' must be 
compensated by a smaller or larger centrifugal forces respectively,
as compared to the Keplerian case. Equation~(\ref{eq:omega-circle}) 
represents a modification of the third Kepler law due to the 
cosmological expansion. In principle this is measurable, but it is 
an effect of order $(R_0/R_c)^3$ and hence very small indeed; 
e.g.~smaller than $10^{-17}$ for a planet in the Solar System.

Instead of adjusting the initial angular velocity as 
in~(\ref{eq:omega-circle}), we can ask how one has to modify 
$r_0$ in order to get a circular orbit with the angular velocity 
$\omega_0=2\pi/T_K$. This is equivalent to searching the minimum 
of the effective potential~(\ref{eq:potential-x}) for $\lambda=1$. 
This condition leads to the fourth order equation 
$\alpha x^4 - x + 1 = 0$ with respect to $x$. Its solutions can be 
exactly written down using Ferrari's formula, though this is not 
illuminating. For our purposes it is more convenient to solve it 
approximately, treating $\alpha$ as a small perturbation. 
Inserting the ansatz $x_{\mathrm{min}}=c_0+c_1\alpha+\Ord(\alpha^2)$ we get
$c_0=c_1=1$. This is really a minimum since 
$u''_{1,\alpha}(x_{\mathrm{min}})=1+\Ord(\alpha)>0$. Hence we have
\begin{equation}\label{eq:r-min}
R_{\rm min} = R_0
  \left( 1 + \mathrm{sign}(A)\left( \frac{R_0}{R_c} \right)^3 
           + \Ord\Big( (R_0/R_c)^6 \Big) 
  \right)
\end{equation}
This tells us that in the accelerating (decelerating) case the radii of 
the circular orbits with $\omega_0=2\pi/T_K$ becomes bigger (smaller), 
again according to physical expectation. As an example, the deviation 
in the radius for an hypothetical spacecraft orbiting around the Sun 
at 100~AU would be just of the order of 1~mm. Since it grows with the 
fourth power of the distance, the deviation at 1000~AU would be of 
the order of 10 meters.

\section{General-relativistic treatment for electromagnetically-bounded systems}
\label{sec:DickePeebles}
In this section we show how to arrive at (\ref{eq:impr-N-eqs})
from a relativistic treatment of an electromagnetically 
bounded two-body system embedded (without back-reaction) into 
an expanding (spatially flat) universe.
This implies solving Maxwell's equations in the cosmological 
background~(\ref{eq:FlatFLRW1}) for an electric point charge (the 
proton) and then integrate the Lorentz equations for the motion of
a particle (electron) in a bound orbit (cf.~\cite{Bonnor:1999}). 
Equation~(\ref{eq:impr-N-eqs}) then appears in an appropriate 
slow-motion limit. However, in order to relate this straightforward 
method to a famous argument of Dicke \& Peebles, we shall proceed 
by taking a slight detour which makes use of the conformal properties 
of Maxwell's equations.

\subsection{The argument of Dicke and Peebles}
In reference~\cite{Dicke.Peebles:1964} Dicke \& Peebles presented 
an apparently very general and elegant argument that purports
to show the insignificance of any dynamical effect of cosmological 
expansion on a local system that is either bound by electromagnetic or 
gravitational forces and which should hold true \emph{at any scale}. 
Their argument involves a rescaling of spacetime 
coordinates, $(t,\vec x)\mapsto (\lambda t,\lambda\vec x)$ and 
certain assumptions on how other physical quantities, most prominently
mass, behave under such scaling transformations. For example, they 
assume mass to transform like $m\mapsto\lambda^{-1}m$. However, their 
argument is really independent of such assumptions, as we shall 
show below. We work from first principles to clearly display all 
assumptions made. 

We consider the motion of a charged point particle in an 
electromagnetic field. The whole system, i.e.~particle plus 
electromagnetic field, is placed into a cosmological FLRW-spacetime 
with flat ($k=0$) spatial geometry. The spacetime metric reads 
\begin{equation}
\label{eq:FlatFLRW1}
\g = c^2\, \ed t^2 - a^2(t) (\ed r^2 + r^2\,\gStwo )\,,
\end{equation}
where 
\begin{equation}\label{eq:Def-gStwo}
  \gStwo = \ed\theta^2 + \sin^2\theta\,\ed\varphi^2
\end{equation}
denotes the metric on the unit two-sphere in standard coordinates.
We introduce conformal time, $t_c$, via 
\begin{equation}\label{eq:ConfTime} 
  t_c=f(t):=\int^t_k\frac{dt'}{a(t')}\,,
\end{equation}
by means of which we can write (\ref{eq:FlatFLRW1}) in a
conformally flat form
\begin{equation}\label{eq:FlatFLRW2}
  \g = a_c^2(t_c) (c^2\,\ed t_c^2-\ed r^2-r^2\,\gStwo ) 
     = a_c^2(t_c)\,\bfeta \,,
\end{equation}
where $\bfeta$ denotes the flat Minkowski metric.
Here we wrote $a_c$ to indicate that we now expressed the expansion 
parameter $a$ as function of $t_c$ rather than $t$, i.e. 
\begin{equation}\label{eq:A-c}
  a_c:=a\circ f^{-1}\,.
\end{equation}

The electromagnetic field is characterized by the tensor
$F_{\mu\nu}$, comprising electric and magnetic fields: 
\begin{equation}\label{eq:EM-Components}
  F_{\mu\nu}=
  \begin{pmatrix}
    0      & E_n/c\\
    -E_m/c & -\varepsilon_{mnj}B_j
  \end{pmatrix}\,.
\end{equation}
In terms of the electromagnetic four-vector potential, 
$A_\mu=(\varphi/c,-\vec A)$, one has 
\begin{equation}\label{eq:FourPotential} 
  F_{\mu\nu}=\partial_\mu A_\nu-\partial_\nu A_\mu
            =\nabla_\mu A_\nu-\nabla_\nu A_\mu\,,
\end{equation}
so that, as usual, $\vec E=-\vec\nabla\phi-\dot{\vec A}$. 
The expression for the four-vector of the Lorentz-force of a particle 
of charge $e$ moving in the field $F_{\mu\nu}$ is 
$e\,F^\mu_{\phantom{\mu}\nu}u^\nu$, where $u^\mu$ is the 
particle's four velocity. 

The equations of motion for the system Particle + EM-Field
follow from an action which is the sum of the action of the 
particle, the action for its interaction with the electromagnetic 
field, and the action for the free field, all placed in the 
background (\ref{eq:FlatFLRW1}). Hence we write:  
\begin{equation}\label{eq:ActionSum}
  S=S_P+S_I+S_F\,,
\end{equation} 
where
\begin{subequations}
\begin{alignat}{2}
\label{eq:P-Action}
&S_P &&\,=\,-mc^2\int_z d\tau
       \,=\,-mc\int\sqrt{g(z',z')}\,d\lambda\,,\\
&S_I &&\,=\,-\,e\int_zA_\mu\,dx^\mu
       \,=\,-\,e\int A_\mu(z(\lambda))z'^\mu\,d\lambda\nonumber\\
\label{eq:I-Action}
&    &&\,=\,-\,\int d^4x\,A_\mu(x)\int\,d\lambda\ e\ 
            \delta^{(4)}(x-z(\lambda))\ z'^\mu\,,\\
&S_F &&\,=\,\frac{-1}{4}\int d^4x\,\sqrt{- g}\;
          g^{\mu\alpha}g^{\nu\beta}\,F_{\mu\nu}F_{\alpha\beta}\nonumber\\
\label{eq:F-Action}
& &&\,=\,\frac{-1}{4}\int d^4x\,\eta^{\mu\alpha}
       \eta^{\nu\beta}\,F_{\mu\nu}F_{\alpha\beta}\,.
\end{alignat}
\end{subequations}
Here $\lambda$ is an arbitrary parameter along the worldline 
$z:\lambda\mapsto z(\lambda)$ of the particle, and $z'$
the derivative $dz/d\lambda$. The differential of the 
proper time along this worldline is 
\begin{equation}\label{eq:DefEigentime}
  d\tau=\sqrt{g(z',z')}\,d\lambda
       =\sqrt{g_{\mu\nu}(z(\lambda))\tfrac{dz^\mu}{d\lambda}
        \tfrac{dz^\nu}{d\lambda}}\,d\lambda\,.
\end{equation}
It is now important to note that 1)~the background metric $g$
does not enter (\ref{eq:I-Action}) and that (\ref{eq:F-Action})
is conformally invariant (in 4 spacetime dimensions only!). Hence 
the expansion factor, $a(t_c)$, does not enter these two 
expressions. For this reason we could write (\ref{eq:F-Action})
in terms of the flat Minkowski metric, though it should be kept 
in mind that the time coordinate is now given by conformal time
$t_c$. This is \emph{not} the time read by standard clocks that 
move with the cosmological observers, which rather show the 
cosmological time $t$ (which is the proper time along the 
geodesic flow of the observer field $\bpartial/\bpartial t$). 

The situation is rather different for the action (\ref{eq:P-Action})
of the particle.  Its variational derivative with respect to
$z(\lambda)$ is 
\begin{equation}\label{eq:VarDerS_p1}
  \frac{\delta S_p}{\delta z^\mu(\lambda)}= -mc\,\left\{
  \frac{\tfrac{1}{2}g_{\alpha\beta,\mu}\,z'^\alpha z'^\beta}{\sqrt{g(z',z')}} 
 -\frac{d}{d\lambda}\left[\frac{g_{\mu\alpha}z'^\alpha}{\sqrt{g(z',z')}}
  \right]\right\} \, .
\end{equation} 
We now introduce the \emph{conformal proper time}, $\tau_c$, via 
\begin{equation}\label{eq:CobfPropTime}
  d\tau_c = (1/c)\,\sqrt{\eta(z',z')}\,d\lambda
          = (1/ca)\,\sqrt{g(z',z')}\,d\lambda\,.
\end{equation}
We denote differentiation with respect to $\tau_c$ by an overdot, 
so that e.g.~$z'/\sqrt{g(z',z')}=\dot z/ca$. Using this to replace 
$z'$ by $\dot z\sqrt{g(z',z')}/ca$ and also $g$ by $a^2\eta$ in 
(\ref{eq:VarDerS_p1}) gives 
\begin{equation}\label{eq:VarDerS_p2}
  \frac{\delta S_p}{\delta z^\mu(\lambda)}= 
  \frac{\sqrt{g(z',z')}}{ac}\,ma\,\left\{
  \eta_{\mu\alpha}\ddot z^\alpha - P^\alpha_\mu\phi_{,\alpha}\right\}
\end{equation} 
where we set 
\begin{equation}\label{eq:Abbrev}
  a=:\exp(\phi/c^2)\quad\text{and}
  \quad
  P^\alpha_\mu := \delta^\alpha_\mu - 
  \frac{\dot z^\alpha\dot z^\nu}{c^2}\eta_{\nu\mu}\,.
\end{equation}
Recalling that 
$\delta S_P=\int\frac{\delta S_p}{\delta z^\mu(\lambda)}\delta z^\mu d\lambda=
\int\frac{\delta S_p}{\delta z^\mu(\tau_c)}\delta z^\mu d\tau_c$ and 
using (\ref{eq:CobfPropTime}), (\ref{eq:VarDerS_p2}) is equivalent to 
\begin{equation}\label{eq:VarDerS_p3}
  \frac{\delta S_p}{\delta z^\mu(\tau_c)}= 
  ma\bigl(\ddot z^\alpha - P^\alpha_\mu\phi_{,\alpha}\bigr)\,,
  \end{equation} 
where from now on we agree to raise and lower indices using the 
Minkowski metric, i.e.~$\eta_{\mu\nu}=\text{diag}(1,-1,-1,-1)$ in Minkowski 
inertial coordinates.
 
Writing (\ref{eq:I-Action}) in terms of the conformal proper time 
and taking the variational derivative with respect to $z(\tau_c)$ 
leads to $\delta S_I/\delta z^\mu(\tau_c)=-eF_{\mu\alpha}{\dot z}^\alpha$, 
so that 
\begin{equation}\label{eq:VarDerSwrtZ}
  \frac{\delta S}{\delta z^\mu(\tau_c)}
  =ma\bigl(\ddot z_\mu - P^\alpha_\mu\phi_{,\alpha}\bigr)
  -e\,F_{\mu\alpha}{\dot z}^\alpha\,.
\end{equation}

The variational derivative of the action with respect to the 
vector potential $A$ is 
\begin{equation}\label{eq:VarDerSwrtA}
  \frac{\delta S}{\delta A_\mu(x)}
  =\partial_{\alpha}F^{\mu\alpha}(x)
  -e\,\int d\tau_c\,\delta(x-z(\tau_c))\,\dot z^\mu(\tau_c)\,. 
\end{equation}

Equations~(\ref{eq:VarDerSwrtZ}) and~(\ref{eq:VarDerSwrtA}) show that the 
fully dynamical problem can be treated as if it were situated in static 
flat space. The field equations that follow from (\ref{eq:VarDerSwrtA}) 
are just the same as in Minkowski space. Hence we can calculate the Coulomb 
field as usual. On the other hand, the equations of motion receive two 
changes from the cosmological expansion term: the first is that the mass 
$m$ is now multiplied with the (time-dependent!) scale factor $a$, the 
second is an additional scalar force induced by $a$. Note that all spacetime 
dependent functions on the right hand side are to be evaluated at the 
particle's location $z(\tau_c)$, whose fourth component corresponds to 
$ct_c$. Hence, writing out all arguments and taking into account that the 
time coordinate is $t_c$, we have for the equation of motion 
\begin{subequations}\label{eq:DP-Motion}
\begin{alignat}{1}
\ddot z^\mu
\ =\ &\frac{e}{ma_c(z^0/c)}\ F^\mu_{\phantom{\mu}\alpha}(z)\dot z^\alpha
\nonumber\\
\label{eq:DP-Motion1}
     & - \bigl(-c^2\eta^{\mu\alpha}+\dot z^\mu\dot z^\alpha\bigr)
         \partial_\alpha\ln a_c(z^0/c)\\
\ =\ &\frac{e}{ma_c(z^0/c)}\ F^\mu_{\phantom{\mu}\alpha}(z)\dot z^\alpha
\nonumber\\
\label{eq:DP-Motion2}
     & - \bigl(-c\eta^{\mu 0}+\dot z^\mu\dot z^0/c\bigr)
         \,a'_c(z^0/c)/a_c(z^0/c)\,,  
\end{alignat}    
\end{subequations}
where $a'_c$ is the derivative of $a_c$.

So far no approximations were made. Now we write $\dot z^\mu=\gamma(c,\vec v)$,
where $\vec v$ is the derivative of $\vec z$ with respect to the 
conformal time $t_c$, henceforth denoted by a prime, and 
$\gamma=1/\sqrt{1-v^2/c^2}$. Then we specialize to slow motions, 
i.e.~neglect effects of quadratic or higher powers in $v/c$ (special
relativistic effects). For the spatial part of (\ref{eq:DP-Motion2}) we get
\begin{equation}\label{eq:DP-Motion3}
  \vecpp{z} + \vecp{z}\ (a'_c/a_c)
  =\frac{e}{ma_c}\bigl( \vec E + \vecp{z} \times \vec B \bigr)\,,
\end{equation}
where we once more recall that the spatial coordinates used here are 
the comoving (i.e.~conformal) ones and the electric and 
magnetic fields are evaluated at the particle's position $\vec z(t_c)$. 

From the above equation we see that the effect of cosmological expansion
in the conformal coordinates shows up in two ways: first in a time 
dependence of the mass which scales with $a_c$, and, second, in the
presence a friction term. Dicke \& Peebles neglect the friction term
and simply conclude as follows: In the adiabatic approximation, which 
is justified if typical time scales of the problem at hand are short 
compared to the world-age (corresponding to small $\varepsilon_{\sss 2}$ 
in (\ref{eq:ExpParameters2})), the time-dependent mass term leads to a 
time varying radius in comoving (or conformal) coordinates of 
$r(t_c)\propto 1/a_c(t_c)$. Hence the physical radius (given by the 
cosmological geodesically spatial distance), $r_*=a_c r$, stays 
constant in this approximation. Hence, within \emph{this} 
approximation, electromagnetically bound systems do not feel 
\emph{any} effect of cosmological expansion.

But what does `this approximation' refer to? We will see that it amounts 
to neglecting precisely the leading order contributions. This is easy to
see if we cast (\ref{eq:DP-Motion3}) into  physical coordinates, 
given by the cosmological time $t$ and the cosmological geodesic 
spatial distance $r_*:=a(t)r$. We have $dt_c/dt=1/a$ 
and the spatial geodesic coordinates are $\vec y:= a(t) \vec z$. 
Denoting by an overdot the time derivative with respect to $t$, 
the two terms on the left hand side of~(\ref{eq:DP-Motion3}) become
\begin{subequations}\label{eq:zPrimesInyDot}
\begin{alignat}{2}
\label{eq:zPrimesInyDot-a}
  \vecpp z&\,=&&a(\vec y H^2 - \dot{\vec y}H) + 
                a(\ddot{\vec y} - \vec y\,\ddot a/a)\,,\\
\label{eq:zPrimesInyDot-b}
  \vecp z(a_c'/a_c) &\,=-&&a(\vec yH^2-\dot{\vec y}H)\,,
\end{alignat}
\end{subequations}
where $H=\dot a/a$. This shows that the friction term cancels against the 
first-order derivative terms in $\vec y$ and $a$ that one gets in rewriting 
the left-hand side of~(\ref{eq:DP-Motion3}) in physical coordinates.%
\footnote{Since the friction term cancels, the critical 
remark [27] in \cite{Adkins.etal:2007} regarding its magnitude is based 
on a misunderstanding.} 
The only additional term next to $\ddot{\vec y}$ that survives is precisely 
the familiar acceleration term~(\ref{eq:cosmological-acc}). Inserting 
(\ref{eq:zPrimesInyDot}) into (\ref{eq:DP-Motion3}), whose right-hand side 
we now specialize to a pure electric Coulomb field, 
$\vec E(\vec{z})=Q\vec{z}/|\vec{z}\,|^3$ and $\vec B(\vec{z}) = 0$, we 
arrive at 
\begin{equation}\label{eq:DP-Motion4}
  \ddot{{\vec y}} - \vec{y}\ (\ddot a/a) = \frac{eQ}{m |\vec{y}|^3}\vec{y} \,.
\end{equation}
After introducing polar coordinates in the orbital plane we exactly 
get~(\ref{eq:impr-N-eqs}).

\subsection{Exact condition for non-expanding circular orbits}
\label{sec:nonExpCirOrbitsFLRW}
In~\cite{Bonnor:1999} a necessary and sufficient condition for the 
existence of non-expanding orbits is derived for the electron-proton 
system in a spatially flat FLRW spacetime. Here `non-expanding' is 
defined as of constant areal radius. This condition follows directly from the 
Lorentz equation of motion for the electron in the external electric field of 
the proton, the normalization condition of the electron's four-velocity, and 
the condition of constancy of the areal radius. In our notation, introducing 
the dimensionless quantities $h(t):=RH(t)/c$, $l:=L/Rc$, and $\mu:=R_e/2R$, 
the conditions for the existence of non-expanding circular orbits reads as 
follows:
\begin{equation}\label{eq:circular-orbit-exact}
  \frac{R}{c} \dot h = \frac{(1-h^2)^{3/2}}{(1+l^2)^{1/2}}
  \left( \mu - \frac{l^2+h^2}{\sqrt{(1+l^2)(1-h^2)}} \right) \,.
\end{equation}
Recall that $R_e$ is defined in~(\ref{eq:electron-proton-length}) and $H(t)$ 
and $L$ denote the Hubble function and, respectively, the (conserved) 
electron's angular momentum per unit mass. 
The above condition is a first-order autonomous ODE for the function $h(t)$, 
and hence for the Hubble function $H(t)$. This is the constraint on the 
spacetime (more precisely, on the scale factor $a(t)$) that one gets by 
imposing the existence of non-expanding circular orbits for two oppositely 
charged point masses.  If such orbits exist, (\ref{eq:circular-orbit-exact}) 
amounts to the generalization of Kepler's third law to FLRW spacetimes, which 
here gives a relation between the scale function on one hand and the orbital 
parameters $R$ and $L$ as well as the field-strength parameter $R_e$ on the 
other. Recall that in Newtonian physics the third Kepler law is, in our 
notation, simply given by $l^2=\mu$. 

The easiest solutions of~(\ref{eq:circular-orbit-exact}) are of course the
stationary ones, that is with $h(t)\equiv h_0$, for some constant 
$h_0$. This means that the scale factors is exponentially expanding, 
\begin{equation}\label{eq:circ-orbit-dS}
  a(t)=a_0\exp(H_0 t) \,,
\end{equation}
where $H_0:=h_0 c/R$ and $a_0$ is some positive constant. In other 
words, the spacetime is given by the de\,Sitter solution 
($\Lambda$-dominated universe). In this 
case~(\ref{eq:circular-orbit-exact}) reduces to 
\begin{equation}\label{eq:circular-orbit-exact-stationary}
  \frac{l^2+h_0^2}{\sqrt{(1+l^2)(1-h_0^2)}} = \mu \,.
\end{equation}
Notice that a larger Hubble parameter, hence a larger $h_0$, makes the 
l.h.s.~larger. Consequently, (\ref{eq:circular-orbit-exact-stationary}) 
tells us that with a larger Hubble parameter we must give to the electron 
a \emph{smaller} angular velocity (smaller $l$) in order to keep it on a 
non-expanding circular orbit with the \emph{same} radius. This, according 
to intuition, is in order to compensate the extra cosmological pull 
with a reduced centrifugal term. In case of Minkowski spacetime ($h_0=0$) 
the above relation reads $l^2/\sqrt{1+l^2}=\mu$, hence one can interpret 
the factor $1/\sqrt{1+l^2}$ as a special-relativistic correction to the 
Newtonian relation $l^2=\mu$. 
The largest radius at which, in an FLRW spacetime with exponentially-growing 
scale factor, there is a non-expanding orbit follows 
from~(\ref{eq:circular-orbit-exact-stationary}) in the limit $l\to 0$. 
In this limit the condition reduces to $h_0^2/\sqrt{1-h_0^2}=\mu$, which, 
for small parameters $h_0$, simplifies to $h_0^2 \approx \mu$. Solving for 
$R$ this gives the radius $(R_e R_H^2/2)^{1/3}$, which, taking into account 
that $q_0=-1$ because of~(\ref{eq:circ-orbit-dS}), exactly corresponds to the 
critical radius~(\ref{eq:r-star-elm}). 

The other (non-stationary) solutions of~(\ref{eq:circular-orbit-exact}) 
can also be found. After separation of variables and an elementary 
integration one gets $t$ as function of $h$ in terms of trigonometric 
functions composed with inverse hyperbolic functions.  This exact 
expression is again not very illuminating and cannot generally be explicitly 
inverted so as to obtain $h$ in terms of elementary functions of $t$. 
However, if we make use of the smallness of the parameters $\mu$, 
$l^2$, and $h^2$, a leading-order expansion in these quantities gives a 
much simpler expression for $t(h)$ which can be explicitly inverted. 
In fact, this approximate solution $h(t)$ is obtained much quicker by solving 
(\ref{eq:circular-orbit-exact}) with the right-hand side being replaced 
with its leading-order expansion in the mentioned quantities, that is, 
by solving
\begin{equation}\label{eq:h-ODE-approx}
  \frac{R}{c}\dot h = \mu-l^2 - h^2 \,.
\end{equation}
Here $\mu-l^2$ is a constant which depends on the orbit parameters. 
One must now distinguish between three cases: 
(a)~$\mu-l^2=:\kappa^2>0$ for some positive $\kappa$,
(b)~$\mu-l^2=:-\nu^2<0$ for some positive $\nu$, and 
(c)~$\mu-l^2=0$. Recalling the Newtonian relation $l^2=\mu$, orbits in 
the three cases have an angular momentum which is, respectively, smaller, 
bigger, and equal to the Newtonian one. Integrating~(\ref{eq:h-ODE-approx})
we get, putting w.l.o.g.~$t_0=0$, 
$h(t)=\kappa \tanh(\kappa c t/R)$, 
$h(t)=-\nu \tan(\nu c t/R)$, and 
$h(t)=R/ct$, for the cases $(a)$, $(b)$, and $(c)$, respectively.
Then, integrating once and exponentiating the result, we get the corresponding 
scale functions: 
\begin{enumerate}[(a)]
\item Case $\mu-l^2=:\kappa^2>0$ 
(non-expanding orbits have sub-Newtonian angular momentum)
\begin{subequations}\label{eq:circ-orbit-approx-a-sol}
\begin{alignat}{3}
  &a(t) = a_0 \cosh\left( \frac{\kappa c t}{R} \right)\,, \qquad{  }
  &&t\in [0,\infty)\,.
\intertext{\item Case $\mu-l^2=:-\nu^2<0$ 
(non-expanding orbits have super-Newtonian angular momentum)}
  &a(t) = a_0 \cos\left( \frac{\nu c t}{R} \right)\,, \qquad{ }
  &&t\in \left[ 0,\frac{\pi R}{2\nu c} \right)\,.
\intertext{\item Case $\mu-l^2=0$ 
(non-expanding orbits have Newtonian angular momentum)}
  &a(t) = a_0 t \,, \qquad{  }
  &&t\in (0,\infty)\,.
\end{alignat}
\end{subequations}
\end{enumerate}
In all three cases (a), (b), and (c) $a_0$ is a positive constant and the 
acceleration term $\ddot a/a$ is a constant which is positive, negative, 
and zero, respectively. Hence, as one would intuitively expect, the 
non-expanding orbits have an angular momentum which is smaller, larger, or 
equal the Newtonian one, depending on whether the acceleration factor 
$\ddot a/a$ is positive, negative, or zero.

\section{General-relativistic treatment for gravitationally-bounded systems}
\label{sec:McVittie}
As advertised in Section~\ref{sec:ExactSolutions}, we now wish to discuss 
exact solutions that may represent quasi-isolated spherically-symmetric 
gravitating systems `embedded' into cosmological spacetimes. As regards 
the meaning of `embedded' we distinguish between the strategies of 
`matching' and `melting', as outlined in Sect.\,\ref{sec:ExactSolutions}.

\subsection{Spherically-symmetric matchings}
\label{sec:spher-symm-match}
The complexity and non-linearity of Einstein's equations make it 
a very difficult task to construct a suitable variety of exact 
solutions which serve as realistic models for actual physical 
situations. Often exact solutions are only known for highly 
idealized situations, typically with high degrees of symmetry, 
in which the field equations sufficiently simplify. One way to 
construct new solutions (in a suitable sense, see below) from 
old ones is to glue them across suitably chosen hypersurfaces 
along which the matter distribution may become singular due to 
surface layers. This approach was pioneered by Lanczos in the 
early 1920s~\cite{Lanczos:1924} and put into geometric form 
by Darmois~\cite{Darmois:1927} and \cite{Israel:1966}; see 
also \S\,21.13 of~\cite{Misner.Thorne.Wheeler:Gravitation}. 
In this section, under the assumption of spherically symmetry, 
we present a new alternative set of conditions which are 
\emph{equivalent} to the old ones. The new conditions only
involve scalar quantities, are easy to verify, and have good 
physical interpretations. More details are contained 
in~\cite{Carrera:2009}.

Here we shall restrict to piecewise continuous matter distributions 
without singular ($\delta$-distribution like) surface layers, as e.g.~in 
the presence of stars with sharply defined surfaces. Einstein's 
equations can the be satisfied for piecewise twice continuously 
differentiable fields, if the field equations at the location of the 
matching hypersurface are replaced by their one-dimensional 
$\varepsilon$-interval integrals in normal direction to the 
hypersurface. The condition that two twice continuously 
differentiable solutions (in the ordinary sense) can be matched 
into a piecewise twice continuously differentiable solution 
(in the re-interpreted sense just explained) is then simply 
given by the so-called 
\begin{Darmois}\label{thm:DJC}
For a non-null matching hypersurface $\varGamma$, $(i)$~the induced metric 
$\gG$ and $(ii)$~the extrinsic curvature $\bK_\varGamma$ shall be continuous 
through~$\varGamma$.
\end{Darmois}
Let us pause for a moment to say a few more words about the notion 
of `continuity through $\varGamma$'. Gluing together two pieces of
spacetimes means the following: Initially one has two spacetimes, 
say $(\M^+,\g^+)$ and $(\M^-,\g^-)$, with oriented boundaries 
$\varGamma^+$ and $\varGamma^-$, respectively. Given a diffeomorphism 
$\phi:\phi:\varGamma^+\rightarrow \varGamma^-$ between the boundaries, 
the glued spacetime is the quotient of the disjoint union of $\M^+$ 
and $\M^-$ under the identification of each point of $p\in \varGamma^+$ 
with $\phi(p)\in\varGamma^-$. The \emph{matching hypersurface $\varGamma$} 
is now the common image of $\varGamma^+$ and $\varGamma^-$ after 
identification in the quotient spacetime. Now, a tensor field $\bT$ 
is said to be \emph{continuous through $\varGamma$} if $\bT|_{\varGamma^+}$ 
equals $\bT|_{\varGamma^-}$ under the push-forward action of the 
diffeomorphism $\phi$, hence if 
$\phi_*(\bT|_{\varGamma^+})=\bT|_{\varGamma^-}$.

Let us now return to the DJC and, in particular,  their physical 
interpretation. If $\bn$ is a continuous choice of unit normal 
of $\varGamma$, it implies that 
\begin{equation}\label{eq:matter-momentum-flow-cont}
  \bT(\bn,\cdot) \; \text{ is continuous through } \varGamma\,.
\end{equation}
This follows directly from the expressions~(\ref{eq:Ein-jc}) of the 
Einstein tensor given in Appendix~\ref{sec:submanifolds}. If $\varGamma$ 
is timelike and hence $\bn$ spacelike, (\ref{eq:matter-momentum-flow-cont}) 
just states the continuity of the normal components for the 
energy-momentum flux-densities, whereas their tangential components 
together with the energy density may jump across $\varGamma$. 
In the absence of surface layers this continuity condition is just 
a physically obvious consequence of local energy-momentum conservation,
whereas jumps in, say, the energy-density must clearly be allowed for. 
For completeness we note that for a spacelike matching surface 
(\ref{eq:matter-momentum-flow-cont}) states the continuity of the 
densities of energy and momentum as measured by an observer moving 
along $\bn$ (taken to be future pointing), whereas the corresponding 
currents may jump.

Let now restrict our attention to spherical symmetric spacetimes 
glued along hypersurfaces of spherical symmetry. This means that 
the latter are left invariant, as set, under the action of $SO(3)$. 
We recall that the structure of a spherically symmetric spacetime 
is that of a warped product $\M=\B\times_R\Stwo$ of a two-dimensional 
Lorentzian manifold $\B$ and the two-sphere by means of the warping 
function, $R:\B\rightarrow\mathbb{R}_+$, called the \emph{areal radius}.
The matching hypersurfaces are then of the form 
$\varGamma=\gamma\times_R\Stwo$, where $\gamma$ is the projection 
of $\varGamma$ into $\B$ and is called the \emph{matching curve}. The 
DJC should then reduce to appropriate conditions along the curve 
$\gamma$. Indeed, Theorem~\ref{thm:equiv-jc} below shows that, 
in the spherically symmetric case, the DJC are equivalent to the 
following

\begin{SSJC}\label{thm:SSJC}
Let $\varGamma$ be a smooth, non-null, spherically symmetric matching 
hypersurface between two spherically symmetric spacetimes and $\bn$ a 
continuous choice of unit normal vector field on $\varGamma$. Denote 
with $\gamma$ the projection of $\varGamma$ onto $\B$. Moreover, let 
$\bv$ the (unique up to a sign) spherically symmetric, unit vector 
field on $\varGamma$ orthogonal to $\bn$. The following four functions 
\begin{compactenum}[$\hspace{3ex} (i)$]
\item the arc-length of $\gamma$,
\item the extrinsic curvature of $\gamma$ in $\B$: $\g(\bn,\bnabla_\bv\bv)$,
\item the areal radius $R$, 
\item the Misner--Sharp energy $E$,
\end{compactenum}
shall be continuous through the matching curve $\gamma$.
\end{SSJC}

The Misner--Sharp (MS) energy~\cite{Misner.Sharp:1964} 
and~\cite{Hernandez.Misner:1966}, which we will denote by $E$, 
is a concept of quasi-local mass that can be defined in presence of 
spherical symmetry, and which proves useful for computational and 
interpretational purposes. It is a function defined in purely 
geometrical manner as follows. Given a point $p$ of spacetime, 
compute the sectional curvature $K$ of the plane tangent to the 
two-dimensional $SO(3)$-orbit through $p$ and multiply this with minus%
\footnote{The minus sign here is just a relict of our signature 
choice.} one-half of the third power of the areal radius: 
\begin{equation}\label{eq:MS-energy-original-def}
  E := -\tfrac{1}{2} R^3 K \,.
\end{equation}
From (\ref{eq:sectional-curv-tangent-S2}) we immediately read 
off 
\begin{equation}\label{eq:MS-energy}
  E = \frac{R}{2} \big( 1 + \sprod{\ed R}{\ed R} \big) \,,
\end{equation}
which provides a convenient expression for the computation of the MS 
energy. We shall show in Appendix~\ref{sec:spherical-symmetry} that 
the MS energy is the charge of a conserved current and how it depends on the 
energy-momentum tensor for the matter. There we will also briefly discuss
its Newtonian limit. This allows 
to interpret it as amount of active gravitational energy contained in the 
interior of the sphere of symmetry ($SO(3)$-orbit) through $p$. There we 
also show that the MS energy at $p$ is equal to the Hawking quasi-local 
mass of the two-sphere of symmetry ($SO(3)$-orbit) through $p$ and hence 
converges to the Bondi-mass at null infinity and, in an asymptotically 
flat spacetime, to the ADM mass at spatial infinity (for the latter two 
issues see \cite{SzabadosLivingReviews:2004} and also~\cite{Hayward:1996}). 
Moreover, we will give the decomposition of the MS energy in its Ricci and 
Weyl parts; see~(\ref{eq:MSE-decomposition}). 

The name `Misner--Sharp energy' seems now to be established in the 
literature, however one should say that this mass concept goes back at 
least to~\cite{Lemaitre:1933}%
\footnote{For an English translation see~\cite{Lemaitre:1933trad}.}%
, which gives a coordinate expression for it. Its 
geometric definition~(\ref{eq:MS-energy-original-def}) was first given 
in~\cite{Hernandez.Misner:1966} and its interpretation as the charge of a 
conserved current was first derived in~\cite{Kodama:1980}. Later, an 
alternative definition was given in~\cite{Zannias:1990}: There it is showed 
that the MS energy can defined in terms of the norm of the Killing 
fields generating the isometry group $SO(3)$, leading directly 
to~(\ref{eq:MS-energy}). Further relevant studies of the MS energy 
are~\cite{Cahill.McVittie:1970a,Cahill.McVittie:1970b} and, more 
recently,~\cite{Hayward:1996,Hayward:1998,Burnett:1991}.

Some comment are needed on the above SSJC.
First we note that, since $\bn$ and $\bv$ are spherically symmetric and 
hence tangent to $\B$ we have in view of~(\ref{eq:connection-warped}) 
and~(\ref{eq:extrinsic-curv-form}) that $\g(\bn,\bnabla_\bv\bv)
= \gB(\bn_\sB,\!\LCB_{\bv_\sB}\bv_\sB) 
= \varepsilon(\bn_\sB)\,\bK_\gamma^\sB(\bv_\sB,\bv_\sB)$.
Hence, the quantity in $(ii)$ is indeed (up to a possible sign) the 
extrinsic curvature of the curve $\gamma$ in $\B$. Second, note that this
quantity, being quadratic in $\bv$, does not depend on the sign choice of
$\bv$. 
Third, since the matching hypersurface has the structure 
$\gamma\times\Stwo$, the words `continuous through the curve~$\gamma$' can 
be interchanged with the words `continuous through the hypersurface~%
$\varGamma$', depending on ones preference to think four- or two-dimensional.

A great advantage of the SSJC is that they are very easy to verify: one 
simply has to impose continuity on four \emph{scalars} along the matching 
curve in the two-dimensional base manifold $\B$. Dealing with scalars, since 
their value is independent on the particular coordinate choice, one does
not need to worry about introducing new coordinates in both spacetimes 
to be glued, in order to get the different metrics in a form which is 
comparable. This is indeed an ingrate task: in general, these coordinates 
are only needed in order to check if the junction conditions are satisfied, 
and for nothing more. In presence of spherical symmetry all this can be 
circumvented by using our new junction conditions.

Furthermore, the SSJC have a good physical interpretation: The continuity 
condition of both, the areal radius as well as of the MS energy, can be 
read as equilibrium condition for the gravitational pull acting from 
opposite directions onto (fictitious) test masses at the location of the 
matching surface. Concerning the continuity of the extrinsic curvature of 
the matching curve, we note the following: In the case where the matching 
hypersurface $\varGamma$ is timelike, let $\gamma=\pi(\varGamma)$ be a 
timelike curve in $\B$ and $\bv$ is future-pointing tangent. 
One can think of $\gamma$ as the `matching observer's' worldline. 
Hence, in the timelike case, the extrinsic curvature is nothing but 
the acceleration of the matching observer. In the spatial case, on 
the contrary, $\bn$ is timelike, $\gamma$ is spacelike (and $\bv$ 
tangent to it). One can choose $\bn$ to be future-pointing and then 
think of it as an observer field defined along the spatial (1-dimensional) 
slice $\gamma$. In view of~(\ref{eq:extrinsic-curv-form-SDn}) one 
then sees that the extrinsic curvature of $\gamma$ is exactly%
\footnote{Recall that, because of our signature choice, the restriction 
of the metric to spacelike directions is \emph{negative} definite.}
the shear-expansion of $\bn$ in direction of $\bv$ `radial direction'.

In Appendix~\ref{sec:equiv-junct-cond} we prove the following 

\begin{thm}[Equivalence of the junction conditions]\label{thm:equiv-jc}
Let $\varGamma$ be a smooth, non-null, spherically symmetric matching 
hypersurface between two spherically symmetric spacetimes and $\bn$ a 
continuous choice of unit normal vector field on $\varGamma$.
Assume, moreover, that the areal radii of the two spacetimes are $C^1$
functions in an open neighborhood of the matching hypersurfaces.
Then the DJC are equivalent to the SSJC.
\end{thm}

Now let us suppose we are faced with the following situation: We are given
two spherically symmetric solutions of Einstein equation and we want to
know if they can be matched together at all, and if so, how to characterize
the curve $\gamma$ (respectively, the hypersurface $\varGamma$) along which 
this is possible. Answers to these questions will be provided by the 
junction conditions SSJC. Note that a timelike or spacelike curve in the 
two-dimensional base manifold $(\B,\g_\sB)$ can be described simply by a 
function $R(\tau)$, where $R$ is the areal radius and $\tau$ the curve's 
arc-length which, in the timelike case, corresponds to the matching 
observer's proper time. 
The conditions $(i)$ and $(iii)$ of the SSJC are then equivalent to the 
condition that the functional dependence $R(\tau)$ must be the same 
(up to a trivial translation in $\tau$) in both spacetimes to be
matched.

\subsection{The Eisenstaedt theorem}
\label{sec:Eisenstaedt-theorem}
Perhaps the simplest attempt to model a compact body (star) in an expanding 
universe is trying to inglobate it in a FLRW spacetime and to assume, for 
simplicity, that the body is spherically symmetric. A direct consequence of 
the SSJC is the following intuitive appealing theorem due to 
Eisenstaedt~\cite{Eisenstaedt:1977}:

\begin{thm}[Eisenstaedt, 1977]\label{thm:Eisenstaedt}
Excise the full world-tube $\mathcal{W}_{r_0}$ of a comoving ball of comoving 
radius $r_0$ from a FLRW spacetime and insert instead a spherical symmetric 
inhomogeneity (hence a piece of a spherically symmetric spacetime together 
with a related matter model, satisfying Einstein's equations). 
Then a necessary condition for the resulting spacetime to satisfy Einstein's 
equation is that the MS energy of the inserted inhomogeneity equals 
that of the excised ball.
\end{thm}

\noindent
This says that the mean energy density (measured with the 
MS energy) of spherically-symmetric inhomogeneities must be the same as the 
one of the FLRW spacetime. That the Eisenstaedt Theorem is a consequence 
of the Darmois junction conditions was already pointed out in~\cite{Hartl:2006}.

\subsection{The Einstein--Straus vacuole revisited}
\label{sec:ESS-vacuole}
As another application of the above described matching procedure we revisit 
the Einstein--Straus solution~%
\cite{Einstein.Straus:1945,Einstein.Straus:1946,Schuecking:1954}, 
which originally consists on a Schwarzschild spacetime (called `vacuole') 
matched to a dust FLRW universe with zero cosmological constant. Later, 
this model was generalized also to the case of a non-vanishing cosmological 
constant~\cite{Balbinot.etal:1988}. We treat here the general case of an 
arbitrary cosmological constant and show that the SSJC allow substantial 
simplifications of the computations. This technique can also be applied 
to Bonnor's vacuole construction in LTB spacetimes. 

Notice that the matching condition~(\ref{eq:matter-momentum-flow-cont}) 
implies, in particular, that the pressure must be continuous through the 
matching hypersurface. Since the interior is a vacuum spacetime, it follows 
that the pressure must vanishes also on the exterior part of the matching 
hypersurface and hence everywhere on the FLRW spacetime. That is why one 
has to restrict to dust FLRW spacetimes.

Since we leave the cosmological constant $\Lambda$ arbitrary (it may be 
positive, negative, or zero) the inner region is given, respectively, by a 
Schwarzschild--de\,Sitter, Schwarzschild--anti-de\,Sitter%
\footnote{The Schwarzschild--(anti-)\,de\,Sitter metric (\ref{eq:SdS-metric}) 
is often called the Kottler solution, after Friedrich Kottler, who was the 
first to write down this metric in~\cite{Kottler:1918}. More details on 
its analytic and global structure may be found in~\cite{Geyer:1980}.}, 
or Schwarzschild spacetime (all abbreviated henceforth by SdS). Recall 
that the SdS spacetime is given by the vacuum solution to Einstein equation 
with cosmological term 
\begin{subequations}\label{eq:SdS-spacetime}
\begin{equation}\label{eq:SdS-metric}
  \g^\ind{SdS} = V(R)\,\ed T^2 - V(R)^{-1}\ed R^2 - R^2 \gStwo \,,
\end{equation}
where
\begin{equation}\label{eq:SdS-g00}
  V(R) = 1 - \frac{2m}{R} - \frac{\Lambda}{3}R^2 \,.
\end{equation}
\end{subequations}
Above, $\gStwo$ denotes the metric on the unit two-sphere~(\ref{eq:Def-gStwo})
and $m$ is a constant which represents the central mass. 

A dust FLRW spacetime is given by the metric 
\begin{equation}\label{eq:FLRW-metric}
  \g^\ind{FLRW} = \ed t^2 
  -a(t)^{2}\left( \frac{\ed r^2}{1-kr^2} + r^2\gStwo \right) 
\end{equation}
together with the matter energy-momentum tensor%
\footnote{Throughout we denote the metric-dual (1-form) of a vector 
$\bu$ by underlining it, that is, \ul{$\bu$}\,$:=\bg(\bu,\,\cdot\,)$ is 
the 1-form metric-dual to vector $\bu$. In local coordinates we have 
$\bu=u^\mu\bpartial_\mu$ and \ul{$\bu$}\,$=u_\mu \ed x^\mu$, where 
$u_\mu:=g_{\mu\nu}u^\nu$.}
$\bT=\varrho\,\ul\bu\otimes\ul\bu$, where $\bu=\bpartial/\bpartial t$ 
is the (geodesic) velocity field of the cosmological dust and $\varrho$ 
is the matter energy density, which depends on $t$ only. Here $r$ 
is the comoving radial coordinate and $k$ is a constant which takes 
the values $0,-1,+1$, depending on whether the spatial slices have 
zero, negative, or positive curvature, respectively. The Einstein 
equation is then equivalent to the following set given by the 
Friedmann equation and a `conservation equation':
\begin{subequations}
\begin{align}
  \label{eq:dust-FLRW-Friedmann}
  &\left( \frac{\dot a}{a} \right)^2 + \frac{k}{a^2} 
   - \frac{C}{a^3} - \frac{\Lambda}{3} = 0 \,, \\
  \label{eq:dust-FLRW-cons-eq}
  &\;\varrho a^3 = \text{constant},
\end{align}
\end{subequations}
where the constant $C:=8\pi\varrho_\ind{0}a^3_\ind{0}/3$ depends on the 
initial conditions $a_\ind{0}:=a(t_\ind{0})$ and 
$\varrho_\ind{0}:=\varrho(t_\ind{0})$ at some `initial' time $t_\ind{0}$.
Here, the dot denotes differentiation with respect to $t$ or, which is the 
same, along $\bu$. 

The central question is now the following: How shall we cut hypersurfaces 
$\varGamma_\ind{SdS}  = \gamma_\ind{SdS}  \times\Stwo$ and 
$\varGamma_\ind{FLRW} = \gamma_\ind{FLRW} \times\Stwo$ in the 
spacetimes SdS and, respectively, FLRW in order that the resulting 
pieces can be matched? In order to apply the SSJC we have to compute 
the MS energy for both spacetimes. For the FLRW spacetime one has 
$R_\ind{FLRW}(t,r)=a(t)r$ and hence 
$\ed R_\ind{FLRW} = \dot a r \ed t + a \ed r$ and 
$\sprod{\ed R}{\ed R}_\ind{FLRW}=g^{\mu\nu}R_{,\mu}R_{,\nu}
=r^2(k+\dot{a}^2)-1$. From 
the definition~(\ref{eq:MS-energy}) one then gets, using Friedmann's equation, 
\begin{equation}\label{eq:MSE-FLRW}
  E_\ind{FLRW} = \frac{4\pi}{3}R_\ind{FLRW}^3(\varrho + \varrho_\Lambda) \,,
\end{equation}
where $\varrho_\Lambda:=\Lambda/8\pi$ is the energy density associated with 
the cosmological constant. Notice that this expression, as well its 
derivation, is completely independent of the specific equation of state of 
the fluid and does not depend on the spatial curvature $k$. 
For the SdS case one has $R_\ind{SdS}=R$ and thus 
$\sprod{\ed R}{\ed R}_\ind{SdS}=-V(R)$ and
\begin{equation}\label{eq:MSE-SdS}
  E_\ind{SdS} = m + \frac{4\pi}{3}R_\ind{SdS}^3 \varrho_\Lambda \,.
\end{equation}

Now, the last two conditions of the SSJC, that is the continuity of areal 
radius and MS energy across the matching hypersurface (yet to be 
determined), are equivalent to the continuity of the areal radius 
$R_\ind{FLRW}=R_\ind{SdS}=:R$, together with the suggestive relation 
\begin{equation}\label{eq:ESS-junction-cond}
  m = \frac{4\pi}{3}R^3\varrho \,.
\end{equation}
This two conditions already determine the matching hypersurface. Indeed, 
inserting $R=R_\ind{FLRW}=a(t)r$ in~(\ref{eq:ESS-junction-cond}) and 
using the relation $\varrho(t)a^3(t)=\text{const.}$ valid for dust FLRW
models, one obtains the matching radius in terms of the FLRW comoving 
radial coordinate: 
\begin{equation}\label{eq:ESS-matching-curve-FLRW}
  r = r_0 
   := \left( \frac{m}{(4\pi/3)a_\ind{0}^3\varrho_\ind{0}} \right)^{1/3} 
    =\text{constant}\,.
\end{equation}
Here $a_\ind{0}:=a(t_\ind{0})$, and similarly for $\varrho$, where 
$t_\ind{0}$ is some fixed `initial' time.
This means that the matching observer moves, in the FLRW spacetime, 
along the integral curve of $\bu=\bpartial/\bpartial t$ with initial 
condition $(t_\ind{0},r_\ind{0})$ and hence is comoving with the cosmological
matter. 

So far we used the last two of the SSJC. As discussed above, the continuity 
of the areal radius and the arc-length (the proper time, in the timelike 
case) of the matching curve are equivalent to the equality of the functional 
dependencies $R(\tau)$ (up to a possible trivial translation in $\tau$) which 
describe the matching curves in the two spacetimes to be matched. 
Now, because of~(\ref{eq:ESS-matching-curve-FLRW}), the matching curve 
(worldline) in the FLRW spacetime is simply 
\begin{equation}\label{eq:FLRW-matching-worldline}
  R(\tau) = a(\tau)r_\ind{0} \,,
\end{equation}
where $a$ is the (unique) solution of the Friedmann 
equation~(\ref{eq:dust-FLRW-Friedmann}) with initial condition $a_\ind{0}$ at 
$\tau_\ind{0}=t_\ind{0}$. (Recall that in FLRW the proper time of an observer
moving along an integral line of $\bu$ equals the cosmological time, hence 
$\tau=t$.) From what we said above, the same functional relation $R(\tau)$ 
must hold also in the SdS---provided we identify $R$ with the areal radius 
and $\tau$ with the matching observer's proper time, both referred to the 
SdS spacetime. This determines the matching curve in SdS. 

Finally we need to show that the junction condition~$(ii)$ is satisfied, 
hence that the matching observer's accelerations coincide. Looking at the 
matching worldline from the FLRW spacetime, it is immediately clear that 
it is geodesic, hence its acceleration vanishes. 
To conclude the matching procedure, we just have to check that this is also 
true for the matching worldline in the SdS spacetime. For this, one has just 
to check that the function defined in~(\ref{eq:FLRW-matching-worldline}) 
satisfies the geodesic equation for a radial motion. The latter is given by 
\begin{equation}\label{eq:SdS-geod-radial}
  \dot{R}^2 + V(R) = e^2 \,,
\end{equation}
where $e:=\g^\ind{SdS}(\bpartial/\bpartial T,\bv)=\text{constant}$ and 
$\bv=\dot{T}~\bpartial/\bpartial T+\dot{R}~\bpartial/\bpartial R$ is the 
matching observer in the SdS spacetime. (Equation~(\ref{eq:SdS-geod-radial}) 
can be quickly derived from the fact that 
$\bv(\g^\ind{SdS}(\bpartial/\bpartial T,\bv))=0$, since 
$\bpartial/\bpartial T$ is Killing and $\bv$ geodesic. Inserting 
$e:=\g^\ind{SdS}(\bpartial/\bpartial T,\bv)=V(R)\dot T$ in the normalization 
condition $1=\g^\ind{SdS}(\bv,\bv)=V(R)\dot{T}^2-\dot{R}^2/V(R)$ one arrives 
immediately at~(\ref{eq:SdS-geod-radial}).)
Now, inserting~(\ref{eq:FLRW-matching-worldline}) 
with~(\ref{eq:ESS-matching-curve-FLRW}) in~(\ref{eq:SdS-geod-radial}) and 
using the Friedmann equation~(\ref{eq:dust-FLRW-Friedmann}), one gets
$\dot{R}^2 + V(R) = 1-k r_\ind{0}^2$. Hence, the geodesic 
equation~(\ref{eq:SdS-geod-radial}) is satisfied (with $e^2=1-kr_\ind{0}^2$)
and herewith all the four junction conditions.

\subsection{The McVittie model}
\label{sec:McVittieModel}
Among all models discussed in the literature which represent a 
quasi-isolated spherically-symmetric gravitating system melted into a 
cosmological spacetime, the one that is presumably best understood as 
regards its analytical structure as well as its physical assumptions is 
that of McVittie~\cite{McVittie:1933}, thanks to the careful analysis 
of Nolan~\cite{Nolan:1998,Nolan:1999a,Nolan:1999b}. Here we shall 
restrict to the `flat' or $k=0$ model, which interpolates between an 
exterior Schwarzschild solution, describing a local mass, and a spatially 
flat (i.e.~$k=0$) ambient FLRW universe. For simplicity we shall from now 
on refer to this model simply as \emph{the} McVittie model. 
The cosmological constant is assumed to be zero, although this 
assumption is not essential (see the last paragraph of 
Section~\ref{sec:McVittieInterpretation}). 

This is not to say that this model is to be taken at face value 
in all its aspects. Its problems lie in the region very close to 
the central object, where the basic assumptions on the behavior 
of matter definitely turn unphysical. However, at radii much 
larger than (in geometric units) the central mass (to be defined 
below) the $k=0$ McVittie solution seems to provide a viable 
approximation for the transition between a homogeneous cosmological 
spacetime and a localized mass immersed in it. We will now briefly 
discuss this model and look at its geodesic equations, showing 
that they reduces to~(\ref{eq:impr-N-eqs}) in an appropriate 
weak-field and slow-motion limit. This provides another and more 
solid justification for the Newtonian approach we carried out in 
Section~\ref{sec:NewtonianApproach}. 

The characterization of the McVittie model is made through two sets
of \emph{a\,priori} specifications. The first set concerns the metric 
(left side of Einstein's equations) and the second set the matter 
(right side of Einstein's equations). The former consists in an 
ansatz for the metric, which can formally be described as follows:
Write down the Schwarzschild metric for the mass parameter $m$ in 
isotropic coordinates, add a conformal factor $a^2(t)$ to the 
spatial part, and allow the mass parameter $m$ to depend on time. 
Hence the metric reads    
\begin{equation}\label{eq:McVittieAnsatz}
\begin{split}
\g =&\left( \frac{1-m(t)/2r}{1+m(t)/2r} \right)^2 \ed t^2\\
   -&\left( 1+\frac{m(t)}{2r} \right)^4 a^2(t)\ (\ed r^2 + r^2 \gStwo)\,,
\end{split}
\end{equation}
where $\gStwo$ is given by~(\ref{eq:Def-gStwo}). The metric 
(\ref{eq:McVittieAnsatz}) is obviously spherically symmetric with 
the spheres of constant radius $r$ being the orbits of the rotation 
group. We will discuss below what this ansatz actually entails. 
For later convenience we also introduce the orthonormal tetrad 
$\{\be_\mu\}_{\mu=0,\cdots, 3}$ with respect to 
(\ref{eq:McVittieAnsatz}), where%
\footnote{We write $\norm{\bv}:=\sqrt{\vert\bg(\bv,\bv)\vert}$.} 
\begin{equation}\label{eq:DefOrthnFrame}
  \be_\mu := \norm{\bpartial/\bpartial x^\mu}^{-1}\,\bpartial/\bpartial x^\mu
\end{equation}
and $\{x^\mu\}=\{t,r,\theta,\varphi\}$. 

The second set of specifications, concerning the matter, is as follows: 
The matter is a perfect fluid with density $\varrho$ and isotropic 
pressure $p$. Hence its energy-momentum tensor is given by
\begin{equation}\label{er:EMTensorForMcVittie}
  \bT = \varrho\,\ul\bu\otimes\ul\bu +p\,(\ul\bu\otimes\ul\bu-\bg)\,.
\end{equation}
Furthermore, and this is where the two sets of specifications make 
contact, the motion of the matter is given by  
\begin{equation}\label{eq:u-McVittie}
  \bu = \be_0 \,.
\end{equation}
No further assumptions are made. In particular, an equation of state, 
like $p=p(\varrho)$, is \emph{not} assumed. The reason for this 
will become clear soon. 

Note that the vector field (\ref{eq:u-McVittie}) is \emph{not} 
geodesic for the metric (\ref{eq:McVittieAnsatz}) (unlike for the 
FLRW and Gautreau metrics), which immediately implies that 
the pressure cannot be constant. Being spherically symmetric, $\bu$ is 
automatically vorticity free. The last property is manifest from its 
hypersurface orthogonality, which is immediate from 
(\ref{eq:McVittieAnsatz}).  Moreover, $\bu$ is also shear free.
This, too, can be immediately read off (\ref{eq:McVittieAnsatz}) 
once one takes into account the result that for spherically 
symmetric metrics vanishing shear for a spherically symmetric 
vector field is equivalent to the corresponding spatial metric 
being conformally related to a spherically symmetric flat metric. 
This is obviously the case here. 

The non-vanishing components of the Einstein tensor with respect to 
the orthonormal basis (\ref{eq:DefOrthnFrame}) are:
\begin{subequations}\label{eq:McV-Ein-tensor}
\begin{alignat}{3}
  &\Ein(\be_0,\be_0) &&= 3 F^2 \,,
   \label{eq:McV-Ein-00}\\
  &\Ein(\be_i,\be_j) &&= -\left( 3F^2 + 2\tfrac{A}{B}\dot{F} \right)\delta_{ij}
  \,, \label{eq:McV-Ein-ij}\\
  &\Ein(\be_0,\be_1) &&= 
   \tfrac{2}{R^2}\left(\tfrac{A}{B}\right)^2
   (a \, m)\!\dot{\phantom{I}}\! \,,
   \label{eq:McV-Ein-01}
\end{alignat}
\end{subequations}
where an overdot denotes differentiation along 
$\bpartial/\bpartial t$. Before explaining the functions $A$, $B$, 
$R$, and $F$, we make the important observation that the Einstein 
tensor is \emph{spatially isotropic}, where `spatially' refers to the 
directions orthogonal to  $\be_0$. By this we mean that 
$\Ein(\be_i,\be_j)\propto\delta_{ij}$ or, expressed more 
geometrically, that the spatial restriction of the Einstein tensor 
is proportional to the spatial restriction of the metric. 

In (\ref{eq:McV-Ein-tensor}) and in the following we set: 
\begin{equation}\label{eq:def-AB}
  A(t,r) := 1+m(t)/2r \,, \quad{  } B(t,r) := 1-m(t)/2r \,,
\end{equation}
and
\begin{equation}\label{eq:McVittieArealRadius}
  R(t,r)=\left( 1+\frac{m(t)}{2r} \right)^2 a(t)\,r \,,
\end{equation}
where $R$ is the areal radius for the McVittie 
ansatz~(\ref{eq:McVittieAnsatz}), and also
\begin{equation}\label{eq:def-F}
  F := \frac{\dot a}{a} + \frac{1}{rB}\frac{(a\,m)^{\bdot}}{a} \,.
\end{equation}
In passing we note that $F$ has the geometric interpretation of being 
one third the expansion of the vector field $\be_0$, that is, 
$F=\Div(\be_0)/3$. Hence (\ref{eq:McV-Ein-00}) could also be written 
in the form $\Ein(\be_0,\be_0)=(\Div(\be_0))^2/3$. We will see later 
that the product $a\,m$ which appears in~(\ref{eq:McV-Ein-01}) also 
has a geometric meaning: it is just the Weyl part of the MS energy; 
see~(\ref{eq:McV-MSE-Weyl}).

Now, the non-vanishing components of the energy momentum 
tensor~(\ref{er:EMTensorForMcVittie}) with~(\ref{eq:u-McVittie}) are:
\begin{equation}\label{eq:T-e0-components}
  \bT(\be_0,\be_0) = \varrho \,, \quad 
  \bT(\be_i,\be_j) = p\,\delta_{ij}\,. 
\end{equation}
The $(\be_0,\be_1)$ component of Einstein's equation therefore implies 
$(a\,m)\!\dot{\phantom{I}}\! = 0$, which means that the Weyl part of 
the MS energy is constant. Physically this can be interpreted as 
saying that the central object does not accrete any energy from 
the ambient matter. Using the constancy of $a\,m$ in~(\ref{eq:def-F}) 
we immediately get: 
\begin{equation}\label{eq:F-McVittie}
  F = \frac{\dot a}{a} =: H \,.
\end{equation}
Hence Einstein's equation is equivalent to the following three 
relations between the four functions $m(t), a(t), \varrho(t,r)$, 
and $p(t,r)$:
\begin{subequations}\label{eq:McV-Einstein}
\begin{alignat}{2}
\label{eq:McV-Einstein-1}
  &(a \, m)\!\dot{\phantom{I}}\! &&\,=\, 0 \,,\\
\label{eq:McV-Einstein-2}
  &8\pi \varrho    &&\,=\, 3 \left( \frac{\dot a}{a} \right)^2 \,,\\
\label{eq:McV-Einstein-3}
  &8\pi p          &&\,=\, - 3 \left( \frac{\dot a}{a} \right)^2
  - 2 \left( \frac{\dot{a}}{a} \right) \!\!\!\!\dot{\phantom{\frac{I}{I}}}
  \left( \frac{1+m/2r}{1-m/2r} \right) \,.
\end{alignat}
\end{subequations}
Note that here Einstein's equation has only three independent 
components (as opposed to four for a general spherically 
symmetric metric), which is a consequence of the fact, 
already stresses above, that the Einstein tensor for the McVittie 
ansatz~(\ref{eq:McVittieAnsatz}) is spatially isotropic. 

Equation~(\ref{eq:McV-Einstein-1}) can be immediately integrated:
\begin{equation}\label{eq:mIntegration}
  m(t) = \frac{m_0}{a(t)} \,,
\end{equation}
where $m_0$ is an integration constant. Below we will show that this 
integration constant is to be interpreted as the mass of the central
body. We will call the metric~(\ref{eq:McVittieAnsatz}) together
with condition~(\ref{eq:mIntegration}) the \emph{McVittie metric}. 

Clearly the system (\ref{eq:McV-Einstein}) is under-determining. 
This is expected since no equation of state has yet been imposed. 
The reason why we did not impose such a condition can now be easily 
inferred from (\ref{eq:McV-Einstein}): whereas  (\ref{eq:McV-Einstein-2}) 
implies that $\varrho$ only depends on $t$, (\ref{eq:McV-Einstein-3}) 
implies that $p$ depends on $t$ \emph{and} $r$ iff $(\dot a/a)\dot{}\ne 0$. 
Hence a non-trivial relation $p=p(\varrho)$ is simply incompatible with the 
assumptions made so far. 
The only possible ways to specify $p$ are $p=0$ or $\varrho+p=0$. 
In the first case (\ref{eq:McV-Einstein-3}) implies that $\dot a/a=0$
if $m_0\ne 0$ (since then the second term on the right-hand side is
$r$ dependent, whereas the first is not, so that both must vanish
separately), which corresponds to the exterior Schwarzschild 
solution, or $a(t)\propto t^{2/3}$ if $m_0=0$, which leads to the 
flat FLRW solution with dust. In the second case the fluid just acts 
like a cosmological constant $\Lambda=8\pi\varrho$ (using the equation 
of state $\varrho+p=0$ in $\Div\bT=0$ it implies $\ed p=0$ and this, 
in turn, using again the equation of state, implies $\ed \varrho=0$) so 
that this case reduces to the Schwarzschild--de\,Sitter solution. 
To see this explicitly, notice first that 
(\ref{eq:McV-Einstein-2},\ref{eq:McV-Einstein-3}) imply the constancy 
of $H=\dot a/a=\sqrt{\Lambda/3}$ and hence one has 
$a(t)=a_0\exp\bigl(t\,\sqrt{\Lambda/3}\bigr)$. With such a scale-factor 
the McVittie metric~(\ref{eq:McVittieAnsatz}) with~(\ref{eq:mIntegration}) 
turns into the Schwarzschild--de\,Sitter metric~(\ref{eq:SdS-spacetime}) 
in disguise. 
The explicit formulae for the coordinate transformation relating 
the two can be found in Section\,5 of~\cite{Robertson:1928} and also in 
Section\,7 of~\cite{Klioner.Soffel:2005}. Finally, note from 
(\ref{eq:McV-Einstein-1}) that constancy of one of the functions $m$ and 
$a$ implies constancy of the other. In this case 
(\ref{eq:McV-Einstein-2},\ref{eq:McV-Einstein-3}) imply $\varrho=p=0$, so 
that we are dealing with the exterior Schwarzschild spacetime. 

A specific McVittie solution can be obtained by \emph{choosing} a function 
$a(t)$, corresponding to the scale function of the FLRW spacetime which 
the McVittie model is required to approach at spatial infinity, and the 
constant $m_0$, corresponding to the `central 
mass'. Relations~(\ref{eq:McV-Einstein-2},\ref{eq:McV-Einstein-3}), 
and~(\ref{eq:mIntegration}) are then used to determine $\varrho$, 
$p$, and $m$, respectively. Clearly this `poor man's way' to solve 
Einstein's equation holds the danger of arriving at unrealistic 
spacetime dependent relations between $\varrho$ and $p$. This must 
be kept in mind when proceeding in this fashion. For further 
discussion of this point we refer to~\cite{Nolan:1998,Nolan:1999a}.

\subsubsection{Interpretation of the McVittie model}
\label{sec:McVittieInterpretation}
In this section we discuss the interpretation of the McVittie model, its 
singularities, trapped regions, symmetry properties, and also the motion 
of the matter. In doing this, we shall take care to isolate those properties 
which are intrinsic to the ansatz~(\ref{eq:McVittieAnsatz}) independent of 
the imposition of Einstein's equation. The analysis can then also be applied 
to all generalizations which maintain the ansatz~(\ref{eq:McVittieAnsatz}). 
Generalizations in this sense have recently been discussed 
in~\cite{Faraoni.Jacques:2007}, on which we will comment at the end 
of this section. 

According to what has just been said, we wish to regard the McVittie 
solution as a candidate model for an isolated mass $m_0$ in an `otherwise' 
flat FLRW universe with scale function $a(t)$. As already emphasized 
in the introduction, this requires specific justification in view of 
the fact that simple superpositions of solutions are disallowed by the 
nonlinearities. A set of criteria for when a solution represents a 
localized mass immersed in a flat FLRW background have been proposed 
and discussed in detail in~\cite{Nolan:1998}. The basic idea is to 
employ the MS energy (in a spherically symmetric context,
where it is equivalent to the Hawking mass) in order to detect 
localized sources of gravity. We will follow this approach and for 
this purpose we compute the Ricci and the Weyl part of the MS energy. 

This we now do for the class of metrics (\ref{eq:McVittieAnsatz}),
without at first making any use of Einstein's equation. The geometric 
definition of the MS energy in terms of the sectional curvature, 
together with formula~(\ref{eq:DefRicci-Ein}) for its Ricci part 
specialized to metrics with spatially isotropic Einstein tensor, 
implies that the Ricci part of the MS energy is given by 
\begin{equation}\label{eq:McV-MSE-Ricci}
  E_\ind{R} = \tfrac{1}{6}R^3 \Ein(\be_0,\be_0)\,.
\end{equation}
The Weyl part is then obtained as the difference between the full 
MS energy and (\ref{eq:McV-MSE-Ricci}). We use the expression 
(\ref{eq:MS-energy}) for the former and write 
$\sprod{\ed R}{\ed R}=\bigl(\be_0(R)\bigr)^2-\bigl(\be_1(R)\bigr)^2$. 
The part involving $\be_0(R)$ equals (\ref{eq:McV-MSE-Ricci}), due to the 
relation $\Ein(\be_0,\be_0)=3(\ed R(\be_0)/R)^2$, which, e.g., follows 
from the comment below Eq.~(\ref{eq:def-F}) and (\ref{eq:shear-scalar}) 
(for vanishing shear). The Weyl part of the MS energy is therefore given 
by $(R/2)\bigl(1-(\be_1(R))^2\bigr)$. From (\ref{eq:McVittieArealRadius}) 
we calculate $\be_1(R)$ and hence obtain for the Weyl part of the MS energy: 
\begin{equation}\label{eq:McV-MSE-Weyl}
  E_\ind{W} = a\,m\,.
\end{equation}

Now we invoke Einstein's equation with source~(\ref{er:EMTensorForMcVittie}) 
and four-velocity (\ref{eq:u-McVittie}). Then the Ricci and Weyl 
contributions to the MS energy can be written in the following form,
also taking into account~(\ref{eq:mIntegration}), 
\begin{subequations}\label{eq:McV-MSE-wEeq}
\begin{alignat}{3}
  &E_\ind{R} &&= \frac{4\pi}{3}R^3 \varrho \,, 
  \label{eq:McV-MSE-Ricci-wEeq} \\
  &E_\ind{W} &&= m_0 \,.
  \label{eq:McV-MSE-Weyl-wEeq}
\end{alignat}
\end{subequations}
Identifying the gravitational mass of the central object with the Weyl 
part of the MS energy, its constancy means that no energy is accreted 
from the ambient matter. As regards the Ricci part, note that the 
factor $(4\pi/3) R^3$ in~(\ref{eq:McV-MSE-Ricci-wEeq}) is smaller than 
the proper geometric volume within the sphere of areal radius $R$. 
This can be attributed to the gravitational binding energy that 
diminishes the gravitational mass of a lump of matter below the value 
given by the proper space integral of $\bT(\be_0,\be_0)$. This is shown 
in more detail in Appendix~\ref{sec:ss-perfect-fluids}, in particular 
in the exact equation~(\ref{eq:MSE-int-expr}) and its leading order 
approximation~(\ref{eq:MSE-int-expr-approx}). 

It is also important to note that the central gravitational mass in 
McVittie's spacetime may be modeled by a shear-free perfect-fluid star 
of positive homogeneous energy density~\cite{Nolan:1992}. The matching 
is performed along a world-tube comoving with the cosmological fluid, 
across which the energy density jumps discontinuously. This means that 
the star's surface is comoving with the cosmological fluid and hence, 
in view of~(\ref{eq:expansion-McVittie}), that it geometrically expands 
(or contracts). This feature, however, should be merely seen as an artifact 
of the McVittie model (in which the relation~(\ref{eq:expansion-McVittie}) 
holds), rather than a general property of compact objects in any cosmological 
spacetimes. Positive pressure within the star seems to be only possible if 
$2a\ddot a+{\dot a}^2<0$ (see Eq.~(3.27) in~\cite{Nolan:1992} with 
$a=\exp(\beta/2)$), that is, for deceleration parameters $q>1/2$. 

Next we comment on the singularity properties of the McVittie model. 
From~(\ref{eq:McV-Einstein-3}) it is clear that, unless $\dot{a}/a$ is 
constant (the Schwarzschild--de\,Sitter case) or $m=0$ (FLRW case), the 
pressure diverges at $r=m/2$ (that is at $R=2m_0=R_S$). In fact, this 
corresponds to a genuine curvature singularity which is built into 
the McVittie ansatz~(\ref{eq:McVittieAnsatz}) independently of
any further assumption. To see that $r=m/2$ (corresponding to 
$R=2am=2E_\ind{W}$) is a singularity it suffices to consider the 
scalar curvature (i.e.~the Ricci scalar) of~(\ref{eq:McVittieAnsatz}), 
\begin{equation}\label{eq:Scal-McV}
  \Scal = - 12F^2 - 6\tfrac{A}{B}\dot{F} \,, 
\end{equation}
which is readily computed from~(\ref{eq:McV-Ein-tensor}). 
In~\cite{Carrera.Giulini:2009a} we show that this becomes singular in the 
limit $r \to m/2$, with the only exceptions being the following three 
special cases: 
$(i)$~$m=0$ and $a$ arbitrary (FLRW spacetimes), 
$(ii)$~$a$ and $m$ constant (Schwarzschild spacetime), and 
$(iii)$~$(a\,m)^{\bdot}=0$ and $(\dot a/a)^{\bdot}=0$ 
(Schwarzschild--de\,Sitter spacetime). 
This means that, as long as we stick to the 
ansatz~(\ref{eq:McVittieAnsatz}), at $r=m/2$ there will always (with the 
only exceptions listed above) be a singularity in the Ricci part of the 
curvature and thus, assuming Einstein's equation is satisfied, also in 
the energy momentum tensor, irrespectively of the details of its underlying 
matter model. Hence any attempt to eliminate this singularity by 
maintaining the ansatz~(\ref{eq:McVittieAnsatz}) and merely modifying 
the matter model is doomed to fail. In particular, this is true for the 
generalizations presented in \cite{Faraoni.Jacques:2007}, contrary to 
what is claimed in that work. We also remark that it makes no sense to 
absorb the singular factors $1/B$ in front of the time derivatives by 
writing $(A/B)\bpartial/\bpartial t$ as $\be_0$ and then argue, as was 
done in \cite{Faraoni.Jacques:2007}, that this eliminates the singularity. 
The point is simply that then $e_0$ applied to any continuously 
differentiable function diverges as $r\rightarrow m/2$. 
Below we will show that this singularity lies within a trapped region. 
Turning back to the McVittie model, recall that in this case it is assumed 
that the fluid moves along the integral curves of $\bpartial/\bpartial t$, 
which become lightlike in the limit as $r$ tends to $m/2$. Their acceleration 
is given by the gradient of the pressure, which necessarily diverges in the 
limit $r\rightarrow m/2$, as one explicitly sees 
from~(\ref{eq:accel-McVittie}). For a more detailed study of the geometric 
singularity at $r=m/2$, see~\cite{Nolan:1999a,Nolan:1999b}. 

For spherically symmetric spacetimes the Weyl part of the curvature has 
only a single independent component (see~(\ref{eq:Weyl-warped-ss-1})) 
which, by its very definition, is $-2/R^3$ times the Weyl part of the 
MS energy (see~(\ref{eq:MSE-Weyl-warped})). The square of the Weyl tensor 
for the ansatz (\ref{eq:McVittieAnsatz}) may then be conveniently expressed 
as (see~(\ref{eq:McV-MSE-Weyl}) and~(\ref{eq:Weyl-square-MSE})) 
\begin{equation}\label{eq:Weyl-norm-McV}
  \sprod{\Weyl}{\Weyl} = 48\frac{(a\,m)^2}{R^6} \,.
\end{equation}
This shows that $R=0$ also corresponds to a genuine curvature singularity, 
though this is not part of the region covered by our original coordinate 
system, for which $r>m/2$ (that is $R>2E_\ind{W}$). 

It is instructive to also determine the trapped regions of McVittie 
spacetime. We do this just using the McVittie ansatz~(\ref{eq:McVittieAnsatz}) 
and making no further assumptions. Recall that a spacelike two-sphere $S$ 
is said to be \emph{trapped, marginally trapped}, or \emph{untrapped} if 
the product $\theta^+\theta^-$ of the expansions (defined below 
Eq.~(\ref{eq:Hawking-mass})) for the ingoing and outgoing future-pointing 
null vector fields normal to $S$  is positive, zero, or negative. Taking 
$S$ to be $S_R$, that is, a sphere of symmetry with areal radius $R$, it 
immediately follows from~(\ref{eq:expansions-product}) that $S_R$ is 
trapped, marginally trapped, or untrapped iff $\sprod{\ed R}{\ed R}$ is 
positive, zero, or negative, respectively. This corresponds to timelike, 
lightlike, or spacelike $\ed R$, or, equivalently, in in view 
of~(\ref{eq:MS-energy}), to $2E-R$ being positive, zero, or negative, 
respectively. Using~(\ref{eq:McV-MSE-Ricci}) together 
with~(\ref{eq:McV-Ein-00}), the MS energy for the McVittie ansatz can 
be written as $E = E_\ind{W} + R^3F^2/2$, so that 
\begin{equation}\label{eq:polynome-trapped}
  2E - R = F^2R^3 - R + R_S \,,
\end{equation}
where $R_S\!:=\!2E_\ind{W}$ denotes the `generalized' Schwarzschild radius. 
Note that in general $F$ depends itself on the radial coordinate---except 
for the McVittie case, in which one has $F=H=:1/R_H$ ($R_H$ denotes the 
Hubble radius). For computational simplicity we specialize in the following 
to the McVittie case, referring to \cite{Carrera.Giulini:2009a} for the 
general case. Doing this, (\ref{eq:polynome-trapped}) becomes a cubic 
polynomial in $R$ which is positive for $R=0$ and tends to $\pm\infty$ for 
$R\rightarrow\pm\infty$. Hence it always has a negative zero (which does 
not interest us) and two positive zeros iff 
\begin{equation}\label{eq:discrim-condition}
  R_S/R_H < 2/3\sqrt{3} \approx 0.38 \,.
\end{equation}
This clearly corresponds to the physical relevant case where the 
Schwarzschild radius is much smaller than the Hubble radius. One zero lies 
in the vicinity of the Schwarzschild radius and one in the vicinity of the 
Hubble radius, corresponding to two marginally trapped spheres. The exact 
expressions for the zeros can be easily written down, but are not very 
illuminating. In leading order in the small parameter $R_S/R_H$, they are 
approximated by 
\begin{subequations}\label{eq:McV-trapped}
\begin{align}
  &R_1 \approx R_S \left( 1 + (R_S/R_H)^2 \right)\,, 
\label{eq:McV-trapped-1}\\
  &R_2 \approx R_H \left( 1 - R_S/2R_H \right)  \,.  
\label{eq:McV-trapped-2}
\end{align}
\end{subequations}
From this one sees that for the McVittie model the radius of the marginally 
trapped sphere of Schwarzschild spacetime ($R_S$) increases and that of the 
FLRW spacetime ($R_H$) decreases. The first feature can be understood as an 
effect of the presence of cosmological matter, whereas the latter is an 
effect of the presence of a central mass abundance. 
All the spheres with $R<R_1$ or $R>R_2$ are trapped and those with 
$R_1<R<R_2$ are untrapped. In particular, the singularity $r=m/2$, that is 
$R=2E_\ind{W}=R_S$, lies within the inner trapped region. 

Another aspect concerns the global behavior of the McVittie 
ansatz~(\ref{eq:McVittieAnsatz}). We note that each hypersurface of constant 
time $t$ is a complete Riemannian manifold, which, besides the rotational 
symmetry, admits a discrete isometry given in $(r,\theta,\varphi)$ 
coordinates by 
\begin{equation}\label{eq:McVittieIsometry}
  \phi(r,\theta,\varphi) = 
  \bigl((m/2)^2\,r^{-1}\,,\,\theta\,,\,\varphi\bigr)\,.
\end{equation} 
This corresponds to an inversion at the two-sphere $r=m/2$ and shows that 
the hypersurfaces of constant $t$ can be thought of as two isometric 
asymptotically-flat pieces joined together at the totally geodesic (being 
a fixed-point set of an isometry and hence also minimal) two-sphere $r=m/2$. 
Except for the time-dependent factor $m(t)$, this is analogous to the
geometry of the  $t=\mathrm{const.}$ slices in the Schwarzschild metric 
(the difference being that (\ref{eq:McVittieIsometry}) does not extend to 
an isometry of the spacetime metric unless $\dot m=0$). This means, in 
particular, that the McVittie metric cannot literally be interpreted as 
corresponding to a point particle sitting at $r=0$ ($r=0$ is at infinite 
metric distance) in a flat FLRW universe, just like the Schwarzschild metric 
does not correspond to a point particle sitting at $r=0$ in Minkowski space. 
Unfortunately, McVittie seems to have interpreted his solution in this 
fashion~\cite{McVittie:1933} which even until recently gave rise to some 
confusion in the literature 
(e.g.~\cite{Gautreau:1984b,Sussman:1988,Ferraris.etal:1996}). 
A clarification was given by~\cite{Nolan:1999a}. 

We now briefly discuss the basic properties of the motion of cosmological 
matter for the McVittie model. We already mentioned that the vorticity and 
shear of the four-velocity $\bu$ vanish identically. On the other hand, the 
expansion (divergence of $\bu$) is 
\begin{equation}\label{eq:expansion-McVittie}
  \theta = 3 H\,,
\end{equation}
just as in the FLRW case (recall that here $H:=\dot a/a$ is defined as in 
the FLRW case, see~(\ref{eq:F-McVittie})). In particular, the 
expansion of the cosmological fluid is \emph{homogeneous in space}. 
The expression for the variation of the areal radius along the 
integral lines of $\bu$ (that is the velocity of cosmological 
matter measured in terms of its proper time and the areal 
radius) is also just as in the FLRW case:  
\begin{equation}\label{eq:Hubble-law-McVittie}
  \bu(R) = HR \,,
\end{equation}
which is nothing but Hubble's law. The acceleration of $\bu$, which in 
contrast to the FLRW case does not vanish, is given by 
\begin{equation}\label{eq:accel-McVittie}
  \bnabla_\bu\bu = \frac{m_0}{R^2} \left(\frac{1+m/2r}{1-m/2r}\right) \be_1\,.
\end{equation}
In leading order in $m_0/R$ this corresponds to the acceleration of the 
observers moving along $\bpartial/\bpartial T$ in Schwarzschild spacetime 
(see Eq.~(\ref{eq:acceleration-u-SdS}) with $\Lambda=0$). 

We conclude this subsection by commenting on the attempts to generalize the 
McVittie model. The first obvious generalization consists in allowing a 
non-vanishing cosmological constant. This is however trivial, since it is 
equivalent to the substitution $\varrho \to \varrho+\varrho_\Lambda$ and 
$p\to p+p_\Lambda$ in~(\ref{eq:McV-Einstein}), where 
$\varrho_\Lambda:=\Lambda/8\pi$ and $p_\Lambda:=-\Lambda/8\pi$ are, 
respectively, the energy-density and pressure associated to the cosmological 
constant $\Lambda$. 
In \cite{Faraoni.Jacques:2007}, attempts to generalize the McVittie model 
have focused on keeping the ansatz~(\ref{eq:McVittieAnsatz}) and relaxing 
the conditions on the matter in various ways. More precisely, they show that 
the McVittie case may be generalized to allow for radial fluid motions 
relative to the $\be_0$ observer field (that is relaxing 
condition~(\ref{eq:u-McVittie})), provided one also allows for a 
non-vanishing radial heat flow. Both energy flows are necessary in order to 
get new solutions consistent with the ansatz~(\ref{eq:McVittieAnsatz}), 
even though the two radial energy flows do not cancel in the energy balance. 
As a result, the Weyl part of the MS energy will now change in time so that 
the new solutions correspond to inhomogeneities of variable strength due to 
accretion or loss of energy.  For further analysis of these solutions we refer 
to~\cite{Carrera.Giulini:2009a}. Another generalization of the McVittie model,
this time away from the ansatz (\ref{eq:McVittieAnsatz}), is given in 
\cite{Sultana.Dyer:2005}. It is constructed by applying 
a particular (time-dependent) conformal transformation to the Schwarzschild 
spacetime. As we will show in~\cite{Carrera.Giulini:2009a}, the metric so
obtained cannot be written in the McVittie form~(\ref{eq:McVittieAnsatz}), 
contrary to what is suggested in~\cite{Faraoni.Jacques:2007}. 
The corresponding energy-momentum tensor (obtained via Einstein's 
equation) can be interpreted as a sum of two contributions, one due to a 
perfect fluid and the other to a null fluid.

\subsubsection{Motion of a test particle in McVittie spacetime}
\label{sec:MotionInMcVittie}
We are interested in the motion of a test particle (idealizing a 
planet or a spacecraft) in McVittie's spacetime. In \cite{McVittie:1933} 
it was concluded within a slow-motion and weak-field approximation
that Keplerian orbits do not expand as measured with the `cosmological 
geodesic radius' $r_*=a(t)r$. Later~\cite{Pachner:1963} and 
\cite{Noerdlinger.Petrosian:1971} argued for the presence of the 
acceleration term~(\ref{eq:cosmological-acc}) proportional to 
$\ddot{a}/a$ within this approximation scheme, hence arriving 
at~(\ref{eq:r-eq}). In the following we shall show how to 
arrive at (\ref{eq:r-eq}) from the exact geodesic equation of the 
McVittie metric by making clear the approximations involved. 
Related recent discussions were given in~\cite{Bolen.etal:2001},
where the effects of cosmological expansion on the periastron 
precession and eccentricity are discussed for constant Hubble 
parameter $H:=\dot{a}/a$. 

We will again work with the areal radius $R$. Note that for 
fixed $t$ the map $r\mapsto R(t,r)$ is 2-to-1 and that $R\geq 2m_0$, 
where $R=2m_0$ corresponds to $r=m_0/2a$. Hence we restrict the 
coordinate transformation (\ref{eq:McVittieArealRadius}) to the 
region $r> m_0/2a$ where it becomes a diffeomorphism onto the 
region $R> 2m_0$. (The region $R<2m_0$ was investigated in 
\cite{Nolan:1999b}.) Reintroducing factors of $c$, McVittie's 
metric assumes the (non-diagonal) form in the region $R>2m_0$ 
(i.e.~$r>m_0/2a(t)$)
\begin{equation}\label{eq:McVittieMetric2}
\begin{split}
\g =& \left( 1 - 2\mu(R) - h(t,R)^2  \right)c^2 \ed t^2\\
    & +\frac{2h(t,R)}{\sqrt{1-2\mu(R)}} c\,\ed t\,\ed R-\frac{dR^2}{1-2\mu(R)} 
     - R^2 \gStwo \,,
\end{split}
\end{equation}
where we put 
\begin{equation}\label{eq:Def-f-h}
  \mu(R) := \frac{m_0}{R} \,, \qquad{ } h(t,R) := \frac{H(t)R}{c} 
\end{equation}
with $H(t):= (\dot a/a)(t)$, as usual. 

The equations for a timelike geodesic (i.e.~parametrized with respect 
to proper time), $\tau \mapsto z^\mu(\tau)$ with $g(\dot z,\dot z)=c^2$, 
follows via variational principle from the Lagrangian 
$\mathcal{L}(z,\dot z)=(1/2)g_{\mu\nu}(z){\dot z^\mu}{\dot z^\nu}$.
Spherical symmetry implies conservation of angular momentum. 
Hence we may choose the particle's orbit to lie in the 
equatorial plane $\theta=\pi/2$. The constant modulus of angular 
momentum is  
\begin{equation}\label{eq:ConsAngMom}
  R^2\dot{\varphi} = L \,.
\end{equation}
The remaining two equations are then coupled second-order ODEs for 
$t(\tau)$ and $R(\tau)$. However, we may replace the first one by its 
first integral that results from $\g(\dot \bz,\dot \bz)=c^2$: 
\begin{equation}\label{eq:TimeEquation}
\begin{split}
  &\left( 1 - 2\mu(R) - h^2(t,R) \right)c^2 \dot t^2\\
  &+\frac{2h(t,R)}{\sqrt{1-2\mu(R)}} c\,\dot t\,\dot R
   -\frac{\dot R^2}{1-2\mu(R)}
   - (L/R)^2 =c^2\,.\\
\end{split}
\end{equation}
The remaining radial equation is given by 
\begin{equation}\label{eq:McV-R-sec}
\begin{split}
\ddot{R} \;
&-\left( 1-2\mu(R)-h^2(t,R) \right)\frac{L^2}{R^3}
\\
&+\frac{m_0\,c^2}{R^2}\big( 1-2\mu(R) \big)\,\dot{t}^2
\\
&-R\Big( \dot{H}(t)\,\big(1-2\mu(R)\big)^{1/2}
\\
&\phantom{-R\Big(}\;+H(t)^2\left( 1-\mu(R)-h^2(t,R) \right)
    \Big)\,\dot{t}^2
\\
&-\frac{\left( \mu(R)-h^2(t,R) \right)}
  {1-2\mu(R)}\,\frac{\dot R}{R}^2
\\
&+\frac{2\left( \mu(R)- h^2(t,R) \right)}
   {\sqrt{1-2\mu(R)}}\, c\,H(t)\,(\dot{R}/c)\,\dot{t} \;\; = \; 0 \,.
\end{split}
\end{equation}
Recall that $m_0=GM/c^2$, where $M$ is the mass of the central star in 
standard units.

Equations~(\ref{eq:TimeEquation},\ref{eq:McV-R-sec}) are exact.
We are interested in orbits of slow-motion (compared with the speed 
of light) in the region where
\begin{equation}\label{eq:R-region}
  R_S \ll R \ll R_H \,.
\end{equation}
Recall that $R_S$ and $R_H$ are the Schwarzschild and the Hubble 
radius, respectively 
(see~(\ref{eq:DefSchwarzschildRadius}),(\ref{eq:DefHubbleRadius})). 
The latter condition clearly covers all situations of practical 
applicability in the Solar System, since the Schwarzschild radius 
$R_S$ of the Sun is about 3~km $=2\cdot 10^{-8}$AU and the Hubble 
radius $R_H$ is about $13.7\cdot 10^{9}$\,ly = $8.7\cdot 10^{14}$\,AU.

The approximation now consists in considering small perturbations of 
Keplerian orbits. Let $T$ be a typical time scale of the problem, like 
the period for closed orbits or else $R/v$ with $v$ a typical velocity. 
The expansion is then with respect to the following two parameters: 
\begin{subequations}\label{eq:ExpParameters}
\begin{equation}\label{eq:ExpParameters1}
  \varepsilon_{\sss 1} 
  \approx\, \frac{v}{c} 
  \approx  \left(\frac{m_0}{R}\right)^{\frac{1}{2}}\,,
\end{equation}
corresponding to a slow-motion and weak-field approximation, and 
\begin{equation}\label{eq:ExpParameters2}
  \varepsilon_{\sss 2} \approx\, HT\,,
\end{equation}
\end{subequations}
corresponding to the approximation for small ratios of characteristic-times 
to the age of the universe. In order to make the expression to be 
approximated dimensionless, we multiply (\ref{eq:TimeEquation}) by 
$1/c^2$ and (\ref{eq:McV-R-sec}) by $T^2/R$. Then we expand the right 
hand sides in powers of the parameters (\ref{eq:ExpParameters}), using 
the fact that $h:=(HR/c)\approx\varepsilon_{\sss 1}\varepsilon_{\sss 2}$. 
From this and (\ref{eq:ConsAngMom}) we obtain (\ref{eq:impr-N-eqs}) if we 
keep only terms to zero-order in $\varepsilon_{\sss 1}$ and leading 
(i.e.~quadratic) order in $\varepsilon_{\sss 2}$, where we also 
re-express $R$ as function of $t$. Note that in this approximation 
the areal radius $R$ is equal to the spatial geodesic distance on 
the $t=\text{const.}$ hypersurfaces. 

We already mentioned that in the special case of constant $H=\dot a/a$
the McVittie solution turns into the Schwarzschild--de\,Sitter 
metric~(\ref{eq:SdS-spacetime}), which also describes the spacetime 
inside the Einstein--Straus vacuoles in case of non-vanishing 
cosmological constant. Recently, exact expressions in terms of 
hyper-elliptic integrals for the integrated geodesic equation in 
Schwarzschild--de\,Sitter spacetimes where derived 
in~\cite{Hackmann.Laemmerzahl:2008a,Hackmann.Laemmerzahl:2008b}. 
Moreover, a general discussion of Solar-System effects in 
Schwarzschild--de\,Sitter, like gravitational redshift, light 
deflection, time delay, perihelion precession, geodetic precession, 
and effects on Doppler tracking, has been given 
in~\cite{Kagramanova.etal:2006}. For example, it was found 
that a non-vanishing $\Lambda$ could account for the anomalous Pioneer 
acceleration if its value was $-10^{-37}\,\mathrm{m^{-2}}$, 
which is \emph{minus} $10^{15}$ times the current most probable 
value. That value would also give rise to a perihelion precession 
four orders of magnitude larger than the accuracy to which this effect 
has been measured today. 

We conclude by commenting on the geodesic equation in the generalizations 
of McVittie's model given in~\cite{Faraoni.Jacques:2007} and 
in~\cite{Sultana.Dyer:2005}. An essential feature which distinguish 
these solutions from the McVittie one, is that in the former the 
Weyl part of the MS energy $E_\ind{W}=a\,m$ is not a constant as for McVittie 
but varies in time, meaning that there is an accretion of cosmological 
matter by the inhomogeneity (see~\cite{Carrera.Giulini:2009a}). 
In view of the fact that the combination 
\begin{equation}\label{eq:m-over-r-McV}
  m/r = A^2E_\ind{W}/R \approx E_\ind{W}/R 
\end{equation}
contained in the McVittie ansatz gives (minus) the `Newtonian' part of the 
potential in the slow-motion and weak-field approximation (see 
Section~\ref{sec:MotionInMcVittie}), we deduce that in order to get 
the geodesic equation for the generalized McVittie models it suffices 
to substitute $m_0$ with $E_\ind{W}$ in the equation of motion derived in 
Section~\ref{sec:MotionInMcVittie}. This means that the strength of the 
central attraction varies in time, leading to an in- or out-spiraling of 
the orbits if $E_\ind{W}$ is increasing or decreasing, respectively.

\subsubsection{Exact condition for non-expanding circular orbits in McVittie spacetime}
\label{sec:nonExpCirOrbitsMcVittie}
Analogously to Section~\ref{sec:nonExpCirOrbitsFLRW}, where we ask whether 
there exist non-expanding circular orbits (i.e.~of constant areal radius) 
of the electron-proton system in a FLRW spacetime, we now ask whether there 
exist non-expanding circular orbits of an (uncharged) test particle around 
the central mass. The necessary and sufficient condition for this to happen 
follows from inserting $R=\textrm{const}$ in the radial part of the geodesic 
equation~(\ref{eq:McV-R-sec}) and using the normalization 
condition~(\ref{eq:TimeEquation}) of the four-velocity in order to eliminate 
$\dot t$. In terms of the dimensionless quantities $h(t):=RH(t)/c$, 
$l:=L/Rc$, and $\mu:=m_0/R$, the condition for the existence of non-expanding 
circular orbits can be given the following form: 
\begin{equation}\label{eq:circular-orbit-exact-McVittie}
  \frac{R}{c}\dot h = 
  \frac{\left( 1-2\mu-h^2 \right)\left( \mu(1+3l^2)-l^2-h^2 \right) }
       {(1+l^2)\sqrt{1-2\mu}} \,.
\end{equation}
As for the electron-proton system in an FLRW spacetime 
(see~(\ref{eq:circular-orbit-exact})) this is a first-order autonomous 
ODE for $h(t)$ and therefore the Hubble function. In the present case 
the ODE is even simpler since it has the elementary form 
$\dot h=p(h^2)$, where $p$ is a polynomial of degree two with 
constant coefficients. From~(\ref{eq:circular-orbit-exact-McVittie}), 
to leading order in the small quantities $\mu$, $l^2$, and $h^2$, 
we get the same approximate ODE~(\ref{eq:h-ODE-approx}) and hence 
the same approximate solutions~(\ref{eq:circ-orbit-approx-a-sol}). 
Hence, the same conclusions as drawn for the electron-proton system 
in FLRW apply here. 

From~(\ref{eq:circular-orbit-exact-McVittie}) it follows that stationary 
solutions $h(t)=\textrm{const}=:h_0$, corresponding to an exponentially-%
growing scale factor~(\ref{eq:circ-orbit-dS}) (and hence leading 
to a Schwarzschild--de\,Sitter spacetime), are those where $h_0$ satisfies 
\begin{equation}\label{eq:circular-orbit-exact-McVittie-stationary}
  \frac{l^2+h_0^2}{(1+3l^2)} = \mu \,,
\end{equation}
where we used that the first factor on the numerator  of the right-hand side 
of (\ref{eq:circular-orbit-exact-McVittie}) is nonzero, as can be immediately 
inferred from the normalization condition~(\ref{eq:TimeEquation}). 
Notice that for a vanishing Hubble parameter (that is for $h_0=0$) the above 
condition reduces to the third Kepler law in Schwarzschild spacetime, as 
expected. The effect of a non-vanishing Hubble parameter is again that we 
must provide the orbiting particle with a \emph{smaller} angular velocity 
(smaller $l$) in order to keep it on a non-expanding circular orbit with the 
\emph{same} radius. The largest radius at which in a McVittie spacetime 
with exponentially-expanding scale factor (that is a 
Schwarzschild--de\,Sitter spacetime) there is a non-expanding circular 
orbit follows from~(\ref{eq:circular-orbit-exact-McVittie-stationary}) in 
the limit $l\to 0$. Then the condition reduces to $h_0^2=\mu$ 
which, solving for $R$, gives $(R_S R_H^2/2)^{1/3}$. This, exactly 
corresponds to the critical radius~(\ref{eq:r-star-grav}), taking into 
account that $q_0=-1$ for an exponentially-growing scale factor.

\section{Kinematical effects}
\label{sec:KinematicalEffects}
In this section we discuss the influence of cosmic expansion 
upon measurements of relative distances, velocities, and 
accelerations. These kinematical notions loose their a\,priori
meaning in general spacetimes, in particular in time-dependent 
ones. Hence it is of utmost importance to carefully reconsider 
statements concerning such notions and their precise relations 
to locally observable quantities.

\subsection{Einstein- versus cosmological simultaneity}
\label{sec:SimultaneityComparison}
Misidentifications in the notion of simultaneity can give rise to 
apparent anomalies in velocities and acceleration. Such an effect 
has e.g.~been suggested in \cite{Rosales.Sanchez-Gomez:1998} and 
again in \cite{Rosales:2002} to be able to account for the PA.
Their argument says that in a spatially flat FLRW universe the 
mismatch between adapted cosmological coordinates on the one hand
and radar coordinates on the other just amount to an apparent 
difference in radial acceleration of magnitude (\ref{eq:HubbleTimesC}). 
We agree on the existence and conceptual importance of such an effect 
but we disagree on the magnitude, which seems to have been grossly 
overestimated as we will show below.  

The cause of such effects lies in the way one actually measures 
spatial distances and determines the clock readings they are 
functions of (a trajectory is a `distance' for each given `time'). 
The point is this: equations of motions give us, for example, 
simultaneous (with respect to cosmological time) spatial geodesic 
distances as functions of cosmological time. This is what we implicitly 
did in the Newtonian analysis. But, in fact, spacecraft ranging is done 
by exchanging electromagnetic signals. The notions of spatial distance 
and simultaneity thereby implicitly used are \emph{not} the same as 
those we referred to above. Hence the analytical expression of the 
`trajectory' so measured will be different. 

We first recall the local version of Einstein simultaneity in 
general spacetimes $(\M,\g)$. We take $ds=g_{\mu\nu}dx^\mu dx^\nu$ 
to carry the unit of length so that $d\tau=ds/c$ carries the 
unit of time. In general coordinates $\{x^\mu\}=\{t,x^i\}$, where $x^0=t$ 
denotes the timelike coordinate, the metric reads 
\begin{equation}\label{eq:MetricGen}
  ds^2 = g_{\mu\nu}dx^\mu\,dx^\nu
       = g_{tt}dt^2 + 2g_{ti}dt\,dx^i + g_{ij}dx^i\,dx^j\,.
\end{equation}
The observer at fixed spatial coordinates is given by the vector 
field (normalized to $\g(\bu,\bu)=c^2$)
\begin{equation}  \label{eq:observer}
  \bu = c\,\norm{\bpartial/\bpartial t}^{-1} \bpartial/\bpartial t
      = \frac{c}{\sqrt{g_{tt}}} \bpartial/\bpartial t \,.
\end{equation}
Consider the light cone with vertex $p\in\M$; one has $ds^2=0$, which 
allows to solve for $dt$ in terms of the $dx^i$ (all functions $g_{\mu\nu}$
are evaluated at $p$, unless noted otherwise): 
\begin{equation}\label{eq:LightConeDiff}
  dt_{1{,}2} = -\,\frac{g_{ti}}{g_{tt}}\,dx^i \pm 
  \sqrt{\left( \frac{g_{ti}g_{tj}}{g^2_{tt}}-\frac{g_{ij}}{g_{tt}} \right)
  dx^i\,dx^j}\,.
\end{equation}
The plus sign corresponds to the future light-cone at $p$, 
the negative sign to the past light cone. An integral line of $\bu$ 
in a neighborhood of $p$ cuts the light cone in two points,
$q_+$ and $q_-$. If $t_p$ is the time assigned to $p$, then 
$t_{q_+}=t_p+dt_1$ and $t_{q_-}=t_p+dt_2$. The coordinate-time 
separation between these two cuts is $t_{q_+}-t_{q_-}=dt_1-dt_2$,
corresponding to a proper time $\sqrt{g_{tt}}(dt_1-dt_2)/c$ for 
the observer $\bu$. This observer will associate 
a \emph{radar-distance} $dl_*$ to the event $p$ of $c/2$ times that 
proper time interval, that is: 
\begin{equation}\label{eq:RadarDist}
  dl_*^2 = h =
  \left(\frac{g_{ti}g_{tj}}{g_{tt}} - g_{ij}\right) dx^i\,dx^j\,.
\end{equation}

The event on the integral line of $\bu$ that the observer will 
call Einstein-synchronous with $p$ lies in the middle between 
$q_+$ and $q_-$. Its time coordinate is in first-order approximation 
given by $\tfrac{1}{2}(t_{q_+}+t_{q_-})=t_p+\tfrac{1}{2}(dt_1+dt_2)=t_p+dt$,
where  
\begin{equation}\label{eq:EinstSynShift}
  dt := \tfrac{1}{2}(dt_1+dt_2) = -\frac{g_{ti}}{g_{tt}}\,dx^i\,.
\end{equation}

This means the following: The integral lines of $\bu$ are parametrized 
by the spatial coordinates $\{x^i\}_{i=1,2,3}$. Given a point $p$, 
specified by the orbit-coordinates $x^i_p$ and the time-coordinate 
$t_p$, we consider a neighboring orbit of 
$\bu$ with orbit-coordinates $x^i_p+dx^i$. The event on the latter which 
is Einstein synchronous with $p$ has a time coordinate $t_p+dt$, where 
$dt$ is given by (\ref{eq:EinstSynShift}), or equivalently 
\begin{equation}\label{eqSynchConnForm}
  \theta := dt+\frac{g_{ti}}{g_{tt}}\,dx^i = 0\,.
\end{equation}
Using a differential geometric language we may say that
Einstein simultaneity defines a \emph{distribution} $\theta=0$. 

The metric (\ref{eq:MetricGen}) can be written in terms of the 
radar-distance metric $h$ (\ref{eq:RadarDist}) and the simultaneity 
1-form $\theta$ as follows: 
\begin{equation}\label{eq:MetricGenAlt}
  ds^2 = g_{\mu\nu}dx^\mu\,dx^\nu = g_{tt}\,\theta^2-h\,,
\end{equation}
showing that the radar-distance is just the same as the 
Einstein-simultaneous distance. A curve $\gamma$ in $\M$ intersects 
the flow lines of $\bu$ perpendicularly iff $\theta(\dot\gamma)=0$, 
which is just the condition that neighboring clocks along 
$\gamma$ are Einstein synchronized. 

We now apply the foregoing to isotropic cosmological metrics. 
In what follows we drop for simplicity the angular dimensions. 
Hence we consider metrics of the form
\begin{equation}\label{eq:CosMetric}
  ds^2 = c^2dt^2 - a(t)^2dr^2\,.
\end{equation}
The comoving observer field,  
\begin{equation}\label{eq:GeodObs}
  \bu = c\,\bpartial/\bpartial t \,,
\end{equation} 
is geodesic and of expansion $3H$. On a hypersurface of constant $t$ 
the radial geodesic distance is given by $a(t)r$. Making this distance 
into a spatial coordinate, $r_*$, we consider the coordinate transformation 
\begin{equation}\label{eq:CoordTrans}
  t\mapsto t_*:=t\,,\quad
  r\mapsto r_*:=a(t)r\,.
\end{equation}
The field $\bpartial/\bpartial t_*$ is given by 
\begin{equation}\label{eq:TstarVF}
  \bpartial/\bpartial t_* =
  \bpartial/\bpartial t - Hr\,\bpartial/\bpartial r\,,
\end{equation}
to which the observer field, 
\begin{equation}\label{eq:ObsFieldStar}
  \bu_* := c\,\norm{\bpartial/\bpartial t_*}^{-1}\,\bpartial/\bpartial t_*
\end{equation}
corresponds. In contrast to (\ref{eq:GeodObs}), whose flow connects 
comoving points of constant coordinate $r$, the flow of 
(\ref{eq:TstarVF}) connects points of constant geodesic distances, 
as measured in the surfaces of constant cosmological time. This could 
be called \emph{cosmologically instantaneous geodesic distance}. 
It is now very important to realize that this notion of distance is not the 
same as the radar distance that one determines by exchanging light signals 
in the usual (Einsteinian) way. Let us explain this in detail: 

From (\ref{eq:CoordTrans}) we have $adr=dr_*-r_*Hdt$, where 
$H:=\dot a/a$ (Hubble parameter). Rewriting the metric 
(\ref{eq:CosMetric}) in terms of $t_*$ and $r_*$ yields
\begin{alignat}{1}
  ds^2
  &\ =\ c^2(1-(Hr_*/c)^2)\,dt_*^2-dr_*^2+2Hr_*\,dt\,dr_*\nonumber\\
  &\ =\ \underbrace{c^2\Bigl\{1-(Hr_*/c)^2\Bigr\}}_{\textstyle g_{t_*t_*}}\
        \Bigl\{\,\underbrace{dt_*+\frac{Hr_*/c^2}{1-(Hr_*/c)^2}\,dr_*}_{%
        \textstyle\theta}\,\Bigr\}^2\nonumber\\
\label{eq:MetricNewCoord}
  &\quad-\ \underbrace{\frac{dr_*^2}{1-(Hr_*/c)^2}}_{\textstyle h}\,.
\end{alignat}
Hence the differentials of radar-distance and  time-lapse for 
Einstein-simultaneity are given by  
\begin{subequations}\label{eq:DiffStarQuant}
\begin{alignat}{2}
\label{eq:DiffRadDist}
  & dl_*&&\,=\,\frac{dr_*}{\sqrt{1-(Hr_*/c)^2}}\,,\\
\label{eq:DiffLapseEinstSim}
  & dt_*&&\,=\,-\,\frac{Hr_*/c^2}{1-(Hr_*/c)^2}dr_*\,.
\end{alignat}
\end{subequations}

Let the distinguished observer (us on earth) now move along the 
geodesic $r_*=0$. Integration of (\ref{eq:DiffStarQuant}) from $r_*=0$
to some value $r_*$ then gives the radar distance $l_*$ as well as 
the time lapse $\Delta t_*$ as functions of the cosmologically 
simultaneous geodesic distance $r_*$: 
\begin{subequations}\label{eq:IntStarQuant}
\begin{alignat}{2}
  & l_*&&\,=\,(c/H)\,\arcsin(H\,r_*/c)\nonumber\\
\label{eq:IntRadDist}
  &    &&\,=\, r_*\bigl\{1+\tfrac{1}{6}(Hr_*/c)^2 
         + \Ord\left((Hr_*/c)^3\right)\bigr\}\\
  & \Delta t_*&&\,=\,(1/2H)\,\ln\bigl(1-(H\,r_*/c)^2\bigr)\nonumber\\
\label{eq:IntLapseEinstSim1}
  &  &&\,=\, (r_*/c)\bigl\{-\tfrac{1}{2}(Hr_*/c) 
         + \Ord\left((Hr_*/c)^2\right)\bigr\}
\end{alignat}
\end{subequations}
Combining both equations in (\ref{eq:IntStarQuant}) allows to 
express the time-lapse in terms of the radar-distance: 
\begin{equation}
\begin{split}\label{eq:IntLapseEinstSim2}
  \Delta t_* &=H^{-1}\,\ln\bigl(\cos(H\,l_*/c)\bigr)\\
  &=(l_*/c)\bigl\{-\tfrac{1}{2}(Hl_*/c) + \Ord\left((Hl_*/c)^2\right)\bigr\}\,.\\
\end{split}
\end{equation}

\begin{figure}
\centering
\includegraphics[width=0.8\linewidth]{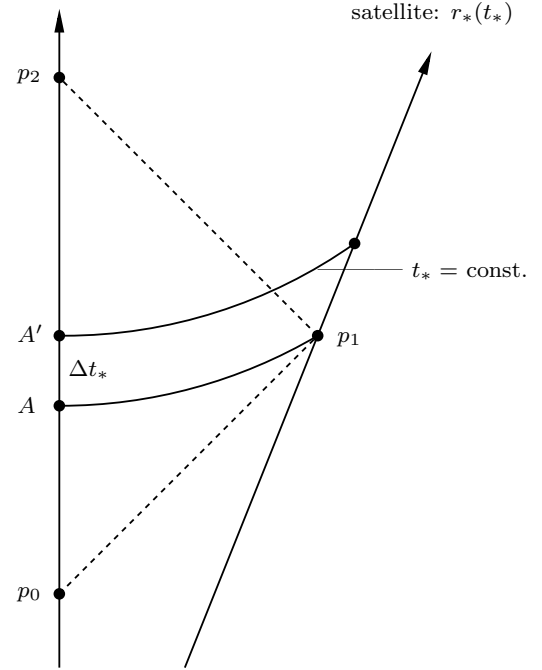}
\caption{\label{fig:SatOrbit}
An observer  moves on the geodesic worldline $r_*=0$ and 
observes a satellite by exchanging electromagnetic signals
(dashed line). 
Event $p_0$ corresponds to the signal emission by the observer, 
$p_1$ to its reflection by the satellite, and $p_2$ to its re-absorption 
by the observer. $t_*$ denotes the cosmological time and the 
two curved lines correspond to hypersurfaces of constant $t_*$. 
On the observer's worldline event $A$ is defined to be simultaneous 
to $p_1$ with respect to $t_*$ and $A'$ is defined to lie half-way 
in proper time between $p_0$ and $p_2$ on the observer's worldline. 
In cosmological time $A'$ is $\Delta t_*$ ahead of $A$. 
In cosmological coordinates, the satellite's trajectory is 
represented by the a function $t_*\mapsto r_*(t_*)$, where $r_*(t_*)$ 
denotes the proper geodesic distance in the hypersurface 
$t_*=\mathrm{const.}$  between its intersection points with both 
worldlines. However, using radar coordinates, the observer takes 
$A'$ to be simultaneous with $p_1$ and uses $l_*$ as measure for 
the satellite's simultaneous distance. Since $r_*$ and $l_*$ are 
related by (\ref{eq:IntRadDist}), it follows that the observer 
uses the function $t_*\mapsto l_*(r_*(t_*-\Delta t_*))$ to characterize 
the satellite's trajectory, which leads to~(\ref{eq:ObserverCurve}).}
\end{figure}

Now, suppose a satellite $S$ moves on a worldline $r_*(t_*)$ 
in the neighborhood of our worldline $r_*=0$. Assume that we measure 
the distance to the satellite by radar coordinates. Then instead of 
the value $r_*$ we would use $l_*$ and instead of the argument $t_*$ 
we would assign the time $t_*-\Delta t_*$ which corresponds to the 
value of cosmological time at that event on our worldline that is 
Einstein synchronous to the event $(t_*,r_*)$; see Fig.~\ref{fig:SatOrbit}. 
%
%
Hence we have 
\begin{subequations}\label{eq:ObserverCurve}
\begin{alignat}{1}
\label{eq:ObserverCurve1}
  l_*(t_*)&\,=\,(c/H)\ \sin^{-1}\bigl\{r_*(t_*+\Delta t_*)H/c\bigr\}\\
\label{eq:ObserverCurve2}
  &\,\approx\, r_*-\tfrac{1}{2}(v/c)(Hc)(r_*/c)^ 2\,,
\end{alignat}
\end{subequations}
where (\ref{eq:ObserverCurve2}) is (\ref{eq:ObserverCurve1}) to leading 
order and all quantities are evaluated at $t_*$. We set $v=\dot r_*$. 

To see what this entails we Taylor expand in $t_*$ around $t_*=0$ 
(just a convenient choice): 
\begin{equation}\label{eq:TaylorExpOrbit1} 
  r_*(t_*) = r_0+v_0t_*+\tfrac{1}{2}a_0t_*^ 2 +\cdots
\end{equation}
and insert in (\ref{eq:ObserverCurve2}). This leads to 
\begin{equation}\label{eq:TaylorExpOrbit2} 
  l_*(t_*) = \tilde r_0+\tilde v_0t_*+\tfrac{1}{2}\tilde a_0t_*^ 2 +\cdots\,,
\end{equation}
where, 
\begin{subequations}\label{eq:TaylorExpOrbitAll}
\begin{alignat}{2}
\label{eq:TaylorExpOrbit3a}
  &\tilde r_0 &&= r_0-(Hc)\ \tfrac{1}{2}(v_0/c)(r_0/c)^2\\
\label{eq:TaylorExpOrbit3b}
  &\tilde v_0 &&= v_0-(Hc)\ (v_0/c)^2(r_0/c)\\
\label{eq:TaylorExpOrbit3c}
  &\tilde a_0 &&= a_0-(Hc)\ \bigl\{(v_0/c)^3+(r_0/c)(v_0/c)(a_0/c)\bigl\} \,. 
\end{alignat}
\end{subequations}
These are, in quadratic approximation, the sought-after relations 
between the quantities measured via radar tracking (tilded) and the 
quantities which arise in the (improved) Newtonian equations of motion 
(not tilded). 

In particular, the last equation (\ref{eq:TaylorExpOrbit3c}) shows that 
there is an apparent inward pointing acceleration, given by $Hc$ times the 
$(v/c)^3+\cdots$ term in curly brackets. As discussed in the introduction,
$Hc$ is indeed of the same order of magnitude as the PA, as was much
emphasized in \cite{Rosales.Sanchez-Gomez:1998,Rosales:2002}. 
However, in contrast to these authors, we also get the additional 
term in curly brackets, which in case of the Pioneer spacecraft suppresses 
the $Hc$ term by 13 orders of magnitude!%
\footnote{Our Eq.~(\ref{eq:IntLapseEinstSim2}) corresponds to Eq.~(10) of 
\cite{Rosales.Sanchez-Gomez:1998}. From it the authors of 
\cite{Rosales.Sanchez-Gomez:1998} and \cite{Rosales:2002} 
immediately jump to the conclusion that there is ``an effective 
residual acceleration directed toward the center of coordinates; 
its constant value is $Hc$''. We were unable to follow this 
conclusion. Likewise, we are unable to follow the conclusion in 
\cite{Fahr.Siewert:2008}.} Hence we conclude that, with respect to the 
PA, there is no significant kinematical effect resulting from the distinct 
simultaneity structures inherent in radar and cosmological coordinates.

\subsection{Doppler tracking in cosmological spacetimes}
\label{sec:DopplerTracking}
Doppler Tracking is a common method of tracking the position of 
vehicles in space. It involves measuring the Doppler shift of 
an electromagnetic signal sent from a spacecraft to a tracking 
station on Earth.  This signal is either coming from an on-board 
oscillator or is coherently transponded by the vehicle in 
response to a signal received from the ground station. Here we 
focus on the second of these modes, which  is more useful for 
navigation, partly because the returning signal is measured 
against the same frequency reference as that of the originally 
transmitted signal and partly because the Earth-based frequency 
reference is also more stable than the oscillator on-board 
the spacecraft.

\subsubsection{Minkowski spacetime}
It is clear that this method will be fundamentally influenced 
if performed within a time varying background geometry. 
Before elaborating on this, we consider the simple case of 
static Minkowski space. 

\begin{figure}[ht]
\centering
\includegraphics[width=0.75\linewidth]{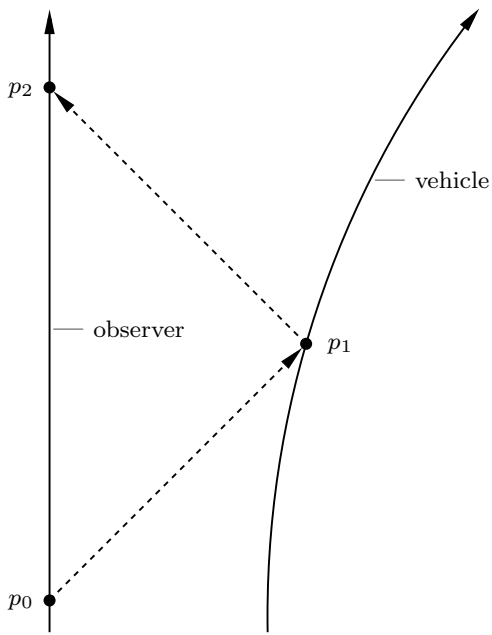}
\caption{\label{fig:SignalExchange}
Exchange of electromagnetic signals (represented by their rays at a 
slope of 45 degrees) between us and the space vehicle. Time runs 
vertically.}
\end{figure}
%
In Fig.~\ref{fig:SignalExchange} we depicted two worldlines, one of the 
observer (straight vertically) and one for the space vehicle. A light 
signal is emitted by the observer at the event $p_0$, reflected by the 
vehicle at event $p_1$, and finally received back by the observer at event 
$p_2$. We choose a global Minkowski frame, that is global coordinates 
$\{x^\mu\}=\{t,x^i\}$ with 
$\g(\bpartial/\bpartial x^\mu,\bpartial/\bpartial x^\nu)=\diag(1,-1,-1,-1)$, 
in which the observer (for simplicity assumed to be inertial) is at rest at 
the origin of the spatial coordinates.  If $\beta:=v/c$ denotes the 
radial velocity of the vehicle in units of $c$, the well known 
special-relativistic Doppler formula (applied twice) says that 
the ratio between the received and the emitted 
frequencies is\footnote{\label{fnote:Doppler}
Note that in Special Relativity the Doppler Formula does, of course, 
not distinguish between moving emitter and moving receiver. 
So (\ref{eq:MinkFrequencyRatio}) is obtained by squaring the frequency 
shift $\sqrt{(1-\beta)/(1+\beta)}$, which is picked up once for the 
ratio $\omega_R(t_R)/\omega_1(t_1)$ (receiver moving relative to the 
Minkowski frame) and once for $\omega_2(t_2)/\omega_1(t_1)$ (emitter 
moving relative to Minkowski frame). Incidentally, exactly the same 
formula would result in non-relativistic physics if the observer is 
taken to be at rest with respect to the wave-guiding medium (e.g., 
the ether), which distinguishes the two states of relative motion. 
Indeed, in this case we have $\omega_1(t_1)/\omega_0(t_0)=(1-\beta(t_1))$ 
and $\omega_2(t_2)/\omega_1(t_1)=1/(1-\beta(t_1))$, whose product 
is again just (\ref{eq:MinkFrequencyRatio}).}
\begin{equation}\label{eq:MinkFrequencyRatio}
  \frac{\omega_2(t_2)}{\omega_0(t_0)}
  = \frac{1-\beta(t_1)}{1+\beta(t_1)}\,.
\end{equation}
Here $t_0,t_1$, and $t_2$ refer to the global Minkowski time attributed to 
$p_0,p_1$, and $p_2$, respectively.  In Doppler tracking 
one is interested in the derivative of this ratio with respect to $t_2$,
which yields a measure for the velocity of the spacecraft. We assume 
$\omega_0$ to be constant in time and note that, given the worldlines 
of the observer and the vehicle, $t_1$ and $t_0$ are uniquely determined 
by $t_2$ (since the events $p_1$ and $p_0$ are determined by $p_2$).  
If $r$ denotes the spatial radius coordinate we have 
$t_2-t_1=r(t_1)/c$. Differentiation with respect to $t_1$ leads to  
\begin{equation}\label{eq:MinkTimeDer}
  \frac{dt_1}{dt_2}=\frac{1}{1+\beta(t_1)}
\end{equation}
and hence 
\begin{equation}\label{eq:MinkFrequencyRatioDer}
\begin{split}
  -\,\frac{1}{2}\,\frac{\dot\omega_2(t_2)}{\omega_0(t_0)}
  & = \dot\beta(t_1)\,\bigl(1+\beta(t_1)\bigr)^{-3}\\
  & = \dot\beta(t_1)\,\bigl(1-3\beta(t_1) + \Ord(\beta^2)\bigr)\,.
\end{split}
\end{equation}
This shows that $-\dot\omega_2/2\omega_0$, namely (minus one-half) the 
derivative of the received to emitted frequency ratio with respect to the 
proper time of the receiving observer, gives the spacecraft's spatial 
acceleration \emph{up to corrections of order} $\beta$. Note that in 
view of note~\ref{fnote:Doppler} it would be inappropriate to call these
corrections `special relativistic'. 

The final goal of this section is to derive the generalization 
of~(\ref{eq:MinkFrequencyRatioDer}) for a cosmological spacetime. 
For this we need two things: First, we need to know what is the 
generalization of the concepts of spatial velocity and spatial acceleration 
in an arbitrary spacetime and, second, we need to know how electromagnetic 
signals propagate in an arbitrary spacetime. This is taken care of in the 
next paragraph.

\subsubsection{General setting}
In order to generalize the notions of spatial velocity and spatial 
acceleration to arbitrary spacetime one needs to introduce a 
fiducial reference `observer-field'; compare, e.g.,~\cite{Bini.etal:1995} 
and also \cite{Carrera:2009}). 
An \emph{observer} at the event $p$ is a future pointing unit timelike 
vector in the tangent space $T_p(\M)$ of $\M$ at $p$. An \emph{observer field} 
is a field of future pointing unit timelike vectors. Any observer 
$\bu$ at $p$ gives rise to an orthogonal split of the tangent space $T_p(\M)$ 
at $p$ in a part parallel to $\bu$ (the local time-axis) and a part orthogonal
to it (the local rest space). Since $\bu$ is not lightlike the two 
orthogonal subspaces are complementary, that is, together they span the 
whole tangent space and intersect only in the zero vector. 
The orthogonal projections of an arbitrary vector $\bX\in T_p(\M)$ 
onto these subspaces are, respectively, given by
\begin{subequations}\label{eq:SplitOperators}
\begin{alignat}{3}
\label{eq:SplitOperators-a}
  &\bQ_\bu(\bX)&&:=\bg(\bX,\bu)\bu \,,\\
\label{eq:SplitOperators-b}
  &\bP_\bu(\bX)&&:=\bX-\bg(\bX,\bu)\bu\,,
\end{alignat}
\end{subequations}
which imply the decomposition identity $\bX=\bQ_\bu(\bX)+\bP_\bu(\bX)$.

If two observers $\bu$ and $\bv$ are defined at the same point, 
the \emph{spatial velocity} (over~$c$) of $\bv$ with respect to $\bu$ 
is given by 
\begin{equation}\label{eq:DefRelVel}
  \bbeta_\bu(\bv)
  :=\frac{\bP_\bu(\bv)}{\norm{\bQ_\bu(\bv)}}
   =\frac{\bv-\bg(\bv,\bu)\bu}{\bg(\bv,\bu)} \,,
\end{equation}
which is an element of the local rest space $\bP_\bu T(\M)$. Its  modulus 
is given by 
\begin{equation}\label{eq:ModulusRelVel}
  \beta_\bu(\bv)
  :=\norm{\bbeta_\bu(\bv)}
   =\sqrt{1-1/\bg(\bu,\bv)^2}\,.
\end{equation}
Note that for the modulus we have $\beta_\bu(\bv)=\beta_\bv(\bu)$, though 
the vectors $\bbeta_\bu(\bv)$ and $\bbeta_\bv(\bu)$ are linearly independent
as they lie in $\bP_uT(\M)$ and $\bP_v T(\M)$, respectively. Note also that 
$\bg(\bu,\bv)=1/\sqrt{1-\beta_\bu^2(\bv)}$ is just the ordinary 
`gamma-factor'. Finally, if $\be\in \bP_\bu T(\M)$ is a unit vector, we define 
the spatial velocity of $\bv$ in direction $\be$ w.r.t.~$\bu$ by 
%
\begin{equation}\label{eq:DefRelVelComp}
  \bbeta^\be_\bu(\bv)
  = -\bg(\be,\bbeta_\bu(\bv))= -\,\frac{\bg(\be,\bv)}{\bg(\bu,\bv)}\,.
\end{equation}

The \emph{spatial acceleration} of a worldline $\gamma$ w.r.t.~a given 
observer field $\bu$ is defined as the rate of change of the spatial velocity 
$\bbeta_{\bu}(\dot\bgamma)$ within the local rest spaces $\bP_\bu T(\M)$ of 
$\bu$ and with respect to the clocks moving along $\bu$. Denoting this 
acceleration (divided by $c$) with $\balpha$, we have 
\begin{equation}\label{eq:DefRelSpatAcc}
  \balpha_\bu(\gamma):=\nabla^\bu_{\dot\bgamma}\bbeta_\bu(\dot\bgamma) \,,
\end{equation}
where we used the following covariant derivative for $\bP_\bu T(\M)$-valued 
vector fields along $\gamma$:
\begin{equation}\label{eq:DefSpatialCovDer}
  \nabla^\bu_{\dot\bgamma}:=\norm{\bQ_\bu\dot\bgamma}^{-1}\
  \bP_\bu\circ\nabla_{\dot\bgamma}\circ \bP_\bu\,.
\end{equation}
Here $\nabla_{\dot\bgamma}$ denotes the ordinary (Levi-Civita) 
covariant derivative along $\gamma$. As an application one can, for example, 
rewrite the geodesic equation, $\bnabla_\gammadot\gammadot=0$, for a 
worldline $\gamma$ in terms of the spatial quantities just introduced. 
One gets (see~\cite{Carrera:2009} or, in a slightly different 
notation,~\cite{Bini.etal:1995}):
\begin{equation}\label{eq:spatial-eom}
  \balpha_\bu = - \bS_{\bbeta_\bu}\bigl[ \ba_\bu + 
  \btheta_\bu(\bbeta_\bu) + \bomega_\bu(\bbeta_\bu) \bigr] \,,
\end{equation}
where for better readability we omitted the arguments $\gamma$ and 
$\gammadot$ in the spatial acceleration and spatial velocity.
Here $\ba_\bu:=\bnabla_\bu\bu$ is the four-acceleration of the observer 
field $\bu$, $\btheta_\bu$ and $\bomega_\bu$ are, respectively, its 
shear-expansion and rotation tensors of rank $(1,1)$ (endomorphism), 
and $\bS_{\bbeta_\bu}:=\bP_\bu+\bbeta_\bu\otimes\ul{\bbeta_\bu}$ is a
rank $(1,1)$ tensor which, in a slow-motion approximation (neglecting 
quadratic and higher terms in $\beta$), reduces to the identity on the 
local rest space of $\bu$. 
Equation~(\ref{eq:spatial-eom}) should be seen as a local version of 
Newton's equation. For example, in Schwarzschild--de\,Sitter 
spacetime~(\ref{eq:SdS-spacetime}), taking $\bu$ to be proportional to 
the timelike Killing field $\bpartial/\bpartial T$, one has 
$\btheta_\bu=\bomega_\bu=0$ (because of the Killing equation and spherical 
symmetry, respectively) and 
\begin{equation}\label{eq:acceleration-u-SdS}
  \ba_\bu = \bnabla_\bu\bu = \frac{1}{\sqrt{V}} 
  \left( \frac{m}{R^2}-\frac{\Lambda}{3} R \right)\be_R\,,
\end{equation}
where $\be_R$ denotes the 
the normalized radial vector field $\bpartial/\bpartial R$ (we use here 
the coordinates and the notation of~(\ref{eq:SdS-spacetime})). Hence, in 
slow-motion and weak-field approximation (that is keeping only linear 
terms in $\beta$, $m/R$, and $\Lambda R^2$), the geodesic equation of 
motion in the form~(\ref{eq:spatial-eom}) reduces to 
\begin{equation}\label{eq:geod-eq-SdS-spatial}
  \balpha_\bu \approx \left( -\frac{m}{R^2}+\frac{\Lambda}{3}R \right)\be_R\,,
\end{equation}
which just gives the `improved' Newtonian equation for geodesic motions 
in Schwarzschild--de\,Sitter spacetime. It has the same form as the 
improved Newtonian equation studied in Section~\ref{sec:NewtonianApproach}. 

We turn now to electromagnetic signals and restrict our attention to 
monochromatic waves in the geometric-optics approximation (i.e.~for 
wave-lengths negligibly small w.r.t.~a typical radius of curvature of the 
spacetime and w.r.t.~a typical length over which amplitude, polarization, 
and frequency vary). In this approximation an electromagnetic signal 
propagates on a lightlike geodesic along which the wave-vector, $\bk$, 
is tangent, future-pointing, and parallelly transported. Recall that 
$\bk$ is so normalized that the frequency measured by an 
observer, say $\bu$, is 
\begin{equation}\label{eq:Def-frequency}
  \omega_\bu(\bk):=\bg(\bu,\bk)\,.
\end{equation}
Given a wave-vector $\bk$ and two observers $\bu,\bv$ at the same spacetime 
point, their observed frequencies are thus $\omega_\bv(\bk)=\bg(\bv,\bk)$ 
and $\omega_\bu(\bk)=\bg(\bu,\bk)$, and their ratio is given by 
\begin{equation}\label{eq:DopplerFormula2}
  \frac{\omega_\bv(\bk)}{\omega_\bu(\bk)}=
  \frac{\bg(\bQ_\bu\bv+\bP_\bu\bv,\bk)}{\bg(\bu,\bk)}=
  g(\bu,\bv)\bigl[ 1-\beta^{\hat\bk}_\bu(\bv) \bigr]\,.
\end{equation}
Here the spacelike unit vector $\hat\bk:=\norm{\bP_\bu(\bk)}^{-1}\bP_\bu(\bk)$ 
defines the direction of $\bk$ in the local rest space of $\bu$. In deriving 
(\ref{eq:DopplerFormula2}) we used (\ref{eq:DefRelVelComp}) 
and $\norm{\bP_\bu(\bk)}=\bg(\bu,\bk)$ to write 
$\g(\bv,\bP_\bu(\bk)) = -\bg(\bv,\bu)\bg(\bu,\bk)\beta^{\hat\bk}_{\bu}(\bv)$.
Equation~(\ref{eq:DopplerFormula2}) is the general form of the 
Doppler formula.  

Let now $\bu$ be an observer field along one integral line 
of which the distinguished observer is moving. The worldline of the 
vehicle is denoted by $\gamma$.  The domain of the field $\bu$ is
assumed to include a neighborhood of $\gamma$. The wave-vector 
$\bk_0$ emitted at $p_0$ suffers three changes:
\begin{compactenum}
\item propagation from $p_0$ to $p_1$: $\bk_0 \rightarrow \bk_1$;
\item reflection at $p_1$:             $\bk_1 \rightarrow \bk'_1$;
\item propagation from $p_1$ to $p_2$: $\bk'_1\rightarrow \bk_2$.
\end{compactenum}
We are interested in the ratio of the received to the emitted 
frequency:
\begin{equation}\label{eq:OmegaRatio}
\frac{\omega_2}{\omega_0}
=\frac{\bg(\bu_2,\bk_2)}{\bg(\bu_0,\bk_0)}
=\left(\frac{\omega_2}{\omega'_1}\right)
 \left(\frac{\omega'_1}{\omega_1}\right)
 \left(\frac{\omega_1}{\omega_0}\right)\,.
\end{equation}

What happens at reflection (the second process: $\bk_1\rightarrow \bk'_1$)?
Well, with respect to the spacecraft moving along $\gamma$ with 
four-velocity $\bv=\dot\bgamma$, the wave vector $\bk_1$ at $p_1$ 
splits according to
\begin{equation}\label{eq:WaveVectorSplitOn Gamma}
  \bk_1 = \bQ_{\dot\bgamma}(\bk_1)+\bP_{\dot\bgamma}(\bk_1)\,.
\end{equation}
A corner-cube reflector transported along $\gamma$ will 
reverse $\bP_{\dot\bgamma}(\bk_1)$ while keeping $\bQ_{\dot\bgamma}(\bk_1)$ 
intact (here we neglect a possible transponder shift which is 
irrelevant for our discussion): 
\begin{equation}\label{eq:CornerCubeReflection}
  \bk_1\mapsto \bk'_1
  = \bQ_{\dot\bgamma}(\bk_1)-\bP_{\dot\bgamma}(\bk_1)
  = 2\bQ_{\dot\bgamma}(\bk_1)-\bk_1\,.
\end{equation}
Hence $\omega_1:=\omega_\bu(\bk_1)=\bg(\bu_1,\bk_1) $ and 
$\omega'_1:=\omega_\bu(\bk'_1)=\bg(\bu_1,\bk'_1)$,
the in- and out-going frequencies measured by the observer 
$\bu$ at $p_1$, are related by  
\begin{equation}\label{eq:ReflectionFrequencyShift}
\frac{\omega'_1}{\omega_1}
  = 2\ \frac{\bg(\bu,\dot\bgamma)\bg(\dot\bgamma,\bk)\vert_{p_1}}%
          {\bg(\bu,\bk)\vert_{p_1}}-1
  = 2\ \frac{1-\bbeta^{\hat\bk}_\bu(\dot\bgamma)\vert_{p_1}}%
          {1-\beta^2_u(\dot\bgamma)\vert_{p_1}}-1 \,,
\end{equation}
where in the last step we just used~(\ref{eq:DopplerFormula2})  
to rewrite the ratio $\bg(\dot\bgamma,\bk)/\bg(\bu,\bk)$. This accounts 
for the middle ratio on the right-hand side of~(\ref{eq:OmegaRatio}).  

To account for the other two ratios in~(\ref{eq:OmegaRatio}), one 
uses the laws of geometric optics in (curved) spacetime to relate 
$\omega_0=\g(\bu_0,\bk_0)$ (at $p_0$) and $\omega_2=\g(\bu_2,\bk_2)$ 
(at $p_2$) to kinematical quantities of $\gamma$ at $p_1$.%
\footnote{In general spacetimes without timelike conformal 
Killing fields these quotients will also explicitly 
depend on time.} For example, if $\bu$ is a Killing field 
(like $\bu=\bpartial/\bpartial t$ in Special Relativity), we have 
$\g(\bu_0,\bk_0)=\g(\bu_1,\bk_1)$ and $\g(\bu_2,\bk_2)=\g(\bu_1,\bk'_1)$, 
so that  
\begin{equation}\label{eq:SRDoubleDopplerFormula}
  \frac{\omega_2}{\omega_0}
  =2\ \frac{1-\beta_\bu^{\hat \bk}(\gammadot)\vert_{p_1}}%
           {1-\beta^2_\bu(\gammadot)\vert_{p_1}} - 1 \,.
\end{equation}
As a trivial application, this includes the generalized form 
of (\ref{eq:MinkFrequencyRatio}), the latter corresponding 
to purely radial motion.

\subsubsection{FLRW spacetimes}
In standard cosmological spacetimes (FLRW), $\bu=\bpartial/\bpartial t$ is 
not Killing, though $\bX=a(t)\bpartial/\bpartial t$ is conformally Killing 
($L_\bX \g = 2\dot a\bg$).
One now has $a_0\bg(\bu_0,\bk_0)=a_1\bg(\bu_1,\bk_1)$ and 
$a_2\bg(\bu_2,\bk_2)=a_1\bg(\bu_1,\bk'_1)$, so that instead of 
(\ref{eq:SRDoubleDopplerFormula}) one gets
\begin{equation}\label{eq:FLRWDoubleDopplerFormula}
  \frac{\omega_2}{\omega_0}
  = \frac{a_0}{a_2}\ 
  \left\{2\ \frac{1-\beta_u^{\hat k}(\dot\gamma)\vert_{p_1}}%
                 {1-\beta^2_u(\dot\gamma)\vert_{p_1}}-1\right\} \,.
\end{equation}

We now want to relate the $t_2$-derivative of 
(\ref{eq:FLRWDoubleDopplerFormula}) to the acceleration of 
$\gamma$. 
In order to calculate the derivative $\dot\omega_2(t_2)/\omega_0(t_0)$
we need to know the derivatives $dt_1/dt_2$ and $dt_0/dt_2$. 
Restricting to the flat FLRW case for simplicity, they follow from 
the law of null propagation: 
\begin{subequations}\label{eq:TimeDerivatigves1}
\begin{alignat}{2}
\label{eq:TimeDerivatives1-a}
  & \int_{t_1(t_2)}^{t_2}\frac{dt}{a(t)}
  &&\,=\,-\frac{1}{c}\int_{r_1(t_1(t_2))}^{r_2}dr\,,\\
\label{eq:TimeDerivatives1-b}
  & \int_{t_0(t_2)}^{t_2}\frac{dt}{a(t)}
  &&\,=\,\frac{1}{c}\left\{
  \int_{r_0}^{r_1(t_1(t_2))}dr\!\!-\int_{r_1(t_1(t_2))}^{r_2}dr\right\}.
\end{alignat}
\end{subequations}
Differentiation with respect to $t_2$ yields, respectively
\begin{subequations}\label{eq:TimeDerivatigves2}
\begin{alignat}{2}
\label{eq:TimeDerivatives2-a}
  & \frac{dt_1}{dt_2}
  &&\,=\,\frac{a(t_1)}{a(t_2)}\,
  \left(1+\beta_\bu^{\hat\bk}(\dot\bgamma)\vert_{p_1}\right)^{-1}\,,\\
\label{eq:TimeDerivatives2-b}
  & \frac{dt_0}{dt_2}
  &&\,=\,\frac{a(t_0)}{a(t_2)}\,
  \frac{1-\beta_\bu^{\hat\bk}(\dot\bgamma)\vert_{p_1}}%
       {1+\beta_\bu^{\hat\bk}(\dot\bgamma)\vert_{p_1}}\,.
\end{alignat}
\end{subequations}

The exact formula for the $t_2$--derivative of the frequency-shift rate 
can now be computed. One obtains
\begin{equation}\label{eq:FLRWShiftRateExact}
\begin{split}
  &-\,\frac{\dot\omega_2(t_2)}{\omega_0(t_0)}=\,\\
  &\frac{a_0}{a_2}\,\Bigg\{
  2\bigl[\alpha^{\hat\bk}-\bg(\bbeta,\nabla^\bu_{\dot\bgamma}\hat\bk)\bigr]
  \frac{a_1}{a_2}\bigl[1+\beta^{\hat k}\bigr]^{-1}\bigl[1-\beta^2\bigr]^{-1}\\
  &+4\bg(\balpha,\bbeta)\frac{a_1}{a_2}
  \left[\frac{1-\beta^{\hat k}}{1+\beta^{\hat k}}\right]
  \bigl[1-\beta^2\bigr]^{-2}\\
  &+\left[\frac{\dot a_2}{a_2}-\frac{\dot a_0}{a_2}
  \left(\frac{1-\beta^{\hat k}}{1+\beta^{\hat k}}\right)\right]
  \left[\frac{1-2\beta^{\hat k}+\beta^2}{1-\beta^2}\right]\Bigg\}\,,\\
\end{split}
\end{equation}
where we suppressed the argument $\dot\bgamma$ and index $\bu$ at $\beta$ 
for better readability. This formula provides an exact relation between 
the time derivative of the observable frequency shift (defined `here') 
and the kinematical quantities of the vehicle (defined `there'), provided 
the scale function $a(t)$ is known. For purely radial motion 
$\nabla^u_{\dot\gamma}\hat \bk=0$ and we obtain the simpler expression 
(now writing $\alpha$ for~$\alpha^{\hat\bk}$)
\begin{equation}\label{eq:FLRWShiftRateExactRadial} 
\begin{split}
  -\,\frac{1}{2}\,\frac{\dot\omega_2(t_2)}{\omega_0(t_0)}
  &=-\,\frac{a_0a_1}{a^2_2}\,\Bigg\{\alpha(1+\beta)^{-3}+\\
  &\frac{1}{2}\,
  \left[ \frac{\dot a_2}{a_1}-\frac{\dot a_0}{a_1}
         \left(\frac{1-\beta}{1+\beta}\right) \right]
  \left[ \frac{1-\beta}{1+\beta} \right]
  \Bigg\}\,.
\end{split}
\end{equation}     
In order to consistently approximate this expression in terms of small 
quantities $\beta$ and $H\Delta t$, where $\Delta t:=(t_2-t_0)/2$, we 
think of (\ref{eq:FLRWShiftRateExactRadial}) as being multiplied with 
$\Delta t$ and regard $\alpha\Delta t$ as being of order $\beta$. 
Then, keeping only quadratic terms in $\beta$, linear terms in 
$H\Delta t$ where $\Delta t:=(t_2-t_0)/2$, and also mixed terms 
$\beta H\Delta t$, we get   
\begin{equation}\label{eq:FLRWShiftRateApproxRadial}
  -\,\frac{1}{2}\,\frac{\dot\omega_2(t_2)}{\omega_0(t_0)}
  \approx \alpha\bigl(1-3\beta-3H\Delta t\bigr)+H\beta\,.
\end{equation}
Hence we see that in this approximation there are two 
modifications, besides the $-3\beta$-term already familiar 
from (\ref{eq:MinkFrequencyRatioDer}), due to cosmic 
expansion: First, there is an additional contribution 
$-3H\Delta t$ acting in the same way as the $-3\beta$-term. 
It can also be interpreted in the same fashion, as its 
corresponds to the velocity (over $c$) of $H\Delta t$ 
that a comoving systems picks up during the time the 
signal went from the observer to the vehicle. 
Second, there is a constant contribution $H\beta$ to 
acceleration/$c$, i.e.~$Hc\beta$ to acceleration, in
a direction parallel to the radial velocity (i.e.~outward 
pointing if the vehicle recedes from the observer). 
Hence it acts opposite to the PA and is smaller in modulus 
by a factor of $\beta$. Applied to the Pioneer spacecrafts, 
the $H\Delta t$-term amounts to a tiny `anomalous' 
acceleration of $\Delta a/a <10^{-12}$, the $Hc\beta$-term 
to $\Delta a/a <10^{-7}$.

A final point must be made regarding the choice of the reference 
observer-field on which the kinematic quantities related to the 
spacecraft (spatial velocity and spatial acceleration) and the 
electromagnetic signal (frequency and spatial propagation direction) 
crucially depend. In the Minkowskian case the reference field 
was just $\bu=\bpartial/\bpartial t$, which is inertial, that is, 
geodesic and of vanishing rotation, shear, and expansion. It is 
clear that in a general spacetime such observer fields do not 
exist and there is no natural choice to replace them. However, 
in the case of spherical symmetry there is, in fact, a 
distinguished observer field, namely that one whose orbits lie 
within the timelike hypersurfaces of constant areal radius and 
there run perpendicular to the orbits of the rotation group. 
This clearly defines  a  non-rotating and `non-expanding' (w.r.t.~the 
areal radius) reference field. It is the normalization of 
the so-called Kodama vector field, which we discuss in detail in 
Appendix~\ref{sec:MS-energy}. In a FLRW spacetime it is just 
given by~(\ref{eq:ObsFieldStar}). Notice that in the present case, 
where the hypersurfaces of constant cosmological time $t$ are flat, 
the areal radius corresponds also to the proper distance. Hence 
the integral curves of $\bu_*$ intersect the hypersurface of constant 
cosmological time at constant spatial geodesic distance.
More precisely, the expansion and the shear scalar of $\bu_*$ are given 
by $\theta_*=R^2H\dot{H}/(1-(RH)^2)^{3/2}$ and $\sigma_*=-\theta_*/3$, 
respectively, showing that they are of order $H^3$ which we neglect. 
In passing we remark that the expansion and shear of $\bu_*$ exactly 
vanish for the de\,Sitter case, whose metric in `static' coordinates 
is given by~(\ref{eq:SdS-spacetime}) for $m=0$. In this case 
$\bpartial/\bpartial t_*=\bpartial/\bpartial T$, that is, $\bu_*$ is 
proportional to the timelike Killing vector field 
$\bpartial/\bpartial T$; see~(\ref{eq:SdS-spacetime}). Coming back 
to the general FLRW case, the acceleration of $\bu_*$ is given by 
$\ba_{\bu_*}=(-R\ddot{a}/a+R^3H^4)/(1-(RH)^2)^{3/2}\be_*$, where 
$\be_*$ is the unit vector field orthogonal to $\bu_*$ and to the 
two-sphere, pointing in positive radial direction. Hence, in the slow-motion 
and weak-field approximation of Eq.~(\ref{eq:FLRWShiftRateApproxRadial}), 
but keeping also quadratic terms in $H$, the geodesic equation in the 
form~(\ref{eq:spatial-eom}) w.r.t.~the observer field $\bu_*$ reads as 
\begin{equation}\label{eq:spatial-g-e-FLRW-ustar}
  \balpha_{\bu_*}(\gamma) \approx 
  \bigg( \, \frac{\ddot a}{a}\, r_* \bigg) \;\be_* \circ \gamma \,.
\end{equation}
This is just an alternative derivation of the acceleration 
term~(\ref{eq:cosmological-acc}). We point out that had we we 
taken~(\ref{eq:GeodObs}) as observer field we would have arrived at 
the equation of motion $\balpha_\bu(\gamma) \approx - H \bbeta_\bu(\gamma)$
instead of~(\ref{eq:spatial-g-e-FLRW-ustar}), that is, no acceleration 
term~(\ref{eq:cosmological-acc}) would have resulted.

In the approximation within which~(\ref{eq:FLRWShiftRateApproxRadial}) 
is derived this equation remains valid if the quantities in it are 
re-interpreted so as to refer to $\bu_*$ instead of $\bu$. Hence we 
may sum up the situation by saying that 
equations~(\ref{eq:FLRWShiftRateApproxRadial}) 
and~(\ref{eq:spatial-g-e-FLRW-ustar}) give, respectively, the 
two-way Doppler-tracking formula and the `Newtonian' equation in a 
FLRW spacetime within the mentioned approximation.

\subsubsection{McVittie spacetime}
The same analysis can be generalized from the spatially flat FLRW 
spacetime~(\ref{eq:FlatFLRW1}) to the spatially flat McVittie 
spacetime~(\ref{eq:McVittieAnsatz}). 
Here the observer moves along $\bpartial/\bpartial t$, which is not 
geodesic. The coordinate $t$ does now not measure proper 
time, denoted by $\tau$,  along the observer's worldline. 
The result corresponding to (\ref{eq:FLRWShiftRateApproxRadial}) 
can now be stated as follows:
\begin{equation}
\label{eq:McVittieShiftRateApproxRadial}
-\,\frac{1}{2}\,\frac{\dot\omega_2(\tau_2)}{\omega_0(\tau_0)}
\approx \alpha\bigl(1-3\beta-3\Delta\tau(H-m_0c/R^2)\bigr)+H\beta\,.
\end{equation}
Here $R$ denotes the areal radius of the observer during the 
measurement. Note that even though it changes along the observer's  
worldline according to (\ref{eq:Hubble-law-McVittie}), 
we do not need to account for the corresponding change in 
$\Delta\tau m_0c/R^2$ of $(-2\Delta\tau m_0c/R^2)(H\Delta\tau)$
which is of subleading order. The additional term in 
(\ref{eq:McVittieShiftRateApproxRadial}) has a straightforward 
interpretation in terms of the acceleration that the observer 
necessarily experiences while keeping a constant radius $R$ 
away from the central inhomogeneity.    

As for the FLRW case, we chose the observer field to which we refer 
the spatial quantities to be proportional to the Kodama vector field 
(along which the areal radius is constant). Putting 
$r_*(t,r):=A^2(t,r)a(t)r$ and $t_*(t,r):=t$, a short computation 
shows that the vector field $\bpartial/\bpartial t_*$ is again 
given by~(\ref{eq:TstarVF}). In the slow-motion and weak-field 
approximation used in Section~\ref{sec:MotionInMcVittie}, the 
geodesic equation in the form~(\ref{eq:spatial-eom}) w.r.t.~the 
observer field $\bu_*$ reads 
\begin{equation}\label{eq:McV-Newton-eq}
  \balpha_{\bu_*}(\gamma) \approx 
  \bigg( \frac{\ddot a}{a}\, r_* - \frac{m_0}{r_*^2} \bigg) \;\be_* 
  \circ \gamma \,,
\end{equation}
where again $\be_*$ denotes the unit outward-pointing vector field 
orthogonal to $\bu_*$ and to the two-spheres of symmetry. This is an 
alternative derivation of the improved Newtonian equation for the 
McVittie spacetime carried out in Section~\ref{sec:MotionInMcVittie}. 
Notice that, again, within the approximations used, 
relation~(\ref{eq:McVittieShiftRateApproxRadial}) remains valid if one 
refers the quantities to $\bu_*$ instead of to $\bu$. Therefore
(\ref{eq:McVittieShiftRateApproxRadial}) and~(\ref{eq:McV-Newton-eq}) 
give the two-way Doppler-tracking formula and the improved 
`Newtonian' equation for the McVittie spacetime within the mentioned 
approximation. 

In the special case of purely radial motion, insertion of 
(\ref{eq:McV-Newton-eq}) into (\ref{eq:McVittieShiftRateApproxRadial})
leads to a formula predicting the two-way Doppler-shift rate in 
linear order in $H\Delta\tau$ and $m_0/r_*$, and quadratic order 
in $\beta_{\bu_*}(\dot{\bgamma})$:
\begin{equation}\label{eq:McVittie-Doppler-tracking-geodesic}
-\,\frac{1}{2}\frac{\dot\omega_2(\tau_2)}{\omega_0(\tau_0)}
=\,-\,\frac{m_0}{r_*^2} \Big( 1 - 3\beta^{\hat{\bk}} \Big)
          +H\beta^{\hat{\bk}}\,.
\end{equation}
Hence there are two corrections to the Newtonian contribution.
One is proportional to $H$ and stems from the cosmological 
expansion, the other, already familiar from the special-relativistic
treatment (\ref{eq:MinkFrequencyRatioDer}), is independent 
of $H$ and merely due to the finiteness of the propagation 
speed of light (recall note~\ref{fnote:Doppler}). Their ratio 
is (up to a factor $\sqrt{3}$) given by the square of the ratio 
of $r_*$ to the geometric mean of the Schwarzschild radius $m_0$ 
and the Hubble radius $c/H$. The latter is of the order of 
$10^{23}\,\mathrm{km}$, so that its geometric mean with a 
Schwarzschild radius of one kilometer is approximately given 
by $2400$ astronomical units. The ratio of the effects is 
therefore of the order $10^{-7}$. Hence the cosmological 
contribution is negligible for any application in the Solar 
System as compared to the $3\beta$--correction. For the Pioneer~10
\& 11 spacecrafts we have a radial velocity of about 12 Km/s. 
This amounts to a $3\beta$--correction of magnitude $4\cdot 10^{-5}$ 
times the Newtonian gravitational acceleration, in an outward-pointing 
direction. This is indeed of the same order of magnitude as the 
PA but directed oppositely.

\section{Summary and outlook}
\label{sec:Summary}
We think it is fair to say that there are no theoretical hints that 
point towards a \emph{dynamical} influence of cosmological expansion 
comparable in size to, say, that of the anomalous acceleration of the 
Pioneer spacecrafts. There seems to be no controversy over this 
point, though for completeness it should be mentioned that there exist 
speculations~\cite{Palle:2005} according to which it might become 
relevant for future missions. But such speculations are often based 
on models which are not easily related to the intended physical 
situation, like that of Gautreau~\cite{Gautreau:1984b}. Rather, as 
the $(\ddot a/a)$--improved Newtonian analysis in 
Section~\ref{sec:NewtonianApproach} together with its justification 
given in the subsequent Sections shows, there is no genuine 
relativistic effect coming from cosmological expansion at the 
levels of precision envisaged here.   

On the other hand, as regards \emph{kinematical} effects, the situation 
is less unanimous. It is very important to unambiguously understand 
what is meant by `mapping out a trajectory', i.e.~how to assign `times' 
and `distances'. Eventually we compare a functional relation between 
`distance' and `time' with observed data. That relation is obtained by 
solving some equations of motion and it has to be carefully checked 
whether the methods by which the tracking data are obtained match the 
interpretation of the coordinates in which the analytical problem is 
solved. In our way of speaking, dynamical effects really influence the 
worldline of the object in question whereas kinematical effects 
change the way in which one and the same worldline is mapped out 
from another worldline representing the observer. Here we have derived 
exact results concerning the influence of cosmic expansion on this 
mapping procedure, which allow to reliably estimate upper bounds 
on their magnitude.  They turn out to be too small to be of any relevance 
in current  satellite trackings, which is in accord with naive 
expectation but in contrast to some statements found in the 
literature. 

At this point it is useful to recall once more the general philosophy 
behind such statements: From the Einstein--Straus solution it is clear that 
local overdensities inhibit cosmic expansion, or at least that part of 
it which is not due to a cosmological constant. Also, as already mentioned 
before, the effect of anisotropies is also to diminish the effect 
of global expansion (see, e.g., \cite{Dominguez.Gaite:2001}). 
Hence calculating such an effect in simple models like the improved 
Newtonian equation discussed in Section~\ref{sec:ImprovedNewtonian} 
(backed up by the various justifications we discussed in detail) 
clearly means to overestimate the impact of cosmic expansion in a 
realistic situation, where the single overdensity (e.g.~representing 
the Sun) is surrounded by more overdense structures (the Solar-System 
environment, the Galaxy, etc.) with less symmetry. If this 
overestimation gives an already insignificant upper bound for the 
envisaged effect, we can conclude that it becomes even more 
insignificant in more realistic models.

Satellite navigation is clearly not the only potential source of 
interest in the question of how local inhomogeneities affect
cosmological expansion. Many predictions concerning cosmological 
data rely on computations within the framework of the standard 
homogeneous and isotropic models, without properly estimating the 
possible effects of local inhomogeneities. Such an estimation would 
ideally be based on an exact inhomogeneous solution to Einstein's 
equations, or at least a fully controlled approximation to such a 
solution.  The dynamical and kinematical impact of local 
inhomogeneities might essentially influence our interpretation of 
cosmological observations. As an example we mention recent serious 
efforts to interpret the same data that are usually taken to prove the 
existence of a positive cosmological constant $\Lambda$ in a context 
with realistic inhomogeneities~%
\cite{Buchert:2000,Rasanen:2006,Wiltshire:2007}, 
i.e.~taking into account that cosmological parameters are 
dressed~\cite{Buchert.Carfora:2003}. See also \cite{Buchert:2008}
for a recent review. One might speculate that the measured $\Lambda$ 
can eventually be \emph{fully} reduced to the action of inhomogeneities, 
as suggested in in \cite{Wiltshire:2007,Wiltshire:2008}. 
For an earlier advance in this direction, see~\cite{Celerier:2000}.

\begin{acknowledgments}
This work was partially supported by the European Space Agency (ESA) under 
the Ariadna scheme of the Advanced Concepts Team, contract 18913/05/NL/MV. 
We are grateful to the ESA and the Albert-Einstein-Institute in Golm for 
their support and hospitality. D.G. acknowledges support from the QUEST 
Excellence Cluster. 
We also thank Claus L\"ammerzahl and Hartmann R\"omer for useful discussions 
and pointing out relevant references as well as the referees for also 
suggesting improvements and references. 
\end{acknowledgments}


\appendix

\section{Notation, conventions, and generalities}
A model for \emph{spacetime} consists of a tuple $(\M,\g)$, where $M$ is 
a four-dimensional manifold and $\g$ a Lorentzian metric whose signature 
we take to be $(+,-,-,-)$, i.e.~we use the `mostly minus' convention. 
Throughout we denote geometric objects, like tensor fields and covariant 
derivative operators, by bold-faced letters or words. The unique metric 
preserving and torsion-free covariant derivative associated with $\g$ will 
be denoted by $\bnabla$ and the covariant derivative in the direction 
of a vector $\bX$ by $\bnabla_\bX$. For a smooth tensor field $\bT$ on 
$M$ its covariant derivative $\bnabla\bT$ defines a linear map, 
$\bX\mapsto\bnabla_\bX\bT$, from the tangent space to the tensor space at 
each point of $\M$ where $\bT$ is defined. Since $\bnabla\bT$ is 
again a tensor field (of rank $(p,q+1)$ if the rank of $\bT$ was 
$(p,q)$) we can form $\bnabla\bnabla\bT:=\bnabla(\bnabla\bT)$. 
Note that $(\bnabla\bnabla\bT)(\bX,\bY)=\bnabla_\bX\bnabla_\bY\bT
-\bnabla_{\bnabla_\bX\bY}\bT$. For a scalar function $f$ on $\M$ 
we have $\bnabla f=\ed f$, the ordinary exterior differential, and 
$\bnabla\bnabla f=\Hess(f)$, the Hessian of $f$.  
The metric $\g$ allows to uniquely associate to any vector 
$\bX$ a linear form $\ul\bX:=\g(\bX,\,\cdot\,)$, called the 
dual (with respect to $\g$) of $\bX$. The inverse of this map will be 
denoted by an overline. The gradient of a function $f$ 
is then $\Grad f = \overline{\ed f}$. The metricity of $\bnabla$ 
implies that the latter commutes with the maps $\ul{\,\cdot\,}$ and 
$\overline{\,\cdot\,}$, e.g.~it holds: $\bnabla_\bX\ul\bY=\ul{\bnabla_\bX\bY}$. 
The scalar product induced by $\g$ on the tensor bundle will be denoted 
by $\sprod{\cdot}{\cdot}$. 

Associated with any two linearly independent vectors $\bX,\bY$ at 
a point $p\in\M$ is a curvature endomorphism, $\R(\bX,\bY)$, of 
the tangent space at $p$:
\begin{equation}\label{eq:CurvEnd}
\begin{split}
  \!\!\!\R(\bX,\bY)\bZ
  &=(\bnabla\bnabla\bZ)(\bX,\bY)-(\bnabla\bnabla\bZ)(\bY,\bX)\\
  &=\bnabla_\bX\bnabla_\bY\bZ-\bnabla_\bY\bnabla_\bX\bZ
   -\bnabla_{[\bX,\bY]}\bZ\,.\\
\end{split}
\end{equation}
The \emph{Riemann}- or \emph{curvature tensor}, $\Riem$, is then defined by
\begin{equation}\label{eq:DefRiem}
  \Riem(\bW,\bZ,\bX,\bY) := \bg\bigl(\bW,\R(\bX,\bY)\bZ\bigr)\,.
\end{equation}
It is antisymmetric under the exchange $\bX\leftrightarrow\bY$ or
$\bW\leftrightarrow\bZ$ and symmetric under the slotwise exchange 
of pairs $(\bW,\bZ)\leftrightarrow(\bX,\bY)$. Moreover, the 
antisymmetrization over any three slots vanishes (first Bianchi
identity). 
The \emph{Ricci tensor}, $\Ric$, is defined by the trace of the 
following endomorphism 
\begin{equation}\label{eq:DefRic}
  \Ric(\bY,\bZ):=\tr\bigl(\bX\mapsto\Riem(\bX,\bY)\bZ\bigr)\,,
\end{equation} 
which is symmetric under exchange $\bY\leftrightarrow\bZ$. 
The \emph{scalar curvature} is defined by taking the 
trace of $\Ric$, also called the Ricci scalar, with respect 
to $\g$ (since $\Ric$ is not an endomorphism, we need the 
metric to define its trace)
\begin{equation}\label{eq:DefScal}
  \Scal=\tr_\g(\Ric)\,.
\end{equation} 
Finally, the \emph{Einstein tensor} is the following combination 
of $\Ric$ and $\Scal$:
\begin{equation}\label{eq:DefEin}
  \Ein := \Ric - \tfrac{1}{2}\Scal\,\g \,.
\end{equation} 

Associated to any spacelike or timelike two-dimensional plane $\Plane$ 
in the tangent space at $p\in\M$ is the \emph{sectional curvature}. Its 
geometric interpretation is just that of the ordinary Gaussian curvature 
at $p$ of the two-dimensional surface in $\M$ that is spanned by the 
geodesic curves through $p$ tangent to $\Plane$. In terms of $\Riem$
it reads    
\begin{equation}\label{eq:sectional-curvature}
  k_\Plane := \frac{\Riem(\bX,\bY,\bX,\bY)}{\bQ(\bX,\bY)} \,,
\end{equation}
where $\bX,\bY$ are any two linear independent vectors in $\Plane$ and 
\begin{equation}\label{eq:area-spanned}
\begin{split}
  \bQ(\bX,\bY) 
  :&= \g(\bX,\bX)\g(\bY,\bY) - \g(\bX,\bY)^2 \\
   &= (\g\odot\g)(\bX,\bY,\bX,\bY) \,. \\
\end{split}
\end{equation}
Note that $\abs{\bQ(\bX,\bY)}$ gives the square of the area of the 
parallelogram spanned by $\bX$ and $\bY$ which is nonzero iff the considered
plane is spacelike or timelike (non-degenerate).

In (\ref{eq:area-spanned}) we introduced the product $\odot$, 
which is called the \emph{Kulkarni--Nomizu} product. It is a symmetric 
bilinear map from the space of symmetric $(0,2)$ tensors to 
the space of $(0,4)$ tensors with the same algebraic symmetries 
as $\Riem$. Its general definition is as follows:
\begin{equation}\label{eq:DefKulkNomProd}
\begin{split}
  (\ba\odot\bb) &(\bW,\bZ,\bX,\bY) := \\
  \frac{1}{2}\Big( &\ba(\bW,\bX)\bb(\bZ,\bY)-\ba(\bW,\bY)\bb(\bZ,\bX) \\
                  +&\bb(\bW,\bX)\ba(\bZ,\bY)-\bb(\bW,\bY)\ba(\bZ,\bX) \Big)\,.
\end{split}
\end{equation}
This can be used to conveniently write down the $\g$-orthogonal 
decomposition of the curvature tensor into the Ricci- and the 
Weyl part:
\begin{equation}\label{eq:DecompRiemann}
  \Riem = \Ricci + \Weyl\,.
\end{equation}
In four spacetime dimensions one has 
\begin{subequations}\label{eq:DefRicci}
\begin{align}
\label{eq:DefRicci-Ric}
  \Ricci :
  &= \left( \Ric-\tfrac{1}{6}\Scal\,\g \right) \odot\g \\
\label{eq:DefRicci-Ein}
  &= \left(\Ein-\tfrac{1}{3}\tr_\g(\Ein)\,\g \right)\odot\g \,.  
\end{align}
\end{subequations}
Inserting this into (\ref{eq:DecompRiemann}) gives the 
definition of $\Weyl$. The definition is such that the Ricci 
part is $\g$-orthogonal to the Weyl part and that the latter 
is totally trace free. Hence the Ricci and the Weyl part 
each contribute 10 independent components to the 20 independent 
components of $\Riem$. The Ricci part may be further decomposed 
according to the decomposition of $\Ric$ into its trace and a 
trace-free part, but this refinement will not be needed here.  

Einstein's equation now express the local determination of 
the Ricci part of the curvature in terms of the energy-momentum 
distribution of matter, the latter being encoded in the 
energy-momentum tensor $\bT$ of the matter. In units where 
Newton's constant $G$ and the velocity of light $c$ equal one%
\footnote{Otherwise the factor  $8\pi$ on the right-hand side 
of (\ref{eq:EinstEq}) should be replaced with $8\pi G/c^4$.}, 
Einstein's equation reads
\begin{equation}\label{eq:EinstEq}
  \Ein = 8\pi\,\bT \,.
\end{equation}
Here we did not write down explicitly a cosmological term,
which can always be thought of as extra contribution to $\bT$
of the form $\g\,\Lambda/8\pi$. Now, assuming that $\g$ satisfies 
Einstein's equation, the Ricci part of the Riemann tensor is 
given in terms of $\bT$ by   
\begin{equation}\label{eq:RicciPartEinEq}
  \Ricci = 8\pi\bigl( \bT-\tfrac{1}{3}\tr_\g(\bT)\,\g \bigr)\odot\g\,.
\end{equation}

\section{Proof of Theorem~\ref{thm:equiv-jc}}
\label{sec:equiv-junct-cond}
In this section we prove Theorem~\ref{thm:equiv-jc}, namely the equivalence, 
in the spherically-symmetric case, of the SSJC with the Darmois junction 
conditions.

\begin{proof}
The proof essentially consists in writing down the induced metric and 
extrinsic curvature for a (non-null) spherically symmetric hypersurface 
in a spherically symmetric spacetime. This is most easily done by introducing 
an adapted orthonormal frame.

We first consider the case where $\varGamma$ is timelike, hence 
$\gamma=\pi(\varGamma)$ is a timelike curve in $\B$.
The following construction shall be carried out in both spacetimes. 
One defines $\bv$ as in the SSJC, hence as the (unique up to a sign) 
spherically symmetric, unit vector field on $\varGamma$ orthogonal to $\bn$. 
That is $\bv$, seen as a vector field on $\B$, is tangent to $\gamma$.
Since $\bn$ is spacelike, $\bv$ is timelike.
The ambient metric can be then written as
\begin{equation}
  \g = \ul\bv\otimes\ul\bv - \ul\bn\otimes\ul\bn - R^2 \gStwo\,,
\end{equation}
so that the induced metric (compare Appendix~\ref{sec:submanifolds} 
and (\ref{eq:induced-metric}) on $\varGamma$ is 
\begin{equation}\label{eq:induced-metric-ss}
  \gG = \ul\bv\otimes\ul\bv - R^2 \gStwo \,.
\end{equation}
In view of (\ref{eq:induced-metric}) note that here $\varepsilon(\bn)=-1$.
For the extrinsic curvature~(\ref{eq:extrinsic-curv-form}), 
using~(\ref{eq:connection-warped}) and the fact that $\bv$ is spherically 
symmetric and hence tangent to $\B$, one has the decomposition
\begin{equation}\label{eq:extrinsic-curv-ss}
  \bK_\varGamma = - \g(\bn,\bnabla_\bv\bv)\,\ul\bv\otimes\ul\bv 
                  - R\,\ed R(\bn)\,\gStwo \,.
\end{equation}

Now, from expressions~(\ref{eq:induced-metric-ss}) 
and~(\ref{eq:extrinsic-curv-ss}) it follows that the DJC, and 
hence the continuity of $\gG$ and $\bK_\varGamma$, are equivalent 
to the continuity of the following four functions: 
\begin{compactenum}[\hspace{3ex} (a)]
\item the arc-length of $\gamma$, 
\item $R$, 
\item $\ed R(\bn)$, and 
\item $\g(\bn,\bnabla_\bv\bv)$.
\end{compactenum}
The statement of the theorem will now follow from the following 
expression of the MS energy~(\ref{eq:MS-energy}):
\begin{equation}\label{eq:MS-energy-ss}
  E = \frac{R}{2} \left( 1 + (\ed R(\bv))^2 - (\ed R(\bn))^2 \right)\,.
\end{equation}
Simply note that if $R$ is continuous through $\varGamma$ (recall the 
definition of this concept below the definition of DJC in 
Section~\ref{sec:spher-symm-match}) the same holds for its 
derivatives tangent to $\varGamma$. In particular, $\ed R(\bv)$ 
is continuous through $\varGamma$ and hence we may substitute 
$\ed R(\bn)$ by the MS energy in the above list (a)--(d). 
This completes the proof for timelike~$\varGamma$.

In the case of spacelike $\varGamma$ the unit normal $\bn$ is 
timelike and $\bv$ is chosen as the unique (up to a sign) 
spherically symmetric unit vector field on $\varGamma$ 
orthonormal to $\bn$. Then $\bv$ is a spacelike `radial' 
unit vector field orthogonal to the $SO(3)$-orbits.
The proof now proceeds analogously to the timelike case.
We merely list the expressions for the ambient metric
\begin{equation*}
  \g = \ul\bn\otimes\ul\bn - \ul\bv\otimes\ul\bv - R^2 \gStwo \,,
\end{equation*}
the induced metric 
\begin{equation}\label{eq:induced-metric-ss-spacelike-case}
  \gG = \ul\bv\otimes\ul\bv + R^2 \gStwo \,,
\end{equation}
the extrinsic curvature
\begin{equation}\label{eq:extrinsic-curv-ss-spacelike-case}
  \bK_\varGamma = \g(\bn,\bnabla_\bv\bv)\,\ul\bv\otimes\ul\bv 
                  + R\,\ed R(\bn)\,\gStwo \,,
\end{equation}
and the MS energy
\begin{equation}\label{eq:MS-energy-ss-spacelike-case}
  E = \frac{R}{2} \left( 1 + (\ed R(\bn))^2 - (\ed R(\bv))^2 \right) \,,
\end{equation}
and conclude exactly as in the timelike case.
\end{proof}

\section{Submanifolds}
\label{sec:submanifolds}
In a Lorentzian manifold $(\M,\g)$ endowed with Levi-Civita connection 
$\bnabla$ consider a smooth submanifold $\varGamma$ of co-dimension 
one and normal vector field $\bn$. We assume $\varGamma$ to be non-null,
that is, either spacelike (then $\bn$ is timelike) or timelike 
(then $\bn$ spacelike). Then $\varGamma$ inherits from the ambient 
manifold $\M$ a (non-degenerate) metric and a connection in a natural 
way. We introduce the orthogonal projectors
\begin{subequations}\label{eq:projectors-n}
\begin{alignat}{2}
  &\bQ_\bn \, &&:= \, \varepsilon(\bn)\,\bn \otimes \ul\bn \label{eq:Q_n}\\
  &\bP_\bn \, &&:= \, \Id - \bQ_\bn         \label{eq:P_n} \,,
\end{alignat}
\end{subequations}
where $\varepsilon(\bn)$ denotes the \emph{indicator}, defined for any 
non-null vector by 
\begin{equation}\label{eq:indicator}
  \varepsilon(\bX) := \frac{\g(\bX,\bX)}{\abs{\g(\bX,\bX)}} = 
  \begin{cases}
    +1& \text{if $\bX$ is timelike}, \\
    -1& \text{if $\bX$ is spacelike}.
  \end{cases}
\end{equation}

The \emph{induced metric} on $\varGamma$ (also called first fundamental form) 
is given by
\begin{equation}\label{eq:induced-metric}
  \g_\varGamma := -\varepsilon(\bn)\bP_\bn \g \,,
\end{equation}
where the sign is just in order to get a positive definite metric in the
case where $\varGamma$ is spacelike. Given two vector fields $\bX,\bY$ 
tangent to $\varGamma$, so that $\bQ_\bn\bX=\bQ_\bn\bY=0$, one 
may decompose the covariant derivative of $\bY$ with respect to $\bX$ 
into its orthogonal components
\begin{align}
  \bnabla_\bX\bY &= \bP_\bn(\bnabla_\bX\bY) + \bQ_\bn(\bnabla_\bX\bY)
                 \nonumber \\
                 &= \LCG_\bX\bY + \bK_\varGamma(\bX,\bY)\bn\,, 
  \label{eq:nabla-decomp-submf}
\end{align}
where 
\begin{equation}\label{eq:induced-LC}
  \LCG_\bX\bY := \bP_\bn(\bnabla_\bX\bY)
\end{equation}
is the \emph{induced connection} on $\varGamma$
and 
\begin{align}\label{eq:extrinsic-curv-form}
  \bK_\varGamma(\bX,\bY) &:= \varepsilon(\bn)\,\g(\bnabla_\bX\bY,\bn) 
                         \nonumber \\
                         &\phantom{:}=-\varepsilon(\bn)\,\g(\bnabla_\bX\bn,\bY)
\end{align}
is the \emph{extrinsic curvature} of $\varGamma$ in $\M$ (also called the
second fundamental form). The second equality sign 
in~(\ref{eq:extrinsic-curv-form}) is an immediate consequence of the metricity 
of $\bnabla$ and the fact that $\bX$ and $\bY$ are orthogonal to $\bn$. 
With this alternative expression for the extrinsic curvature one has 
$\bK_\varGamma = -\varepsilon(\bn)\,\bP_\bn\bnabla\ul{\bn}$ and hence%
\footnote{Here and below $\mathcal{S}$ and $\mathcal{A}$ denote the 
projection operators of full symmetrization and full antisymmetrization, 
respectively.}
\begin{equation}\label{eq:extrinsic-curv-form-SDn}
  \bK_\varGamma = -\varepsilon(\bn)\,\mathcal{S}(\bP_\bn\bnabla\ul{\bn})\,.
\end{equation}
Since $\bn$ is hypersurface orthogonal (by definition) we have 
$\mathcal{A}\bigl(\bP_\bn\bnabla\ul{\bn}\bigr)=0$.  Hence, 
the extrinsic curvature is a symmetric $(0,2)$-tensor field. We recall also 
that the induced connection is the Levi-Civita connection of 
$(\varGamma,\gG)$, as one may easily check. 

The full relations between the curvature of $\M$ and those (intrinsic 
and extrinsic) of $\varGamma$ can be found, e.g.,~in~\cite{Giulini:1998}. 
Here we are only interested in the `Einstein part' of the curvature.
One gets:
\begin{subequations}\label{eq:Ein-jc}
\begin{align}
  &\Ein(\bn,\bn) = \frac{1}{2}\left(
                 -\varepsilon(\bn) \sideset{^\varGamma}{}\Scal 
                 + (\tr\bK)^2 - \norm{\bK}^2 \right) 
   \label{eq:Ein-jc-nn} \\
  &\Ein(\bn,\bP_\bn\cdot\,) = -\varepsilon(\bn)\,
                 \Div_\varGamma \big( \bK - (\tr\bK)\,\gG \big) 
   \label{eq:Ein-jc-ne}\\
  &\Ein(\bP_\bn\cdot\,,\bP_\bn\cdot\,) = \sideset{^\varGamma}{}\Ein \nonumber\\
  &\phantom{\Ein}   
   + \varepsilon(\bn) 
       \bigg(
           \frac{1}{2}\left( (\tr\bK)^2 + \norm{\bK}^2 \right)\gG 
            - (\tr\bK)\bK \nonumber \\ 
  &\phantom{\Ein(\bP_\bn\cdot\,,\bP_\bn\cdot\,) = } 
           + \Lie_\bn \big( \bK - (\tr\bK)\,\gG \big)
       \bigg)\,.
   \label{eq:Ein-jc-ee}
\end{align}
\end{subequations}

\section{Spherical symmetry}
\label{sec:spherical-symmetry}
We recall that the \emph{isometry group},
$\mathrm{Isom}(\M,\g)$, of a spacetime $(\M,\g)$ is the subgroup of the 
diffeomorphism group of $\M$, $\mathrm{Diff}(\M)$, which leaves the metric 
$\g$ invariant: 
$\mathrm{Isom}(\M,\g):=\{ \phi \in \mathrm{Diff(\M)} \,|\, \phi^*\g = \g\}$.
\begin{defi}[Spherical symmetry]\label{defi:spherical-sym}
A four-dimensional Lorentzian manifold $(\M,\g)$ is said to be 
\emph{spherically symmetric} 
if its isometry group, $\mathrm{Isom}(\M,\g)$, contains a subgroup $G$ 
with the following two properties: $(i)$~$G$ is isomorphic to $SO(3)$ and 
$(ii)$~each orbit of $G$ is spacelike and two-dimensional (up to 
some closed proper subset of fixed points). A tensor field $\bT$ on a 
spherically symmetric spacetime is said to be \emph{spherically symmetric}
if it is invariant under $G$, hence if $\phi^*\bT = \bT$ for all 
$\phi\in G$.
\end{defi}
From this definition it follows (excluding the case where the 
orbits of $G$ are diffeomorphic to the two-dimensional 
real-projective space) that a four-dimensional spherically 
symmetric Lorentzian manifold $(\M,\g)$ can, at least locally, 
be expressed as a warped product $\M = \B \times_R S^2$ between a 
two-dimensional Lorentzian manifold $(\B,\g_\B)$, called the 
`base', and the standard unit two-sphere $(S^2,\g_{S^2})$, called 
the `fiber' (see~\cite{Straumann:2004} and~\cite{ONeill:1983}). 
This means that, at least locally, the manifold is a product 
\begin{equation}\label{eq:sph-sym-M-warped}
  \M \mathop{=}^{\rm loc} \B \times S^2
\end{equation}
and the metric is given by
\begin{equation}\label{eq:sph-sym-g-warped}
  \g = \pi^*(\g_\B) - (R \circ \pi)^2 \sigma^*(\g_{S^2}) \,.
\end{equation}
Here, $\pi$ and $\sigma$ are the projections of $\B \times S^2$ onto $\B$
and $S^2$, respectively, and $\pi^*,\sigma^*$ their pull-backs.
The warping function $R$ is nothing but the 
\emph{areal radius}, since, 
for a point $p\in\B$, the area of the fiber $p \times S^2$ is just 
$4\pi R(p)^2$.

In this contest, a vector field $\bX$ on $\M$ at some point 
$(p,q)\in\B\times S^2$ has then a unique decomposition 
$\bX = \tan_\B \bX + \tan_\Stwo \bX$ in a component 
tangent to the `leaves' $\B \times q = \sigma^{-1}(q)$ and a component 
tangent to the `fibers' $p \times S^2 = \pi^{-1}(p)$.
Arbitrary tensor fields on $\B$ and on $\Stwo$ can be lifted to tensor fields
on $\M$ in the standard way. For covariant tensor fields (and hence, in 
particular, for functions) this is achieved via the pull-pack of the 
respective projection: as an example, just look 
at~(\ref{eq:sph-sym-g-warped}). 
For contravariant tensor fields it suffices to consider the special case 
of vector fields.
Let, for instance, $\bX$ be a vector in the tangent space of $\B$ at 
$p$. Then the lift $\tilde\bX$ at $(p,q)$ of $\bX$ is defined 
as the unique vector in the tangent space of $\M$ at $(p,q)$
with $\pi_*(\tilde\bX)=\bX$ and $\sigma_*(\tilde\bX)=0$. 
Since this assignment is smooth, one gets the lifting of a vector field 
via the pointwise lifting just described. 
In this work, we will mainly omit lifts and projections and not explicitly 
distinguish between original and lifted quantities. For example, when 
referring to a vector `tangent to $\B$' we refer to a vector in the 
tangent space of $\B$ or to the lift thereof in the tangent space of $\M$.

If $\bX$ is spherically symmetric, then the component tangent to the fibers 
must vanish: $\tan_\Stwo \bX =0$. Similarly, a spherically symmetric 
one-form $\btheta$ on $\M$ must necessarily be tangent to $\B$ (i.e.~normal 
to $S^2$) and thus it can be written as $\btheta = \pi^*(\btheta_\B)$, where 
$\btheta_\B$ is a one-form on $\B$. Finally, a spherical symmetric function 
is simply the lift of a function on $\B$.

\subsection{Connection and curvature decomposition}
\label{sec:conn-curv-decomp}
In the following we discuss relations which express the curvature of $\M$
in terms of the warping function $R$ and the curvatures of the base $\B$ 
and the fiber $\Stwo$. We start out from the relations between the 
Levi-Civita connection $\LC$ of $(\M,\g)$ and the Levi-Civita connections
of the base and the fiber, denoted by $\LCB$ and $\LCStwo$, respectively. 
These relations can be derived, for example, by means of the Koszul 
formula (see e.g.~\cite{ONeill:1983} Proposition 7.35). Let in the 
following $\bX,\bY,\bZ$ be vector fields tangent to $\B$ and $\bU,\bV,\bW$ 
tangent to $\Stwo$. Suppressing lifts and projections, we have:
\begin{subequations}
\label{eq:connection-warped}
\begin{align}
\label{eq:connection-warped-a}
  &\LC_\bX\bY = \LCB_\bX\bY \\
\label{eq:connection-warped-b}
  &\LC_\bX\bV = \LC_\bV\bX = R^{-1} \bX(R) \bV \\
\label{eq:connection-warped-c}
  &\tan_\Stwo \LC_\bV\bW = \LCStwo_\bV\bW \\
\label{eq:connection-warped-d}
  &\tan_\B \LC_\bV\bW = - \g(\bV,\bW) R^{-1}\LC R \,.
\end{align}
\end{subequations}
Note that, for a function $f$ on $\B$, the lift of the gradient is equal 
to the gradient of the lifted function, that is (suppressing the lifts): 
$\Grad f = \GradB f$. For brevity, we write just $\Grad f$ for it. 
Take care that for the Hessian and the Laplacian this is in general not 
true (see~(\ref{eq:Hessian-warped-ss}) and~(\ref{eq:Laplacian-warped-ss})). 
Therefore we write explicitly the superscripts `$\B$' in $\HessB f$ and 
$\LaplacianB f$ to denote the Hessian and Laplacian of $f$ on $\B$, 
respectively, or the lifts thereof. 

By means of~(\ref{eq:connection-warped}) one can now compute the 
expressions for the Riemann tensor. As the sectional curvature of $S^2$ 
is obviously constant and equal to one, the Riemann tensor, the Ricci 
tensor, and the Ricci scalar of $S^2$ are simply given by 
$\sideset{^\sStwo}{}\Riem = \gStwo\!\odot\gStwo$, 
$\sideset{^\sStwo}{}\Ric = \gStwo$, and 
$\sideset{^\sStwo}{}\Scal = 2$, respectively. Here we made again use 
of the Kulkarni--Nomizu product (\ref{eq:DefKulkNomProd}). Moreover, 
since the basis manifold $\B$ is two-dimensional, one can express its 
curvature tensors in terms of the scalar curvature. The expression for 
the Riemann tensor, Ricci tensor, and Ricci scalar of a spherically 
symmetric Lorentzian manifold 
(\ref{eq:sph-sym-M-warped},\ref{eq:sph-sym-g-warped}) are, respectively, 
\begin{align}
  \Riem &= \frac{\ScalB}{2} \gB\!\odot\gB 
         \nonumber \\
        &- \frac{1}{R^2} \Big( 1 + \sprod{\ed R}{\ed R} \Big)\,
           R^2\gStwo\!\odot R^2\gStwo 
   \label{eq:Riemann-cov-warped-ss} \\
        &+ 2R\,\gStwo\!\odot\!\!\HessB R \,,
         \nonumber
\end{align}
\begin{align}
   \Ric &= \frac{\ScalB}{2}\gB - \frac{2}{R}\HessB R 
       \nonumber \\
        &+ \Big( 1+\sprod{\ed R}{\ed R} + R\LaplacianB R \Big) \gStwo \,,
  \label{eq:Ricci-warped-ss}
\end{align}
and 
\begin{equation}\label{eq:Scal-warped-ss}
  \Scal = \ScalB 
        - \frac{2}{R^2} \Big( 1+\sprod{\ed R}{\ed R} \Big)
        - \frac{4}{R}\LaplacianB R \,.
\end{equation}
Hence, for the Einstein tensor we have the expression 
\begin{align}
  \Ein = & \left( \frac{1}{R^2}\big( 1\!+\!\sprod{\ed R}{\ed R} \big) 
                + \frac{2}{R}\LaplacianB R \right)\gB - \frac{2}{R}\HessB R 
         \nonumber \\
         &+\left( \frac{\ScalB}{2} 
                - \frac{1}{R} \LaplacianB R \right) R^2 \gStwo
         \label{eq:Einstein-warped-ss}
\end{align}
and for the Weyl tensor, using~(\ref{eq:DecompRiemann}) 
and~(\ref{eq:DefRicci-Ric}), the simple expression 
\begin{subequations}\label{eq:Weyl-warped-ss}
\begin{equation}\label{eq:Weyl-warped-ss-1}
  \Weyl = w\,
  ( \gB\odot\gB + \gB\odot R^2\gStwo + R^2\gStwo\odot R^2\gStwo )\,.
\end{equation}
Here we put 
\begin{align}
  \label{eq:Weyl-warped-ss-2}
  w :\!&= \frac{1}{6} 
  \left( \ScalB - \frac{2}{R^2}\big( 1\!+\!\sprod{\ed R}{\ed R} \big) + 
         \frac{2}{R}\LaplacianB R \right) \\
       &= \frac{1}{6}\Scal + \frac{1}{R}\LaplacianB R \,.
  \label{eq:Weyl-warped-ss-3}
\end{align}
\end{subequations}
In the derivation of~(\ref{eq:Weyl-warped-ss}) we made use of the 
formula $\bh_\sB\odot\gB=(1/2)(\tr_{\gB}\bh_\sB)\gB\odot\gB$, valid%
\footnote{To prove this just note that the only independent component 
of this formula is the $(\be_0,\be_1,\be_0,\be_1)$ one, where 
$\{\be_\mu\}$ is an adapted orthonormal basis of $(\M,\g)$ 
such that $\be_0,\be_1$ are tangent to $\B$ and $\be_2,\be_3$ are tangent 
to $S^2$. Then, the equality follows immediately using the 
definition~(\ref{eq:DefKulkNomProd}) of the Kulkarni--Nomizu product.} 
for any symmetric bilinear form $\bh_\sB$ on $\B$, in order to express the 
only term involving $\HessB R$ in terms of the Laplacian of $R$. 
Note that from~(\ref{eq:Weyl-warped-ss-1}) it is immediate that the Weyl 
tensor of a spherically symmetric spacetime has only one independent 
component, as it must be the case due to it being of Petrov-type D.

Comparing expression~(\ref{eq:Riemann-cov-warped-ss}) with the definition 
of sectional curvature~(\ref{eq:sectional-curvature}) one can immediately 
read off that the sectional curvature $K$ of the plane tangential to the 
(two-dimensional) $SO(3)$-orbits at a given point is
\begin{equation}\label{eq:sectional-curv-tangent-S2}
  K = -\frac{1}{R^2} \big( 1+\sprod{\ed R}{\ed R} \big) \,,
\end{equation}
and hence the MS energy, defined as (minus one-half) $K$ times the third 
power of the areal radius, is given by~(\ref{eq:MS-energy}). Using the MS 
energy~(\ref{eq:MS-energy}), we can write the Einstein 
tensor~(\ref{eq:Einstein-warped-ss}) as:
\begin{align}
  \Ein = & \frac{2}{R}\left( \LaplacianB R + \frac{E}{R^2} \right)\gB 
         - \frac{2}{R}\HessB R 
         \nonumber \\
         &+\left( \frac{\ScalB}{2} - \frac{1}{R} \LaplacianB R \right) 
         R^2 \gStwo 
         \label{eq:Einstein-warped-ss-MS} \,.
\end{align}

We conclude giving the decomposition for the divergence of a spherically 
symmetric vector field (that is a vector field $\bX$ tangent to $\B$) and 
for the Hessian and Laplacian of a spherically symmetric function (that 
is a function $f$ on $\B$). First, from~(\ref{eq:connection-warped-a}) 
and~(\ref{eq:connection-warped-b}), we obtain the following decomposition for 
the covariant derivative of $\bX$ (expressed as a $(0,2)$-tensor): 
\begin{equation}\label{eq:cov-deriv-ss}
  \bnabla\ul{\bX} = \LCB \ul{\bX} - \bX(R)R\,\gStwo\,.
\end{equation}
Note that the mixed term ($\B$-$\Stwo$) vanishes---as it should due to 
spherical symmetry. Taking the trace of (\ref{eq:cov-deriv-ss}) one 
obtains the following expression for the divergence:
\begin{equation}\label{eq:div-warped-ss}
  \Div \bX  = \Div_\B \bX + \frac{2}{R} \bX(R) 
            = \frac{1}{R^2}\Div_\B ( R^2 \bX ) \,.
\end{equation}
The decompositions for the Hessian and Laplacian of a function $f$ on $\B$ 
follow from inserting $\bX=\Grad f$ in the above formulae. One gets: 
\begin{equation}\label{eq:Hessian-warped-ss}
  \Hess f = \HessB f - \gB(\Grad f,\Grad R) R\,\gStwo 
\end{equation}
and
\begin{equation}\label{eq:Laplacian-warped-ss}
  \Laplacian f = \LaplacianB f + 2\,\gB(\Grad f,\Grad R)/R \,,
\end{equation}
respectively.

\subsection{Einstein equation in case of spherical symmetry}
\label{sec:Einstein-eq-ss-case}
A general spherically symmetric matter energy-momentum tensor has the form
\begin{equation}\label{eq:matter-e-m-t-ss}
  \bT = \bT_\B + p\,R^2\gStwo \,,
\end{equation}
where $p$ is the spherical part of the pressure. Hence, using the 
decomposition~(\ref{eq:Einstein-warped-ss-MS}) of the Einstein's tensor 
found in Appendix~\ref{sec:spherical-symmetry}, the Einstein equation takes 
the form
\begin{subequations}\label{eq:Einstein-eq-ss-prel}
\begin{align}
  &\frac{2}{R}\left( \frac{E}{R^2} + \LaplacianB R \right)\gB - 
   \frac{2}{R}\HessB R = 8\pi \bT_\B \\
  &\frac{\ScalB}{2} - \frac{1}{R}\LaplacianB R = 8\pi p \,.
\end{align}
\end{subequations}
Using the trace of the first equation,
\begin{equation}\label{eq:trace-Einstein-eq-ss-B}
  \frac{1}{R}\left( \LaplacianB R + \frac{2E}{R^2}\, \right) = 
  4\pi\, \tr \bT_\B \,,
\end{equation}
to eliminate $\LaplacianB R$, one can write~(\ref{eq:Einstein-eq-ss-prel}) 
in the equivalent form
\begin{subequations}\label{eq:Einstein-eq-ss-fin}
\begin{align}
  &\frac{1}{R}\left( \frac{E}{R^2}\gB + \HessB R \right) = 
   -4\pi \star\bT_\B\star 
  \label{eq:Einstein-eq-ss-B}\\
  &\frac{\ScalB}{2} + \frac{2E}{R^2} = 4\pi \big( \tr\bT_\B + 2p ) 
  \label{eq:Einstein-eq-ss-Stwo} \,.
\end{align}
\end{subequations}
Here, and in the following, $\star$ denotes the Hodge-duality map 
for $(\B,\g_\sB)$ (for the definition, see e.g.~\cite{Straumann:2004}). 
In the first equation we used the identity 
$\star\btau\star = \btau - \tr(\btau)\,\g$, which is valid for 
any bilinear form $\btau$ on $\B$, where the first (second) 
star acts on the first (second) slot of $\btau$. 

Finally, the integrability condition $\Div \bT =0$ for the 
energy-momentum tensor~(\ref{eq:matter-e-m-t-ss}) reads 
\begin{equation}\label{eq:div-e-m-tensor-ss}
  \Div_\B(R^2\bT_\B) + p\, \ed(R^2) = 0 \,.
\end{equation}

\subsection{Misner--Sharp energy}
\label{sec:MS-energy}
Let us now turn to the MS energy and its properties. We first show that 
it is the charge of a conserved current. The treatment presented here 
follows mainly~\cite{Hayward:1996}. In a spherically symmetric spacetime one 
defines the \emph{Kodama vector field}~\cite{Kodama:1980} as the (unique up 
to a sign) spherically symmetric vector field orthogonal to, and of the 
same norm as, the gradient of $R$; hence we put 
\begin{equation}\label{eq:Kodama-def}
  \ul\bk := \star \ed R \,.
\end{equation}
With this sign choice $\bk$ is future-pointing if the gradient of $R$ is 
spacelike. The orthogonality between $\bk$ and the gradient of $R$ is 
simply expressed by 
\begin{equation}\label{eq:Kodama-ortho-dR}
  \bk(R) = 0 \,,
\end{equation}
which clearly means that the integral curves of $\bk$ stay at 
constant areal radius. 

An immediate but important property of the Kodama vector field 
is that it is conserved:
\begin{equation}\label{eq:Kodama-conserved}
  \Div\bk = 0 \,.
\end{equation}
Indeed, using~(\ref{eq:div-warped-ss}) and (\ref{eq:Kodama-ortho-dR}), 
one has: 
$\Div\bk = \Div_\B\bk = \bdelta\ul\bk = -\star\ed\star\star\ed R \equiv 0$.

Now, a key point for the study of spherically symmetric spacetimes is the 
following equation relating the MS energy with the matter's energy-momentum
tensor:
\begin{equation}\label{eq:dE}
  \ed E = 4\pi R^2 \star\ul\bj \,,
\end{equation}
where $\bj$ is the so-called \emph{Kodama current} (tangent to the base 
manifold $\B$) defined by 
\begin{equation}\label{eq:j-def}
  \ul\bj := \bT(\bk,\cdot) \,.
\end{equation}
Equation~(\ref{eq:dE}) follows from Einstein's equation; more precisely, it 
is equivalent to its $\B$-part (that is Eq.~(\ref{eq:Einstein-eq-ss-B})) fed 
with $\bnabla R$. To see this, just compute the differential 
of~(\ref{eq:MS-energy}) as follows: 
$\ed E 
= (E/R)\ed R + (R/2)\ed\sprod{\ed R}{\ed R}
= (E/R)\ed R + R \HessB R \bdot \ed R
= R \big( (E/R^2)\gB + \HessB R \big) \bdot \ed R$, 
where the dot denotes here the contraction of the last slot of the tensor 
on the left with the first slot of the tensor on the right of the dot. 
In the second step we use that $\ed(\sprod{\ed R}{\ed R})(\bX)
=\bX(\g(\bnabla R,\bnabla R)) = 2\,\gB(\bnabla_\bX\bnabla R,\bnabla R)
=2\HessB R(\bnabla R,\bX)$ for any $\bX$ tangent to $\B$. 
Then, inserting~(\ref{eq:Einstein-eq-ss-B}) and using that $\star$ is 
skew-adjoint on one-forms, one gets 
$\ed E= -4\pi R^2 \star\bT_\B\star \bdot \ed R
= 4\pi R^2 \star\bT_\B \bdot \star \ed R$ 
and hence, using the definitions~(\ref{eq:Kodama-def}) and~(\ref{eq:j-def}) 
together with the symmetry of $\bT$, one arrives at~(\ref{eq:dE}). 

From~(\ref{eq:dE}) it is clear that 
\begin{equation}\label{eq:j-ortho-dE}
  \bj(E)=0 \,,
\end{equation}
which means that the vector field $\bj$ is tangent to the curves in $\B$ 
(hypersurfaces in $\M$) of constant MS energy. Moreover, (\ref{eq:dE}) 
implies that $\bj$ is also conserved
\begin{equation}\label{eq:j-conserved}
  \Div\bj =0 \,,
\end{equation}
where the divergence is here taken on the spacetime $(\M,\g)$. To see this, 
just compute the divergence of $\bj$ with~(\ref{eq:div-warped-ss}) and using 
the Hodge-dual version of~(\ref{eq:dE}): 
$\Div \bj 
= R^{-2}\Div_\B(R^2\bj)
= R^{-2}\bdelta(R^2\ul\bj)
=(4\pi R^2)^{-1}\bdelta\star\ed E \equiv 0$.

Following~\cite{Hayward:1996}, we can now show that the charges corresponding
to the conserved currents $\bj$ and $\bk$ are, respectively, the MS energy and 
the areal volume. Let $\Sigma$ be some spatial three-dimensional 
hypersurface which, because of spherical symmetry, decomposes as 
$\Sigma=\sigma\times\Stwo$, where $\sigma$ is some spatial curve in $\B$. 
Recall that the charge related to a conserved current $\bX$ is given by 
$Q_\bX(\Sigma):=\int_\Sigma i_\bX\bmu$, where $\bmu$ is the volume 
form on $\M$, and, because of spherical symmetry, the latter decomposes as 
$\bmu=\bmu_\B\wedge R^2\bmu_\Stwo$, where $\bmu_\B$ and $\bmu_\Stwo$ are the
volume forms on $\B$ and on the unit two-sphere, respectively. 
After integration of the spherical part and since $i_\bj\bmu_\B=\star\ul\bj$, 
using~(\ref{eq:dE}) one gets
\begin{equation}\label{eq:j-charge}
  Q_\bj(\Sigma) = \int_\sigma \ed E\,,
\end{equation}
which means that the charge of $\bj$ is the MS energy. 
This justifies the interpretation of the MS energy as a quantity associated 
to the `interior' of the considered sphere of symmetry. In fact, 
due to~(\ref{eq:j-conserved}), the charge does not depend how one 
choose the spatial slice to define the interior. Similarly, since 
$i_\bk\bmu_\B=\star\ul\bk=\ed R$, the charge to $\bk$ is simply 
$\int_\sigma 4\pi R^2 \ed R$ and hence
\begin{equation}\label{eq:Kodama-charge}
  Q_\bk(\Sigma) = \int_\sigma \ed \left( \frac{4\pi}{3}R^3 \right)\,,
\end{equation}
which says that the charge of $\bk$ is the flat-space volume 
computed with the areal radius. 

Incidentally, the Kodama vector can be used to give an elegant 
proof of Birkhoff's theorem, which states that spherically symmetric 
solutions of Einstein's equations are, in fact, static. Indeed, by 
direct computation one shows that in vacuum $\bk$ is Killing and, 
because of spherical symmetry, it is clearly also hypersurface 
orthogonal. 

Next we turn to the relation between the MS energy and the Hawking 
quasi-local mass~\cite{Hawking:1968}. The latter is a quantity 
associated to a spatial two-sphere, $S$, in an arbitrary spacetime.
It is defined by
\begin{equation}\label{eq:Hawking-mass}
  M_\ind{H}(S) := \sqrt{\frac{\textrm{Area}(S)}{16\pi}}
  \left( 1 + \frac{1}{2\pi}\int_S \theta^+\theta^-\bmu_{S} \right) \,.
\end{equation}
Here, $\theta^\pm:=\tr_\Stwo(\bnabla\bl^\pm)/2$ are, respectively, 
the expansions of the outgoing and ingoing future-pointing null 
vector fields $\bl^\pm$ normal to $S$, the latter being 
partially normalized such that $\g(\bl^+,\bl^-)=1$ (there remains
the freedom to rescale $\bl^\pm\rightarrow\alpha^{\pm 1}\bl^\pm$,
where $\alpha$ is a positive real-valued function). In the special case of 
spherical symmetry we take $S$ to be an orbit of the rotation 
group. Then we clearly have $\textrm{Area}(S)=4\pi R^2$. It is also 
obvious that the metric of the base $\B$, evaluated on $S$, can 
simply be written in the form 
\begin{equation}\label{eq:MetricBaseLightVectors}
  \gB = \ul\bl^+\otimes\ul\bl^- + \ul\bl^-\otimes\ul\bl^+\,.
\end{equation}  
Now, for $\bV$ tangent to $S$, (\ref{eq:connection-warped-b}) 
gives $\bnabla_\bV\bl^\pm=R^{-1}\bl^\pm(R)\bV$ so that 
$\theta^\pm=R^{-1}\bl^\pm(R)$. Hence we have: 
\begin{equation}\label{eq:expansions-product}
  2\,\theta^+\theta^- 
  = 2R^{-2}\,\bd R(\bl^+)\bd R(\bl^-)
  = \sprod{\ed R}{\ed R}/R^2 \,,
\end{equation}
where we used (\ref{eq:MetricBaseLightVectors}), or rather its 
contravariant version, in the last step. Equation~(\ref{eq:MS-energy}) 
now establishes the equality between the MS energy at $p$ and the 
Hawking quasi-local mass of $S$, where $p$ is any point on $S$: 
\begin{equation}\label{eq:MSE-Hawking-ql-mass}
  E(p) = M_\ind{H}(S)\,.
\end{equation}

As is the case for the Hawking quasi-local mass, the MS energy can be 
naturally decomposed into a Ricci and a Weyl part: 
\begin{equation}\label{eq:MSE-sum}
  E = E_\ind{R} + E_\ind{W} \,,
\end{equation}
where 
\begin{subequations}\label{eq:MSE-decomposition}
\begin{alignat}{3}
\label{eq:MSE-Ricci}
  &E_\ind{R} &&:= -\tfrac{1}{2}R^3 K_\ind{R} \,, \\
\label{eq:MSE-Weyl}
  &E_\ind{W} &&:= -\tfrac{1}{2}R^3 K_\ind{W} \,.
\end{alignat}
\end{subequations}
Here $K_\ind{R}$ and $K_\ind{W}$ denote, respectively, the Ricci and 
the Weyl parts of the sectional curvature of the plane tangential to 
the $SO(3)$-orbits. These are obtained inserting the decomposition of 
the Riemann tensor~(\ref{eq:DecompRiemann}) in the definition of the 
sectional curvature~(\ref{eq:sectional-curvature}). 
The Ricci part of the MS energy is determined by the local matter distribution 
via Einstein's equation: Using expressions~(\ref{eq:matter-e-m-t-ss}) 
and~(\ref{eq:sph-sym-g-warped}) for an arbitrary spherically symmetric 
energy-momentum tensor and, respectively, metric 
in~(\ref{eq:RicciPartEinEq}) one gets
\begin{equation}\label{eq:MSE-RicciWithEinstein}
  E_\ind{R} = \frac{4\pi}{3}R^3(\tr\bT_\B+p) \,.
\end{equation}
For the Weyl part of the MS energy we have, in view 
of~(\ref{eq:Weyl-warped-ss-1}), that
\begin{equation}\label{eq:MSE-Weyl-warped}
  E_\ind{W} = -\tfrac{1}{2}R^3 w \,,
\end{equation}
where $w$ is given by~(\ref{eq:Weyl-warped-ss-2}) 
or~(\ref{eq:Weyl-warped-ss-3}). Hence, in particular, the Weyl 
tensor vanishes iff $E_\ind{W}$ does. Since the square of the Weyl 
tensor is $\sprod{\Weyl}{\Weyl} 
\equiv W_{\alpha\beta\gamma\delta}W^{\alpha\beta\gamma\delta} =12 w^2$, 
with~(\ref{eq:MSE-Weyl-warped}) we obtain the nice expression 
\begin{equation}\label{eq:Weyl-square-MSE}
  \sprod{\Weyl}{\Weyl} = 48 \frac{E_\ind{W}^2}{R^6} \,.
\end{equation}
From this one sees that, in a spherically-symmetric spacetime, the 
non-vanishing of $E_\ind{W}$ (that is the non-vanishing of the Weyl tensor) 
for $R\to 0$ implies a curvature singularity at $R=0$. 

To gain a better physical understanding of the Weyl part of the MS energy 
we take a look at the equation of geodesic deviation (sometime called Jacobi 
equation). Let $\bu$ be a geodesic observer field and $\bs$ the spatial 
($\g(\bu,\bs)=0$) separation vector between two nearby integral curves 
of $\bu$. Then the equation of geodesic deviation (see 
e.g.~\cite{Straumann:2004}) is 
$\LC_\bu\LC_\bu\bs = \R(\bu,\bs)\bu = \bB_\bu(\bs) + \bC_\bu(\bs)$. In the 
last step we decomposed the endomorphism on the r.h.s.~of the geodesic 
deviation equation in its Ricci- and Weyl-part, denoted here by $\bB_\bu$ 
and $\bC_\bu$, respectively. For an arbitrary spherically symmetric 
spacetime the latter is given by (see~(\ref{eq:Weyl-warped-ss}) 
and~(\ref{eq:MSE-Weyl-warped})) 
\begin{equation}\label{eq:Jacobi-Weyl}
  \bC_\bu = \frac{2E_\ind{W}}{R^3}\bP_\bu^\sB - 
           \frac{E_\ind{W}}{R^3}\bP_\bu^\sStwo \,,
\end{equation}
where $\bP_\bu^\sB$ and $\bP_\bu^\sStwo$ are, respectively, the $\B$ and $\Stwo$ 
parts of the projector $\bP_\bu$. Recall that $\bP_\bu$ projects onto the 
subspace of the tangential space orthogonal to $\bu$ 
(see~(\ref{eq:SplitOperators-b})). Equation~(\ref{eq:Jacobi-Weyl}) is exactly 
the same expression one gets in Newtonian gravity---provided one identifies 
$E_\ind{W}$ with the mass of the central object. The spatial endomorphism 
$\bC_\bu$ just describes the familiar volume-preserving tidal deformation 
which produces an expansion in radial direction and a contraction in 
the orthogonal directions tangential to the $SO(3)$-orbits. 

Concerning the Ricci part $\bB_\bu$, in the case where $\bu$ is the velocity 
field of dust%
\footnote{Recall that dust particles moves along geodesics by the Euler 
equation.},
making use of Einstein's equation (see~(\ref{eq:RicciPartEinEq})) we have: 
\begin{equation}\label{eq:Jacobi-Ricci}
  \bB_\bu = - \frac{4\pi}{3}\varrho\,\bP_\bu \,.
\end{equation}
This just says that the local effect of matter (here given by dust) is an 
isotropic contraction.

\subsection{Spherically symmetric perfect fluids}
\label{sec:ss-perfect-fluids}
We specialize now to a perfect fluid, which is described by a four-velocity 
vector field $\bu$, density $\varrho$, and pressure $p$. In case of spherical 
symmetry $\bu$ is tangent to the basis manifold and the matter energy-momentum 
tensor~(\ref{er:EMTensorForMcVittie}) decomposes as
\begin{equation}\label{eq:EMT-ss-perfect-fluid}
  \bT = \varrho\,\ul\bu\otimes\ul\bu + p\,(\ul\bu\otimes\ul\bu-\gB) 
      + p\,R^2\gStwo \,,
\end{equation}
from which one can read off the part tangent to $\B$:
\begin{equation}\label{eq:EMT-ss-perfect-fluid-B}
\bT_\B = \varrho\,\ul\bu\otimes\ul\bu + p\,(\ul\bu\otimes\ul\bu-\gB) \,.
\end{equation}
Usually, the description is to be completed with the specification of an 
equation of state. We will not assume any equation of state yet, since in 
some cases (e.g.~McVittie spacetime) this happens to be determined by 
Einstein's equation.

Inserting~(\ref{eq:EMT-ss-perfect-fluid-B}) in~(\ref{eq:MSE-RicciWithEinstein})
we get for the Ricci part of the MS energy the simple expression
\begin{equation}\label{eq:MSE-Ricci-PerfectFluid}
  E_\ind{R} = \frac{4\pi}{3}R^3\varrho \,.
\end{equation}
Also the expression~(\ref{eq:dE}) for the differential of the MS energy
simplifies in case of a perfect fluid. Using~(\ref{eq:EMT-ss-perfect-fluid-B}) 
the Kodama current (as one-form) becomes 
\begin{equation}\label{eq:j-perfect-fluid}
  \ul\bj = (\varrho+p)\g(\bk,\bu)\,\ul\bu - p\,\ul\bk \,.
\end{equation}
It is useful to introduce an adapted orthonormal basis $\{\bu,\be\}$ tangent 
to the basis manifold, where $\bu$ is the velocity vector field of the fluid 
and $\be$ is chosen to point in direction of increasing areal radius. Because 
of our choice of orientation we have $\ul\be=\star\ul\bu$ (the volume form 
on $\B$ is simply $\bmu_\B=\ul\bu\wedge\ul\be$). 
Using this expression for $\be$ and the definition of the Kodama vector 
field~(\ref{eq:Kodama-def}) we have $\g(\bk,\bu)=\sprod{\ul\bk}{\ul\bu}
=\sprod{\star\ed R}{\ul\bu}=-\sprod{\ed R}{\star\ul\bu}=-\sprod{\ed R}{\ul\be}
=-\ed R(\be)$ and hence, the Hodge star of the Kodama current becomes
\begin{equation}\label{eq:star-j-perfect-fluid}
  \star\ul\bj = -p\, \ed R(\bu)\ul\bu -\varrho\,\ed R(\be)\ul\be \,,
\end{equation}
which, inserted in~(\ref{eq:dE}), gives the following expression for the 
differential of the MS energy for a perfect fluid:
\begin{equation}\label{eq:dE-pf}
  \ed E = -4\pi R^2 
  \Bigl( p\, \ed R(\bu)\ul\bu +\varrho\,\ed R(\be)\ul\be \Bigr) \,.
\end{equation}
Hence, the variation of the MS energy along $\bu$ and $\be$ is, respectively:
\begin{subequations}\label{eq:dE-pf-u-e}
\begin{alignat}{3}
&\ed E(\bu) &&= - &&4\pi R^2\, p\,       \ed R(\bu) \label{eq:dE-pf-u} \,,\\
&\ed E(\be) &&= + &&4\pi R^2\, \varrho\, \ed R(\be) \label{eq:dE-pf-e} \,.
\end{alignat}
\end{subequations}
These expressions have a good physical interpretation: Since the matter 
moves along $\bu$, (\ref{eq:dE-pf-u}) expresses the fact that the energy 
can only increase (decrease) if the motion along $\bu$ does (releases) work 
against (with) the action of the pressure. Equation~(\ref{eq:dE-pf-e}) 
expresses the almost obvious increase (decrease) of gravitational mass 
with increase (decrease) of volume in the rest system of the matter. 
We said `almost' because $4\pi R^2\ed R(\be)$ is not quite the increment 
of proper volume. The difference accounts for the fact that kinetic 
and gravitational binding energy are themselves gravitationally active. 
To see that this is indeed what (\ref{eq:dE-pf-e}) implies, let $p$ 
be some point in spacetime and $S_p$ the two-sphere of spherical symmetry 
through $p$. Assume $S_p$ to have a regular interior, that is, that 
$S_p$ bounds a 3-ball $B_p$ in the hypersurface $\Sigma$ orthogonal 
to $\bu$. Except for the origin of $B_p$, we can write 
$B_p=\sigma\times\Stwo$, where $\sigma$ is a spacelike curve in $\B$ 
orthogonal to $\bu$, going from the center of symmetry to $\pi(p)$. 
Using the expression $E=(R/2)(1 + (\ed R(\bu))^2 - (\ed R(\be))^2)$ 
for the MS energy to eliminate $\ed R(\be)$ in~(\ref{eq:dE-pf-e}), 
integrating the latter over $\sigma$, and re-expressing the result 
as a volume integral, one gets: 
\begin{equation}\label{eq:MSE-int-expr}
  E(p) = \int_{B_p} \varrho
         \left( 1 + (\ed R(\bu))^2 - \frac{2E}{R} \right)^{1/2} \bmu_\Sigma\,.
\end{equation}
One sees that the MS energy contains the contribution from the 
proper mass contained in the ball $B_p$, 
\begin{equation}\label{eq:proper-mass}
  M(p) = \int_{B_p} \varrho\, \bmu_\Sigma \,,
\end{equation}
as well as contributions from the `kinetic' and `potential' 
energy~\cite{Misner.Sharp:1964,Hayward:1996}. In a Newtonian approximation, 
that is for small `velocity' $\ed R(\bu)$ and weak field (small $E/R$) one 
can expand the square root in~(\ref{eq:MSE-int-expr}) and gets, in leading 
order:
\begin{equation}\label{eq:MSE-int-expr-approx}
  E(p) \approx \int_{B_p}\left( \varrho + \tfrac{1}{2}\varrho(\ed R(\bu))^2 - 
  \frac{\varrho M}{R} \right) \bmu_\Sigma\,.
\end{equation}
In this approximation the MS energy is therefore just the sum of the 
proper mass and the Newtonian kinetic and potential energies contained 
in the ball $B_p$. This provides a sound justification for the 
interpretation of the MS energy as the active gravitational energy. 

At this point we can compute also the differentials of the two 
parts~(\ref{eq:MSE-decomposition}) of the MS energy separately. The 
differential of the Ricci part follows directly 
from~(\ref{eq:MSE-Ricci-PerfectFluid}):
\begin{equation}\label{eq:dE-Ricci-pf}
  \ed E_\ind{R} = 4\pi R^2 
  \bigl( \varrho\,\ed R + \tfrac{1}{3}R\,\ed\varrho \bigr) 
\end{equation}
and the differential of the Weyl part is just the difference of this 
with~(\ref{eq:dE-pf}):
\begin{equation}\label{eq:dE-Weyl-pf}
  \ed E_\ind{W} = -4\pi R^2 
  \Bigl( (\varrho+p)\ed R(\bu)\,\ul\bu + \tfrac{1}{3}R\,\ed\varrho \Bigr) \,.
\end{equation}
Its components in the directions $\bu$ and $\be$ are then
\begin{subequations}\label{eq:dE-Weyl-pf-u-e}
\begin{alignat}{3}
\label{eq:dE-Weyl-pf-u}
&\ed E_\ind{W}(\bu) &&= -4\pi R^2(\varrho+p)\ed R(\bu)
                        -\tfrac{4\pi}{3}R^3\ed\varrho(\bu)\,,
\\
\label{eq:dE-Weyl-pf-e}
&\ed E_\ind{W}(\be) &&= -\tfrac{4\pi}{3}R^3 \ed\varrho(\be)\,.
\end{alignat}
\end{subequations}

It is now instructive to express the variation along $\bu$ of the Ricci and 
Weyl parts of the MS energy in terms of the kinematical properties of the 
fluid velocity $\bu$. Recall that, because of spherical symmetry, the 
rotation tensor vanishes identically and the shear tensor has only one 
independent component. The kinematical quantities reduces thus to two 
scalars: the expansion 
\begin{equation}\label{eq:expansion}
  \theta := \Div \bu
\end{equation}
and the shear scalar 
\begin{equation}\label{eq:shear-scalar}
  \sigma := \frac{\ed R(\bu)}{R}-\frac{1}{3}\theta \,.
\end{equation}
The shear tensor is then given by the trace-free endomorphism 
$\bsigma = \sigma ( \bQ_\Stwo - 2\,\bQ_\be )$, where $\bQ_\Stwo$ 
and $\bQ_\be$ denote, respectively, the projections onto 
the two-dimensional subspace of $T(\M)$ tangential to the two-sphere and onto 
the one-dimensional space parallel to $\be$ (for the latter 
see~(\ref{eq:Q_n})).
We recall that the divergence-freeness of the energy-momentum 
tensor~(\ref{eq:EMT-ss-perfect-fluid}) is equivalent to 
\begin{subequations}\label{eq:divT-ss-pf}
\begin{alignat}{3}
&(\varrho+p)\,\theta &&= - \ed\varrho(\bu) \label{eq:divT-ss-pf-u}\\
&(\varrho+p)\,b &&= -\ed p(\be)\,,         \label{eq:divT-ss-pf-e}
\end{alignat}
\end{subequations}
where $b:=-\g(\bnabla_\bu\bu,\be)$ is the acceleration (scalar) of $\bu$ in
positive radial direction (the minus sign in the latter formula is because 
the metric is negative definite in spatial directions). 

Now, using~(\ref{eq:divT-ss-pf-u}) and~(\ref{eq:shear-scalar}) we get:
\begin{subequations}
\begin{alignat}{3}
  &\ed E_\ind{R}(\bu) &&= \frac{4\pi}{3}R^3(3\varrho\,\sigma-p\,\theta) \\
  &\ed E_\ind{W}(\bu) &&= -\frac{4\pi}{3}R^3 (\varrho+p)3\sigma
\end{alignat}
\end{subequations}

With the equations just derived we can now say when the MS energy, and its 
Ricci and Weyl parts, are temporally or spatially constant. Here, by 
temporally (spatially) constant we mean that the variation in direction of 
$\bu$ ($\be$) vanishes. We collect the results in the following 

\begin{thm}\label{thm:dE-vanishing}
Consider a spherically symmetric fluid with $\varrho+p \neq 0$ and restrict 
to the region where $dR$ is spacelike. Then for the MS energy $E$ and 
its Ricci and Weyl parts $E_\ind{R}$ and $E_\ind{W}$ the following 
statements hold true:
\begin{compactenum}[(i)]
\item $E$ is temporally constant iff $p=0$ or $\ed R(\bu)=0$; 
\item $E$ is spatially constant iff $\varrho=0$;
\item $E_\ind{W}$ is temporally constant iff $\sigma=0$;
\item $E_\ind{W}$ is spatially constant iff $\varrho$ is spatially constant;
\item $E_\ind{R}$ is temporally (spatially) constant iff $R^3\varrho$ is 
      temporally (spatially) constant.
\end{compactenum}
\end{thm}

\noindent
The proof is a straightforward application of the formulae just derived above.
Note that the assumption $\varrho+p \neq 0$ is needed only for $(iii)$. 
The assumption that $\ed R$ is spacelike is needed only for $(ii)$: If 
$\ed R$ is spacelike, then for any spacelike spherically symmetric vector 
$\be$ (hence tangent to the basis manifold $\B$) it holds 
$\ed R(\be) \neq 0$, since in a two-dimensional Lorentzian manifold any 
two spacelike vectors are linearly dependent.


\bibliographystyle{apsrmplong-spires}
\bibliography{COSMOLOGY}

\begin{thebibliography}{107}
\expandafter\ifx\csname natexlab\endcsname\relax\def\natexlab#1{#1}\fi
\expandafter\ifx\csname bibnamefont\endcsname\relax
  \def\bibnamefont#1{#1}\fi
\expandafter\ifx\csname bibfnamefont\endcsname\relax
  \def\bibfnamefont#1{#1}\fi
\expandafter\ifx\csname citenamefont\endcsname\relax
  \def\citenamefont#1{#1}\fi
\expandafter\ifx\csname url\endcsname\relax
  \def\url#1{\texttt{#1}}\fi
\expandafter\ifx\csname urlprefix\endcsname\relax\def\urlprefix{URL }\fi
\providecommand{\bibinfo}[2]{#2}
\providecommand{\eprint}[2][]{\url{#2}}

\bibitem[{\citenamefont{Adkins} \emph{et~al.}(2007)\citenamefont{Adkins,
  McDonnell, and Fell}}]{Adkins.etal:2007}
\bibinfo{author}{\bibnamefont{Adkins}, \bibfnamefont{G.~S.}},
  \bibinfo{author}{\bibfnamefont{J.}~\bibnamefont{McDonnell}}, and
  \bibinfo{author}{\bibfnamefont{R.~N.} \bibnamefont{Fell}},
  \bibinfo{year}{2007}, {``}\bibinfo{title}{Cosmological perturbations on local
  systems},{''} \bibinfo{journal}{Physical Review D}
  \textbf{\bibinfo{volume}{75}}(\bibinfo{number}{6}), \bibinfo{eid}{064011}
  (pages~\bibinfo{numpages}{9}).

\bibitem[{\citenamefont{{Anderson}}
  \emph{et~al.}(1998)\citenamefont{{Anderson}, {Laing}, {Lau}, {Liu}, {Nieto},
  and {Turyshev}}}]{Anderson.etal:1998}
\bibinfo{author}{\bibnamefont{{Anderson}}, \bibfnamefont{J.~D.}},
  \bibinfo{author}{\bibfnamefont{P.~A.} \bibnamefont{{Laing}}},
  \bibinfo{author}{\bibfnamefont{E.~L.} \bibnamefont{{Lau}}},
  \bibinfo{author}{\bibfnamefont{A.~S.} \bibnamefont{{Liu}}},
  \bibinfo{author}{\bibfnamefont{M.~M.} \bibnamefont{{Nieto}}}, and
  \bibinfo{author}{\bibfnamefont{S.~G.} \bibnamefont{{Turyshev}}},
  \bibinfo{year}{1998}, {``}\bibinfo{title}{{Indication, from Pioneer 10/11,
  Galileo, and Ulysses Data, of an Apparent Anomalous, Weak, Long-Range
  Acceleration}},{''} \bibinfo{journal}{Physical Review Letters}
  \textbf{\bibinfo{volume}{81}},  \bibinfo{pages}{2858--2861}.

\bibitem[{\citenamefont{Anderson} \emph{et~al.}(2002)\citenamefont{Anderson,
  Laing, Lau, Liu, Nieto, and Turyshev}}]{Anderson.etal:2002a}
\bibinfo{author}{\bibnamefont{Anderson}, \bibfnamefont{J.~D.}},
  \bibinfo{author}{\bibfnamefont{P.~A.} \bibnamefont{Laing}},
  \bibinfo{author}{\bibfnamefont{E.~L.} \bibnamefont{Lau}},
  \bibinfo{author}{\bibfnamefont{A.~S.} \bibnamefont{Liu}},
  \bibinfo{author}{\bibfnamefont{M.~M.} \bibnamefont{Nieto}}, and
  \bibinfo{author}{\bibfnamefont{S.~G.} \bibnamefont{Turyshev}},
  \bibinfo{year}{2002}, {``}\bibinfo{title}{{Study of the anomalous
  acceleration of Pioneer 10 and 11}},{''} \bibinfo{journal}{Physical Review D}
  \textbf{\bibinfo{volume}{65}}(\bibinfo{number}{8}), \bibinfo{eid}{082004}
  (pages~\bibinfo{numpages}{50}).

\bibitem[{\citenamefont{{Balbinot}}
  \emph{et~al.}(1988)\citenamefont{{Balbinot}, {Bergamini}, and
  {Comastri}}}]{Balbinot.etal:1988}
\bibinfo{author}{\bibnamefont{{Balbinot}}, \bibfnamefont{R.}},
  \bibinfo{author}{\bibfnamefont{R.}~\bibnamefont{{Bergamini}}}, and
  \bibinfo{author}{\bibfnamefont{A.}~\bibnamefont{{Comastri}}},
  \bibinfo{year}{1988}, {``}\bibinfo{title}{{Solution of the Einstein-Strauss
  problem with a {$\Lambda$} term}},{''} \bibinfo{journal}{Physical Review D}
  \textbf{\bibinfo{volume}{38}},  \bibinfo{pages}{2415--2418}.

\bibitem[{\citenamefont{Barnes} \emph{et~al.}(2006)\citenamefont{Barnes,
  Francis, James, and Lewis}}]{Barnes.etal:2006}
\bibinfo{author}{\bibnamefont{Barnes}, \bibfnamefont{L.~A.}},
  \bibinfo{author}{\bibfnamefont{M.~J.} \bibnamefont{Francis}},
  \bibinfo{author}{\bibfnamefont{J.~B.} \bibnamefont{James}}, and
  \bibinfo{author}{\bibfnamefont{G.~F.} \bibnamefont{Lewis}},
  \bibinfo{year}{2006}, {``}\bibinfo{title}{Joining the Hubble Flow:
  Implications for Expanding Space},{''} \bibinfo{journal}{Monthly Notices of
  the Royal Astronomical Society}
  \textbf{\bibinfo{volume}{373}}(\bibinfo{number}{1}),
  \bibinfo{pages}{382--390}.

\bibitem[{\citenamefont{Bini} \emph{et~al.}(1995)\citenamefont{Bini, Carini,
  and Jantzen}}]{Bini.etal:1995}
\bibinfo{author}{\bibnamefont{Bini}, \bibfnamefont{D.}},
  \bibinfo{author}{\bibfnamefont{P.}~\bibnamefont{Carini}}, and
  \bibinfo{author}{\bibfnamefont{R.~T.} \bibnamefont{Jantzen}},
  \bibinfo{year}{1995}, {``}\bibinfo{title}{Relative Observer Kinematics in
  General Relativity},{''} \bibinfo{journal}{Classical and Quantum Gravity}
  \textbf{\bibinfo{volume}{12}},  \bibinfo{pages}{2549--2563}.

\bibitem[{\citenamefont{{Bolen}} \emph{et~al.}(2001)\citenamefont{{Bolen},
  {Bombelli}, and {Puzio}}}]{Bolen.etal:2001}
\bibinfo{author}{\bibnamefont{{Bolen}}, \bibfnamefont{B.}},
  \bibinfo{author}{\bibfnamefont{L.}~\bibnamefont{{Bombelli}}}, and
  \bibinfo{author}{\bibfnamefont{R.}~\bibnamefont{{Puzio}}},
  \bibinfo{year}{2001}, {``}\bibinfo{title}{{Expansion-induced contribution to
  the precession of binary orbits}},{''} \bibinfo{journal}{Classical and
  Quantum Gravity} \textbf{\bibinfo{volume}{18}},  \bibinfo{pages}{1173--1178}.

\bibitem[{\citenamefont{Bol{\'o}s}(2007)}]{Bolos:2007}
\bibinfo{author}{\bibnamefont{Bol{\'o}s}, \bibfnamefont{V.~J.}},
  \bibinfo{year}{2007}, {``}\bibinfo{title}{Intrinsic Definitions of ``Relative
  Velocity'' in General Relativity},{''} \bibinfo{journal}{Communications in
  Mathematical Physics} \textbf{\bibinfo{volume}{273}}(\bibinfo{number}{1}),
  \bibinfo{pages}{217--236}.

\bibitem[{\citenamefont{Bona and Stela}(1987)}]{Bona.Stela:1987}
\bibinfo{author}{\bibnamefont{Bona}, \bibfnamefont{C.}}, and
  \bibinfo{author}{\bibfnamefont{J.}~\bibnamefont{Stela}},
  \bibinfo{year}{1987}, {``}\bibinfo{title}{``Swiss cheese'' models with
  pressure},{''} \bibinfo{journal}{Physical Review D}
  \textbf{\bibinfo{volume}{36}}(\bibinfo{number}{10}),
  \bibinfo{pages}{2915--2918}.

\bibitem[{\citenamefont{{Bonnor}}(1999)}]{Bonnor:1999}
\bibinfo{author}{\bibnamefont{{Bonnor}}, \bibfnamefont{W.~B.}},
  \bibinfo{year}{1999}, {``}\bibinfo{title}{{Size of a hydrogen atom in the
  expanding universe}},{''} \bibinfo{journal}{Classical and Quantum Gravity}
  \textbf{\bibinfo{volume}{16}},  \bibinfo{pages}{1313--1321}.

\bibitem[{\citenamefont{{Bonnor}}(2000)}]{Bonnor:2000b}
\bibinfo{author}{\bibnamefont{{Bonnor}}, \bibfnamefont{W.~B.}},
  \bibinfo{year}{2000}, {``}\bibinfo{title}{{A generalization of the
  Einstein-Straus vacuole}},{''} \bibinfo{journal}{Classical and Quantum
  Gravity} \textbf{\bibinfo{volume}{17}},  \bibinfo{pages}{2739--2748}.

\bibitem[{\citenamefont{Buchert}(2000)}]{Buchert:2000}
\bibinfo{author}{\bibnamefont{Buchert}, \bibfnamefont{T.}},
  \bibinfo{year}{2000}, {``}\bibinfo{title}{On Average Properties of
  Inhomogeneous Fluids in General Relativity: Dust Cosmologies},{''}
  \bibinfo{journal}{General Relativity and Gravitation}
  \textbf{\bibinfo{volume}{32}}(\bibinfo{number}{1}),
  \bibinfo{pages}{105--125}.

\bibitem[{\citenamefont{Buchert}(2008)}]{Buchert:2008}
\bibinfo{author}{\bibnamefont{Buchert}, \bibfnamefont{T.}},
  \bibinfo{year}{2008}, {``}\bibinfo{title}{{Dark Energy from Structure - A
  Status Report}},{''} \bibinfo{journal}{General Relativity and Gravitation}
  \textbf{\bibinfo{volume}{40}},  \bibinfo{pages}{467--527}.

\bibitem[{\citenamefont{Buchert and Carfora}(2003)}]{Buchert.Carfora:2003}
\bibinfo{author}{\bibnamefont{Buchert}, \bibfnamefont{T.}}, and
  \bibinfo{author}{\bibfnamefont{M.}~\bibnamefont{Carfora}},
  \bibinfo{year}{2003}, {``}\bibinfo{title}{Cosmological Parameters Are
  Dressed},{''} \bibinfo{journal}{Physical Review Letters}
  \textbf{\bibinfo{volume}{90}}(\bibinfo{number}{3}), \bibinfo{eid}{031101}
  (pages~\bibinfo{numpages}{4}).

\bibitem[{\citenamefont{Buchert and Dominguez}(2005)}]{Buchert.Dominguez:2005}
\bibinfo{author}{\bibnamefont{Buchert}, \bibfnamefont{T.}}, and
  \bibinfo{author}{\bibfnamefont{A.}~\bibnamefont{Dominguez}},
  \bibinfo{year}{2005}, {``}\bibinfo{title}{{Adhesive Gravitational
  Clustering}},{''} \bibinfo{journal}{Astronomy and Astrophysics}
  \textbf{\bibinfo{volume}{438}},  \bibinfo{pages}{443--460}.

\bibitem[{\citenamefont{{Burnett}}(1991)}]{Burnett:1991}
\bibinfo{author}{\bibnamefont{{Burnett}}, \bibfnamefont{G.~A.}},
  \bibinfo{year}{1991}, {``}\bibinfo{title}{{Incompleteness theorems for the
  spherically symmetric spacetimes}},{''} \bibinfo{journal}{Physical Review D}
  \textbf{\bibinfo{volume}{43}},  \bibinfo{pages}{1143--1149}.

\bibitem[{\citenamefont{Cahill and
  McVittie}(1970{\natexlab{a}})}]{Cahill.McVittie:1970a}
\bibinfo{author}{\bibnamefont{Cahill}, \bibfnamefont{M.~E.}}, and
  \bibinfo{author}{\bibfnamefont{G.~C.} \bibnamefont{McVittie}},
  \bibinfo{year}{1970}{\natexlab{a}}, {``}\bibinfo{title}{{Spherical Symmetry
  and Mass-Energy in General Relativity. I. General Theory}},{''}
  \bibinfo{journal}{Journal of Mathematical Physics}
  \textbf{\bibinfo{volume}{11}}(\bibinfo{number}{4}),
  \bibinfo{pages}{1382--1391}.

\bibitem[{\citenamefont{Cahill and
  McVittie}(1970{\natexlab{b}})}]{Cahill.McVittie:1970b}
\bibinfo{author}{\bibnamefont{Cahill}, \bibfnamefont{M.~E.}}, and
  \bibinfo{author}{\bibfnamefont{G.~C.} \bibnamefont{McVittie}},
  \bibinfo{year}{1970}{\natexlab{b}}, {``}\bibinfo{title}{{Spherical Symmetry
  and Mass-Energy in General Relativity. II. Particular Cases}},{''}
  \bibinfo{journal}{Journal of Mathematical Physics}
  \textbf{\bibinfo{volume}{11}}(\bibinfo{number}{4}),
  \bibinfo{pages}{1392--1401}.

\bibitem[{\citenamefont{Carrera}(2009)}]{Carrera:2009}
\bibinfo{author}{\bibnamefont{Carrera}, \bibfnamefont{M.}},
  \bibinfo{year}{2009}, Ph.D. thesis, \bibinfo{school}{University of Freiburg
  (Germany)}.

\bibitem[{\citenamefont{Carrera and Giulini}(2005)}]{Carrera.Giulini:2005}
\bibinfo{author}{\bibnamefont{Carrera}, \bibfnamefont{M.}}, and
  \bibinfo{author}{\bibfnamefont{D.}~\bibnamefont{Giulini}},
  \bibinfo{year}{2005}, {``}\bibinfo{title}{{On the influence of the global
  cosmological expansion on the local dynamics in the Solar System}},{''},
  \bibinfo{howpublished}{European Space Agency, the Advanced Concepts Team,
  Ariadna Final Report 04-1302}, \eprint{arXiv:gr-qc/0602098}.

\bibitem[{\citenamefont{Carrera and Giulini}(2006)}]{Carrera.Giulini:2006b}
\bibinfo{author}{\bibnamefont{Carrera}, \bibfnamefont{M.}}, and
  \bibinfo{author}{\bibfnamefont{D.}~\bibnamefont{Giulini}},
  \bibinfo{year}{2006}, {``}\bibinfo{title}{On {Doppler} tracking in
  cosmological spacetimes},{''} \bibinfo{journal}{Classical and Quantum
  Gravity} \textbf{\bibinfo{volume}{23}},  \bibinfo{pages}{7483--7492}.

\bibitem[{\citenamefont{Carrera and Giulini}(2009)}]{Carrera.Giulini:2009a}
\bibinfo{author}{\bibnamefont{Carrera}, \bibfnamefont{M.}}, and
  \bibinfo{author}{\bibfnamefont{D.}~\bibnamefont{Giulini}},
  \bibinfo{year}{2009}, {``}\bibinfo{title}{On the generalization of
  {McVittie's} model for an inhomogeneity in a cosmological spacetime},{''}
  \eprint{arXiv:0908.3101}.

\bibitem[{\citenamefont{C{\'e}l{\'e}rier}(2000)}]{Celerier:2000}
\bibinfo{author}{\bibnamefont{C{\'e}l{\'e}rier}, \bibfnamefont{M.-N.}},
  \bibinfo{year}{2000}, {``}\bibinfo{title}{{Do we really see a cosmological
  constant in the supernovae data?}},{''} \bibinfo{journal}{Astronomy and
  Astrophysics} \textbf{\bibinfo{volume}{353}},  \bibinfo{pages}{63--71}.

\bibitem[{\citenamefont{{Cooperstock}}
  \emph{et~al.}(1998)\citenamefont{{Cooperstock}, {Faraoni}, and
  {Vollick}}}]{Cooperstock.etal:1998}
\bibinfo{author}{\bibnamefont{{Cooperstock}}, \bibfnamefont{F.~I.}},
  \bibinfo{author}{\bibfnamefont{V.}~\bibnamefont{{Faraoni}}}, and
  \bibinfo{author}{\bibfnamefont{D.~N.} \bibnamefont{{Vollick}}},
  \bibinfo{year}{1998}, {``}\bibinfo{title}{{The Influence of the Cosmological
  Expansion on Local Systems}},{''} \bibinfo{journal}{Astrophysical Journal}
  \textbf{\bibinfo{volume}{503}},  \bibinfo{pages}{61--66}.

\bibitem[{\citenamefont{Cox}(2007)}]{Cox:2007}
\bibinfo{author}{\bibnamefont{Cox}, \bibfnamefont{D.~P.~G.}},
  \bibinfo{year}{2007}, {``}\bibinfo{title}{How far is {`infinity'}?},{''}
  \bibinfo{journal}{General Relativity and Gravitation}
  \textbf{\bibinfo{volume}{39}}(\bibinfo{number}{2}),
  \bibinfo{pages}{87--104}.

\bibitem[{\citenamefont{Darmois}(1927)}]{Darmois:1927}
\bibinfo{author}{\bibnamefont{Darmois}, \bibfnamefont{G.}},
  \bibinfo{year}{1927}, {``}\bibinfo{title}{{Les {\'e}quations de la
  gravit{\'e} einsteinienne}},{''} \bibinfo{journal}{M{\'e}morial des sciences
  math{\'e}matiques} \textbf{\bibinfo{volume}{XXV}},  \bibinfo{pages}{1--47}.

\bibitem[{\citenamefont{{Dicke} and {Peebles}}(1964)}]{Dicke.Peebles:1964}
\bibinfo{author}{\bibnamefont{{Dicke}}, \bibfnamefont{R.~H.}}, and
  \bibinfo{author}{\bibfnamefont{P.~J.~E.} \bibnamefont{{Peebles}}},
  \bibinfo{year}{1964}, {``}\bibinfo{title}{{Evolution of the Solar System and
  the Expansion of the Universe}},{''} \bibinfo{journal}{Physical Review
  Letters} \textbf{\bibinfo{volume}{12}},  \bibinfo{pages}{435--437}.

\bibitem[{\citenamefont{{Dominguez} and {Gaite}}(2001)}]{Dominguez.Gaite:2001}
\bibinfo{author}{\bibnamefont{{Dominguez}}, \bibfnamefont{A.}}, and
  \bibinfo{author}{\bibfnamefont{J.}~\bibnamefont{{Gaite}}},
  \bibinfo{year}{2001}, {``}\bibinfo{title}{{Influence of the Cosmological
  Expansion on Small Systems}},{''} \bibinfo{journal}{Europhysics Letters}
  \textbf{\bibinfo{volume}{55}}(\bibinfo{number}{4}),
  \bibinfo{pages}{458--464}.

\bibitem[{\citenamefont{{Einstein} and {Straus}}(1945)}]{Einstein.Straus:1945}
\bibinfo{author}{\bibnamefont{{Einstein}}, \bibfnamefont{A.}}, and
  \bibinfo{author}{\bibfnamefont{E.~G.} \bibnamefont{{Straus}}},
  \bibinfo{year}{1945}, {``}\bibinfo{title}{{The Influence of the Expansion of
  Space on the Gravitation Fields Surrounding the Individual Stars}},{''}
  \bibinfo{journal}{Reviews of Modern Physics} \textbf{\bibinfo{volume}{17}},
  \bibinfo{pages}{120--124}.

\bibitem[{\citenamefont{{Einstein} and {Straus}}(1946)}]{Einstein.Straus:1946}
\bibinfo{author}{\bibnamefont{{Einstein}}, \bibfnamefont{A.}}, and
  \bibinfo{author}{\bibfnamefont{E.~G.} \bibnamefont{{Straus}}},
  \bibinfo{year}{1946}, {``}\bibinfo{title}{{Corrections and Additional Remarks
  to our Paper: The Influence of the Expansion of Space on the Gravitation
  Fields Surrounding the Individual Stars}},{''} \bibinfo{journal}{Reviews of
  Modern Physics} \textbf{\bibinfo{volume}{18}},  \bibinfo{pages}{148--149}.

\bibitem[{\citenamefont{{Eisenstaedt}}(1977)}]{Eisenstaedt:1977}
\bibinfo{author}{\bibnamefont{{Eisenstaedt}}, \bibfnamefont{J.}},
  \bibinfo{year}{1977}, {``}\bibinfo{title}{{Density constraint on local
  inhomogeneities of a Robertson-Walker cosmological universe}},{''}
  \bibinfo{journal}{Physical Review D} \textbf{\bibinfo{volume}{16}},
  \bibinfo{pages}{927--928}.

\bibitem[{\citenamefont{Fahr and Siewert}(2008)}]{Fahr.Siewert:2008}
\bibinfo{author}{\bibnamefont{Fahr}, \bibfnamefont{H.~J.}}, and
  \bibinfo{author}{\bibfnamefont{M.}~\bibnamefont{Siewert}},
  \bibinfo{year}{2008}, {``}\bibinfo{title}{Imprints from the Global
  Cosmological Expansion on the Local Spacetime Dynamics},{''}
  \bibinfo{journal}{Naturwissenschaften}
  \textbf{\bibinfo{volume}{95}}(\bibinfo{number}{5}),
  \bibinfo{pages}{413--425}.

\bibitem[{\citenamefont{Faraoni and Jacques}(2007)}]{Faraoni.Jacques:2007}
\bibinfo{author}{\bibnamefont{Faraoni}, \bibfnamefont{V.}}, and
  \bibinfo{author}{\bibfnamefont{A.}~\bibnamefont{Jacques}},
  \bibinfo{year}{2007}, {``}\bibinfo{title}{Cosmological Expansion and Local
  Physics},{''} \bibinfo{journal}{Physical Review D}
  \textbf{\bibinfo{volume}{76}}, \bibinfo{eid}{063510}
  (pages~\bibinfo{numpages}{16}).

\bibitem[{\citenamefont{{Ferraris}}
  \emph{et~al.}(1996)\citenamefont{{Ferraris}, {Francaviglia}, and
  {Spallicci}}}]{Ferraris.etal:1996}
\bibinfo{author}{\bibnamefont{{Ferraris}}, \bibfnamefont{M.}},
  \bibinfo{author}{\bibfnamefont{M.}~\bibnamefont{{Francaviglia}}}, and
  \bibinfo{author}{\bibfnamefont{A.}~\bibnamefont{{Spallicci}}},
  \bibinfo{year}{1996}, {``}\bibinfo{title}{{Associated radius, energy and
  pressure of McVittie's metric, in its astrophysical application}},{''}
  \bibinfo{journal}{Nuovo Cimento} \textbf{\bibinfo{volume}{B111}},
  \bibinfo{pages}{1031--1036}.

\bibitem[{\citenamefont{{Fouqu{\'e}}}
  \emph{et~al.}(2001)\citenamefont{{Fouqu{\'e}}, {Solanes}, {Sanchis}, and
  {Balkowski}}}]{Fouque.etal}
\bibinfo{author}{\bibnamefont{{Fouqu{\'e}}}, \bibfnamefont{P.}},
  \bibinfo{author}{\bibfnamefont{J.~M.} \bibnamefont{{Solanes}}},
  \bibinfo{author}{\bibfnamefont{T.}~\bibnamefont{{Sanchis}}}, and
  \bibinfo{author}{\bibfnamefont{C.}~\bibnamefont{{Balkowski}}},
  \bibinfo{year}{2001}, {``}\bibinfo{title}{Structure, mass and distance of the
  Virgo cluster from a Tolman-Bondi model},{''} \bibinfo{journal}{Astronomy and
  Astrophysics} \textbf{\bibinfo{volume}{375}},  \bibinfo{pages}{770--780}.

\bibitem[{\citenamefont{Gao and Zhang}(2004)}]{Gao.Zhang:2004}
\bibinfo{author}{\bibnamefont{Gao}, \bibfnamefont{C.~J.}}, and
  \bibinfo{author}{\bibfnamefont{S.~N.} \bibnamefont{Zhang}},
  \bibinfo{year}{2004}, {``}\bibinfo{title}{{Reissner-Nordstr\"om} metric in
  the {Friedman-Robertson-Walker} universe},{''} \bibinfo{journal}{Physics
  Letters B} \textbf{\bibinfo{volume}{595}},  \bibinfo{pages}{28--35}.

\bibitem[{\citenamefont{{Gautreau}}(1984)}]{Gautreau:1984b}
\bibinfo{author}{\bibnamefont{{Gautreau}}, \bibfnamefont{R.}},
  \bibinfo{year}{1984}, {``}\bibinfo{title}{{Imbedding a Schwarzschild mass
  into cosmology}},{''} \bibinfo{journal}{Physical Review D}
  \textbf{\bibinfo{volume}{29}},  \bibinfo{pages}{198--206}.

\bibitem[{\citenamefont{Geyer}(1980)}]{Geyer:1980}
\bibinfo{author}{\bibnamefont{Geyer}, \bibfnamefont{K.~H.}},
  \bibinfo{year}{1980}, {``}\bibinfo{title}{{Geometrie der Raum-Zeit der
  Ma{\ss}bestimmung von Kottler, Weyl und Trefftz}},{''}
  \bibinfo{journal}{Astronomische Nachrichten}
  \textbf{\bibinfo{volume}{301}}(\bibinfo{number}{3}),
  \bibinfo{pages}{135--149}.

\bibitem[{\citenamefont{Giulini}(1998)}]{Giulini:1998}
\bibinfo{author}{\bibnamefont{Giulini}, \bibfnamefont{D.}},
  \bibinfo{year}{1998}, {``}\bibinfo{title}{{On the Construction of
  Time-Symmetric Black Hole Initial Data}},{''} \bibinfo{journal}{LNP Vol.~514:
  Black Holes: Theory and Observation} \textbf{\bibinfo{volume}{514}},
  \bibinfo{pages}{224--243}.

\bibitem[{\citenamefont{Hackmann and
  L\"ammerzahl}(2008{\natexlab{a}})}]{Hackmann.Laemmerzahl:2008a}
\bibinfo{author}{\bibnamefont{Hackmann}, \bibfnamefont{E.}}, and
  \bibinfo{author}{\bibfnamefont{C.}~\bibnamefont{L\"ammerzahl}},
  \bibinfo{year}{2008}{\natexlab{a}}, {``}\bibinfo{title}{Complete analytic
  solution of the geodesic equation in {Schwarzschild}--(anti-){de\,Sitter}
  spacetimes},{''} \bibinfo{journal}{Physical Review Letters}
  \textbf{\bibinfo{volume}{100}}, \bibinfo{eid}{171101}
  (pages~\bibinfo{numpages}{4}).

\bibitem[{\citenamefont{Hackmann and
  L\"ammerzahl}(2008{\natexlab{b}})}]{Hackmann.Laemmerzahl:2008b}
\bibinfo{author}{\bibnamefont{Hackmann}, \bibfnamefont{E.}}, and
  \bibinfo{author}{\bibfnamefont{C.}~\bibnamefont{L\"ammerzahl}},
  \bibinfo{year}{2008}{\natexlab{b}}, {``}\bibinfo{title}{Geodesic Equation and
  Theta-Divisor},{''} \bibinfo{journal}{AIP Conference Proceedings}
  \textbf{\bibinfo{volume}{977}}(\bibinfo{number}{1}),
  \bibinfo{pages}{116--133}.

\bibitem[{\citenamefont{Hackmann and
  L\"ammerzahl}(2008{\natexlab{c}})}]{Hackmann:Laemmerzahl:2008c}
\bibinfo{author}{\bibnamefont{Hackmann}, \bibfnamefont{E.}}, and
  \bibinfo{author}{\bibfnamefont{C.}~\bibnamefont{L\"ammerzahl}},
  \bibinfo{year}{2008}{\natexlab{c}}, {``}\bibinfo{title}{Geodesic equation in
  Schwarzschild-(anti-)de\,Sitter space-times: Analytical solutions and
  applications},{''} \bibinfo{journal}{Physical Review D}
  \textbf{\bibinfo{volume}{78}}(\bibinfo{number}{2}), \bibinfo{eid}{024035}
  (pages~\bibinfo{numpages}{22}).

\bibitem[{\citenamefont{Hartl}(2006)}]{Hartl:2006}
\bibinfo{author}{\bibnamefont{Hartl}, \bibfnamefont{M.~G.}},
  \bibinfo{year}{2006}, \emph{\bibinfo{title}{{Untersuchungen {\"u}ber den
  Einfluss der kosmologischen Expansion auf die Dynamik lokaler Systeme}}},
  \bibinfo{type}{Diploma thesis}, \bibinfo{school}{University of Freiburg,
  Germany}.

\bibitem[{\citenamefont{Hawking}(1968)}]{Hawking:1968}
\bibinfo{author}{\bibnamefont{Hawking}, \bibfnamefont{S.~W.}},
  \bibinfo{year}{1968}, {``}\bibinfo{title}{Gravitational Radiation in an
  Expanding Universe},{''} \bibinfo{journal}{Journal of Mathematical Physics}
  \textbf{\bibinfo{volume}{9}}(\bibinfo{number}{4}),
  \bibinfo{pages}{598--604}.

\bibitem[{\citenamefont{{Hayward}}(1996)}]{Hayward:1996}
\bibinfo{author}{\bibnamefont{{Hayward}}, \bibfnamefont{S.~A.}},
  \bibinfo{year}{1996}, {``}\bibinfo{title}{{Gravitational energy in spherical
  symmetry}},{''} \bibinfo{journal}{Physical Review D}
  \textbf{\bibinfo{volume}{53}},  \bibinfo{pages}{1938--1949}.

\bibitem[{\citenamefont{{Hayward}}(1998)}]{Hayward:1998}
\bibinfo{author}{\bibnamefont{{Hayward}}, \bibfnamefont{S.~A.}},
  \bibinfo{year}{1998}, {``}\bibinfo{title}{{Inequalities Relating Area,
  Energy, Surface Gravity, and Charge of Black Holes}},{''}
  \bibinfo{journal}{Physical Review Letters} \textbf{\bibinfo{volume}{81}},
  \bibinfo{pages}{4557--4559}.

\bibitem[{\citenamefont{Hernandez and Misner}(1966)}]{Hernandez.Misner:1966}
\bibinfo{author}{\bibnamefont{Hernandez}, \bibfnamefont{J., Walter~C.}}, and
  \bibinfo{author}{\bibfnamefont{C.~W.} \bibnamefont{Misner}},
  \bibinfo{year}{1966}, {``}\bibinfo{title}{Observer Time as a Coordinate in
  Relativistic Spherical Hydrodynamics},{''} \bibinfo{journal}{Astrophysical
  Journal} \textbf{\bibinfo{volume}{143}},  \bibinfo{pages}{452--464}.

\bibitem[{\citenamefont{Hubble}(1929)}]{Hubble:1929}
\bibinfo{author}{\bibnamefont{Hubble}, \bibfnamefont{E.}},
  \bibinfo{year}{1929}, {``}\bibinfo{title}{A Relation Between Distance and
  Radial Velocity Among Extra-Galactic Nebulae},{''}
  \bibinfo{journal}{Proceedings of the National Academy of Sciences of the
  United States of America} \textbf{\bibinfo{volume}{15}}(\bibinfo{number}{3}),
   \bibinfo{pages}{168--173}.

\bibitem[{\citenamefont{Israel}(1966)}]{Israel:1966}
\bibinfo{author}{\bibnamefont{Israel}, \bibfnamefont{W.}},
  \bibinfo{year}{1966}, {``}\bibinfo{title}{Singular Hypersurfaces and Thin
  Shells in General Relativity},{''} \bibinfo{journal}{Nuovo Cimento}
  \textbf{\bibinfo{volume}{44B}}(\bibinfo{number}{1}),  \bibinfo{pages}{1--14;}
  \bibinfo{note}{{E}rrata ibid {\bf 48B}(2), 463}.

\bibitem[{\citenamefont{Kagramanova}
  \emph{et~al.}(2006)\citenamefont{Kagramanova, Kunz, and
  L\"ammerzahl}}]{Kagramanova.etal:2006}
\bibinfo{author}{\bibnamefont{Kagramanova}, \bibfnamefont{V.}},
  \bibinfo{author}{\bibfnamefont{J.}~\bibnamefont{Kunz}}, and
  \bibinfo{author}{\bibfnamefont{C.}~\bibnamefont{L\"ammerzahl}},
  \bibinfo{year}{2006}, {``}\bibinfo{title}{Solar system effects in
  {Schwarzschild--de\,Sitter} space--time},{''} \bibinfo{journal}{Physics
  Letters B} \textbf{\bibinfo{volume}{634}},  \bibinfo{pages}{465--470}.

\bibitem[{\citenamefont{{Klioner} and {Soffel}}(2005)}]{Klioner.Soffel:2005}
\bibinfo{author}{\bibnamefont{{Klioner}}, \bibfnamefont{S.~A.}}, and
  \bibinfo{author}{\bibfnamefont{M.~H.} \bibnamefont{{Soffel}}},
  \bibinfo{year}{2005}, {``}\bibinfo{title}{{Refining the Relativistic Model
  for Gaia: Cosmological Effects in the BCRS}},{''} in
  \emph{\bibinfo{booktitle}{The Three-Dimensional Universe with Gaia}}, edited
  by \bibinfo{editor}{\bibfnamefont{C.}~\bibnamefont{{Turon}}},
  \bibinfo{editor}{\bibfnamefont{K.~S.} \bibnamefont{{O'Flaherty}}}, and
  \bibinfo{editor}{\bibfnamefont{M.~A.~C.} \bibnamefont{{Perryman}}}
  (\bibinfo{publisher}{ESA}), volume \bibinfo{volume}{576} of
  \emph{\bibinfo{series}{ESA Special Publication}},  \bibinfo{pages}{305--308},
  \bibinfo{note}{proceedings of the Symposium ``The Three-Dimensional Universe
  with Gaia'', 4-7 October 2004, Observatoire de Paris-Meudon, France},
  \eprint{arXiv:astro-ph/0411363}.

\bibitem[{\citenamefont{Kodama}(1980)}]{Kodama:1980}
\bibinfo{author}{\bibnamefont{Kodama}, \bibfnamefont{H.}},
  \bibinfo{year}{1980}, {``}\bibinfo{title}{Conserved Energy Flux for the
  Spherically Symmetric System and the Backreaction Problem in the Black Hole
  Evaporation},{''} \bibinfo{journal}{Progress of Theoretical Physics}
  \textbf{\bibinfo{volume}{63}}(\bibinfo{number}{4}),
  \bibinfo{pages}{1217--1228}.

\bibitem[{{Komatsu} \emph{et~al.}(2009)\citenamefont{{Komatsu}}
  \emph{et~al.}}]{Komatsu.etal:2008}
\bibinfo{author}{\bibnamefont{{Komatsu}}, \bibfnamefont{E.}}, \emph{et~al.},
  \bibinfo{year}{2009}, {``}\bibinfo{title}{Five-Year Wilkinson Microwave
  Anisotropy Probe (WMAP) Observations: Cosmological Interpretation},{''}
  \bibinfo{journal}{Astrophysical Journal, Supplement Series}
  \textbf{\bibinfo{volume}{180}},  \bibinfo{pages}{330--376}.

\bibitem[{\citenamefont{Kottler}(1918)}]{Kottler:1918}
\bibinfo{author}{\bibnamefont{Kottler}, \bibfnamefont{F.}},
  \bibinfo{year}{1918}, {``}\bibinfo{title}{{\"Uber die physikalischen
  Grundlagen der Einsteinschen Gravitationstheorie}},{''}
  \bibinfo{journal}{Annalen der Physik}
  \textbf{\bibinfo{volume}{56}}(\bibinfo{number}{14}),
  \bibinfo{pages}{401--462}.

\bibitem[{\citenamefont{{Krasi{\'n}ski}}(1998)}]{Krasinski:1998}
\bibinfo{author}{\bibnamefont{{Krasi{\'n}ski}}, \bibfnamefont{A.}},
  \bibinfo{year}{1998}, \emph{\bibinfo{title}{{Inhomogeneous Cosmological
  Models}}}, Cambridge Monographs on Mathematical Physics
  (\bibinfo{publisher}{Cambridge University Press},
  \bibinfo{address}{Cambridge}).

\bibitem[{\citenamefont{{Krasinsky} and
  {Brumberg}}(2004)}]{Krasinsky.Brumberg:2004}
\bibinfo{author}{\bibnamefont{{Krasinsky}}, \bibfnamefont{G.~A.}}, and
  \bibinfo{author}{\bibfnamefont{V.~A.} \bibnamefont{{Brumberg}}},
  \bibinfo{year}{2004}, {``}\bibinfo{title}{{Secular increase of astronomical
  unit from analysis of the major planet motions, and its interpretation}},{''}
  \bibinfo{journal}{Celestial Mechanics and Dynamical Astronomy}
  \textbf{\bibinfo{volume}{90}},  \bibinfo{pages}{267--288}.

\bibitem[{\citenamefont{L{\"a}mmerzahl}
  \emph{et~al.}(2006)\citenamefont{L{\"a}mmerzahl, Preuss, and
  Dittus}}]{Laemmerzahl.etal:2006}
\bibinfo{author}{\bibnamefont{L{\"a}mmerzahl}, \bibfnamefont{C.}},
  \bibinfo{author}{\bibfnamefont{O.}~\bibnamefont{Preuss}}, and
  \bibinfo{author}{\bibfnamefont{H.}~\bibnamefont{Dittus}},
  \bibinfo{year}{2006}, {``}\bibinfo{title}{{Is the physics within the Solar
  system really understood?}},{''} \eprint{arXiv:gr-qc/0604052}.

\bibitem[{\citenamefont{Lanczos}(1924)}]{Lanczos:1924}
\bibinfo{author}{\bibnamefont{Lanczos}, \bibfnamefont{K.}},
  \bibinfo{year}{1924}, {``}\bibinfo{title}{{Fl{\"a}chenhafte Verteilung der
  Materie in der Einsteinschen Gravitationstheorie}},{''}
  \bibinfo{journal}{Annalen der Physik}
  \textbf{\bibinfo{volume}{379}}(\bibinfo{number}{14}),
  \bibinfo{pages}{518--540}.

\bibitem[{\citenamefont{{Lema{\^i}tre}}(1933)}]{Lemaitre:1933}
\bibinfo{author}{\bibnamefont{{Lema{\^i}tre}}, \bibfnamefont{G.}},
  \bibinfo{year}{1933}, {``}\bibinfo{title}{{L'Univers en expansion}},{''}
  \bibinfo{journal}{Annales de la Soci{\'e}t{\'e} Scientifique de Bruxelles,
  Series A, Sciences Math{\'e}matiques, Astronomiques et Physiques}
  \textbf{\bibinfo{volume}{53}},  \bibinfo{pages}{51--85}.

\bibitem[{\citenamefont{{Lema{\^i}tre}}(1997)}]{Lemaitre:1933trad}
\bibinfo{author}{\bibnamefont{{Lema{\^i}tre}}, \bibfnamefont{G.}},
  \bibinfo{year}{1997}, {``}\bibinfo{title}{{The expanding universe}},{''}
  \bibinfo{journal}{General Relativity and Gravitation}
  \textbf{\bibinfo{volume}{29}},  \bibinfo{pages}{641--680}
  \bibinfo{note}{traduction of the original article~\cite{Lemaitre:1933}}.

\bibitem[{\citenamefont{{Markwardt}}(2002)}]{Markwardt:2002}
\bibinfo{author}{\bibnamefont{{Markwardt}}, \bibfnamefont{C.~B.}},
  \bibinfo{year}{2002}, {``}\bibinfo{title}{{Independent Confirmation of the
  Pioneer 10 Anomalous Acceleration}},{''} \eprint{arXiv:gr-qc/0208046}.

\bibitem[{\citenamefont{McClure}(2006)}]{McClure:2006}
\bibinfo{author}{\bibnamefont{McClure}, \bibfnamefont{M.~L.}},
  \bibinfo{year}{2006}, \emph{\bibinfo{title}{Cosmological Black Holes as
  models of cosmological inhomogeneities}}, Ph.D. thesis,
  \bibinfo{school}{University of Toronto}.

\bibitem[{\citenamefont{{McVittie}}(1933)}]{McVittie:1933}
\bibinfo{author}{\bibnamefont{{McVittie}}, \bibfnamefont{G.~C.}},
  \bibinfo{year}{1933}, {``}\bibinfo{title}{{The mass-particle in an expanding
  universe}},{''} \bibinfo{journal}{Monthly Notices of the Royal Astronomical
  Society} \textbf{\bibinfo{volume}{93}},  \bibinfo{pages}{325--339}.

\bibitem[{\citenamefont{{Mena}} \emph{et~al.}(2002)\citenamefont{{Mena},
  {Tavakol}, and {Vera}}}]{Mena.etal:2002}
\bibinfo{author}{\bibnamefont{{Mena}}, \bibfnamefont{F.~C.}},
  \bibinfo{author}{\bibfnamefont{R.}~\bibnamefont{{Tavakol}}}, and
  \bibinfo{author}{\bibfnamefont{R.}~\bibnamefont{{Vera}}},
  \bibinfo{year}{2002}, {``}\bibinfo{title}{{Generalization of the
  Einstein-Straus model to anisotropic settings}},{''}
  \bibinfo{journal}{Physical Review D}
  \textbf{\bibinfo{volume}{66}}(\bibinfo{number}{4}), \bibinfo{eid}{044004}
  (pages~\bibinfo{numpages}{13}).

\bibitem[{\citenamefont{{Mena}} \emph{et~al.}(2003)\citenamefont{{Mena},
  {Tavakol}, and {Vera}}}]{Mena.etal:2003}
\bibinfo{author}{\bibnamefont{{Mena}}, \bibfnamefont{F.~C.}},
  \bibinfo{author}{\bibfnamefont{R.}~\bibnamefont{{Tavakol}}}, and
  \bibinfo{author}{\bibfnamefont{R.}~\bibnamefont{{Vera}}},
  \bibinfo{year}{2003}, {``}\bibinfo{title}{{On Modifications of the
  Einstein-Straus Model to Anisotropic Settings}},{''} \bibinfo{journal}{LNP
  Vol.~617: Current Trends in Relativistic Astrophysics}
  \textbf{\bibinfo{volume}{617}},  \bibinfo{pages}{343--347}.

\bibitem[{\citenamefont{{Mena}} \emph{et~al.}(2005)\citenamefont{{Mena},
  {Tavakol}, and {Vera}}}]{Mena.etal:2005}
\bibinfo{author}{\bibnamefont{{Mena}}, \bibfnamefont{F.~C.}},
  \bibinfo{author}{\bibfnamefont{R.}~\bibnamefont{{Tavakol}}}, and
  \bibinfo{author}{\bibfnamefont{R.}~\bibnamefont{{Vera}}},
  \bibinfo{year}{2005}, {``}\bibinfo{title}{{Generalisations of the
  Einstein-Straus model to cylindrically symmetric settings}},{''} in
  \emph{\bibinfo{booktitle}{The Tenth Marcel Grossmann Meeting. On recent
  developments in theoretical and experimental general relativity, gravitation
  and relativistic field theories}}, edited by
  \bibinfo{editor}{\bibfnamefont{M.}~\bibnamefont{{Novello}}},
  \bibinfo{editor}{\bibfnamefont{S.}~\bibnamefont{{Perez Bergliaffa}}}, and
  \bibinfo{editor}{\bibfnamefont{R.}~\bibnamefont{{Ruffini}}}
  (\bibinfo{publisher}{World Scientific, Singapore}),
  \bibinfo{pages}{1749--1751}, \eprint{arXiv:gr-qc/0405043}.

\bibitem[{\citenamefont{{Misner} and {Sharp}}(1964)}]{Misner.Sharp:1964}
\bibinfo{author}{\bibnamefont{{Misner}}, \bibfnamefont{C.~W.}}, and
  \bibinfo{author}{\bibfnamefont{D.~H.} \bibnamefont{{Sharp}}},
  \bibinfo{year}{1964}, {``}\bibinfo{title}{{Relativistic Equations for
  Adiabatic, Spherically Symmetric Gravitational Collapse}},{''}
  \bibinfo{journal}{Physical Review} \textbf{\bibinfo{volume}{136}},
  \bibinfo{pages}{571--576}.

\bibitem[{\citenamefont{Misner} \emph{et~al.}(1973)\citenamefont{Misner,
  Thorne, and Wheeler}}]{Misner.Thorne.Wheeler:Gravitation}
\bibinfo{author}{\bibnamefont{Misner}, \bibfnamefont{C.~W.}},
  \bibinfo{author}{\bibfnamefont{K.~S.} \bibnamefont{Thorne}}, and
  \bibinfo{author}{\bibfnamefont{J.~A.} \bibnamefont{Wheeler}},
  \bibinfo{year}{1973}, \emph{\bibinfo{title}{Gravitation}}
  (\bibinfo{publisher}{W.H. Freeman and Company}, \bibinfo{address}{New York}).

\bibitem[{\citenamefont{Nieto and Turyshev}(2004)}]{Nieto.Turyshev:2004}
\bibinfo{author}{\bibnamefont{Nieto}, \bibfnamefont{M.~M.}}, and
  \bibinfo{author}{\bibfnamefont{S.~G.} \bibnamefont{Turyshev}},
  \bibinfo{year}{2004}, {``}\bibinfo{title}{{Finding the origin of the Pioneer
  anomaly}},{''} \bibinfo{journal}{Classical and Quantum Gravity}
  \textbf{\bibinfo{volume}{21}}(\bibinfo{number}{17}),
  \bibinfo{pages}{4005--4023}.

\bibitem[{\citenamefont{{Nieto}} \emph{et~al.}(2005)\citenamefont{{Nieto},
  {Turyshev}, and {Anderson}}}]{Nieto.etal:2005}
\bibinfo{author}{\bibnamefont{{Nieto}}, \bibfnamefont{M.~M.}},
  \bibinfo{author}{\bibfnamefont{S.~G.} \bibnamefont{{Turyshev}}}, and
  \bibinfo{author}{\bibfnamefont{J.~D.} \bibnamefont{{Anderson}}},
  \bibinfo{year}{2005}, {``}\bibinfo{title}{{Directly measured limit on the
  interplanetary matter density from Pioneer 10 and 11}},{''}
  \bibinfo{journal}{Physics Letters B} \textbf{\bibinfo{volume}{613}},
  \bibinfo{pages}{11--19}.

\bibitem[{\citenamefont{{Noerdlinger} and
  {Petrosian}}(1971)}]{Noerdlinger.Petrosian:1971}
\bibinfo{author}{\bibnamefont{{Noerdlinger}}, \bibfnamefont{P.~D.}}, and
  \bibinfo{author}{\bibfnamefont{V.}~\bibnamefont{{Petrosian}}},
  \bibinfo{year}{1971}, {``}\bibinfo{title}{{The Effect of Cosmological
  Expansion on Self-Gravitating Ensembles of Particles}},{''}
  \bibinfo{journal}{Astrophysical Journal} \textbf{\bibinfo{volume}{168}},
  \bibinfo{pages}{1--9}.

\bibitem[{\citenamefont{Nolan}(1993)}]{Nolan:1992}
\bibinfo{author}{\bibnamefont{Nolan}, \bibfnamefont{B.~C.}},
  \bibinfo{year}{1993}, {``}\bibinfo{title}{Sources for {McVittie's} Mass
  Particle in an Expanding Universe},{''} \bibinfo{journal}{Journal of
  Mathematical Physics} \textbf{\bibinfo{volume}{34}}(\bibinfo{number}{1}),
  \bibinfo{pages}{178--185}.

\bibitem[{\citenamefont{{Nolan}}(1998)}]{Nolan:1998}
\bibinfo{author}{\bibnamefont{{Nolan}}, \bibfnamefont{B.~C.}},
  \bibinfo{year}{1998}, {``}\bibinfo{title}{{A point mass in an isotropic
  universe: Existence, uniqueness, and basic properties}},{''}
  \bibinfo{journal}{Physical Review D}
  \textbf{\bibinfo{volume}{58}}(\bibinfo{number}{6}), \bibinfo{eid}{064006}
  (pages~\bibinfo{numpages}{10}).

\bibitem[{\citenamefont{{Nolan}}(1999{\natexlab{a}})}]{Nolan:1999a}
\bibinfo{author}{\bibnamefont{{Nolan}}, \bibfnamefont{B.~C.}},
  \bibinfo{year}{1999}{\natexlab{a}}, {``}\bibinfo{title}{{A point mass in an
  isotropic universe: II. Global properties}},{''} \bibinfo{journal}{Classical
  and Quantum Gravity} \textbf{\bibinfo{volume}{16}},
  \bibinfo{pages}{1227--1254}.

\bibitem[{\citenamefont{{Nolan}}(1999{\natexlab{b}})}]{Nolan:1999b}
\bibinfo{author}{\bibnamefont{{Nolan}}, \bibfnamefont{B.~C.}},
  \bibinfo{year}{1999}{\natexlab{b}}, {``}\bibinfo{title}{{A point mass in an
  isotropic universe: III. The region $R \leq 2m$}},{''}
  \bibinfo{journal}{Classical and Quantum Gravity}
  \textbf{\bibinfo{volume}{16}},  \bibinfo{pages}{3183--3191}.

\bibitem[{\citenamefont{{Nottale}}(2003)}]{Nottale:2003}
\bibinfo{author}{\bibnamefont{{Nottale}}, \bibfnamefont{L.}},
  \bibinfo{year}{2003}, {``}\bibinfo{title}{The Pioneer anomalous acceleration:
  a measurement of the cosmological constant at the scale of the solar
  system},{''} \eprint{arXiv:gr-qc/0307042}.

\bibitem[{\citenamefont{{O'Neill}}(1983)}]{ONeill:1983}
\bibinfo{author}{\bibnamefont{{O'Neill}}, \bibfnamefont{B.}},
  \bibinfo{year}{1983}, \emph{\bibinfo{title}{{Semi-Riemannian Geometry with
  Applications to Relativity}}} (\bibinfo{publisher}{Academic Press},
  \bibinfo{address}{Orlando}).

\bibitem[{\citenamefont{Pachner}(1963)}]{Pachner:1963}
\bibinfo{author}{\bibnamefont{Pachner}, \bibfnamefont{J.}},
  \bibinfo{year}{1963}, {``}\bibinfo{title}{Mach's Principle in Classical and
  Relativistic Physics},{''} \bibinfo{journal}{Physical Review}
  \textbf{\bibinfo{volume}{132}}(\bibinfo{number}{4}),
  \bibinfo{pages}{1837--1842}.

\bibitem[{\citenamefont{Pachner}(1964)}]{Pachner:1964}
\bibinfo{author}{\bibnamefont{Pachner}, \bibfnamefont{J.}},
  \bibinfo{year}{1964}, {``}\bibinfo{title}{Nonconservation of Energy During
  Cosmic Evolution},{''} \bibinfo{journal}{Physical Review Letters}
  \textbf{\bibinfo{volume}{12}}(\bibinfo{number}{4}),
  \bibinfo{pages}{117--118}.

\bibitem[{\citenamefont{{Palle}}(2005)}]{Palle:2005}
\bibinfo{author}{\bibnamefont{{Palle}}, \bibfnamefont{D.}},
  \bibinfo{year}{2005}, {``}\bibinfo{title}{{On the anomalous acceleration in
  the solar system}},{''} \bibinfo{journal}{Acta Physica Slovaca}
  \textbf{\bibinfo{volume}{55}}(\bibinfo{number}{2}),
  \bibinfo{pages}{237--240}.

\bibitem[{\citenamefont{{Patel} and {Trivedi}}(1982)}]{Patel.Trivedi:1982}
\bibinfo{author}{\bibnamefont{{Patel}}, \bibfnamefont{L.~K.}}, and
  \bibinfo{author}{\bibfnamefont{H.~B.} \bibnamefont{{Trivedi}}},
  \bibinfo{year}{1982}, {``}\bibinfo{title}{{Kerr-Newman metric in cosmological
  background}},{''} \bibinfo{journal}{Journal of Astrophysics and Astronomy}
  \textbf{\bibinfo{volume}{3}},  \bibinfo{pages}{63--67}.

\bibitem[{\citenamefont{Price}(2005)}]{Price:2005}
\bibinfo{author}{\bibnamefont{Price}, \bibfnamefont{R.~H.}},
  \bibinfo{year}{2005}, {``}\bibinfo{title}{In an expanding universe, what
  doesn't expand?},{''} \eprint{arXiv:gr-qc/0508052}.

\bibitem[{\citenamefont{{Rajesh Nayak}}
  \emph{et~al.}(2001)\citenamefont{{Rajesh Nayak}, {MacCallum}, and
  {Vishveshwara}}}]{RajeshNayak.etal:2001}
\bibinfo{author}{\bibnamefont{{Rajesh Nayak}}, \bibfnamefont{K.}},
  \bibinfo{author}{\bibfnamefont{M.~A.} \bibnamefont{{MacCallum}}}, and
  \bibinfo{author}{\bibfnamefont{C.~V.} \bibnamefont{{Vishveshwara}}},
  \bibinfo{year}{2001}, {``}\bibinfo{title}{{Black holes in nonflat
  backgrounds: The Schwarzschild black hole in the Einstein universe}},{''}
  \bibinfo{journal}{Physical Review D}
  \textbf{\bibinfo{volume}{63}}(\bibinfo{number}{2}), \bibinfo{eid}{024020}
  (pages~\bibinfo{numpages}{5}).

\bibitem[{\citenamefont{{Ramachandra}}
  \emph{et~al.}(2003)\citenamefont{{Ramachandra}, {Nayak}, and
  {Vishveshwara}}}]{Ramachandra.etal:2003}
\bibinfo{author}{\bibnamefont{{Ramachandra}}, \bibfnamefont{B.~S.}},
  \bibinfo{author}{\bibfnamefont{K.~R.} \bibnamefont{{Nayak}}}, and
  \bibinfo{author}{\bibfnamefont{C.~V.} \bibnamefont{{Vishveshwara}}},
  \bibinfo{year}{2003}, {``}\bibinfo{title}{{Kerr Black Hole in the Background
  of the Einstein Universe}},{''} \bibinfo{journal}{General Relativity and
  Gravitation} \textbf{\bibinfo{volume}{35}},  \bibinfo{pages}{1977--2005}.

\bibitem[{\citenamefont{{Ramachandra} and
  {Vishveshwara}}(2002)}]{Ramachandra.Vishveshwara:2002}
\bibinfo{author}{\bibnamefont{{Ramachandra}}, \bibfnamefont{B.~S.}}, and
  \bibinfo{author}{\bibfnamefont{C.~V.} \bibnamefont{{Vishveshwara}}},
  \bibinfo{year}{2002}, {``}\bibinfo{title}{Schwarzschild black hole in the
  background of the {Einstein} universe: some physical effects},{''}
  \bibinfo{journal}{Classical and Quantum Gravity}
  \textbf{\bibinfo{volume}{19}},  \bibinfo{pages}{127--141}.

\bibitem[{\citenamefont{{Ranada}}(2005)}]{Ranada:2005}
\bibinfo{author}{\bibnamefont{{Ranada}}, \bibfnamefont{A.~F.}},
  \bibinfo{year}{2005}, {``}\bibinfo{title}{{The Pioneer anomaly as
  acceleration of the clocks}},{''} \bibinfo{journal}{Foundations of Physics}
  \textbf{\bibinfo{volume}{34}},  \bibinfo{pages}{1955--1971}.

\bibitem[{\citenamefont{{R{\"a}s{\"a}nen}}(2006)}]{Rasanen:2006}
\bibinfo{author}{\bibnamefont{{R{\"a}s{\"a}nen}}, \bibfnamefont{S.}},
  \bibinfo{year}{2006}, {``}\bibinfo{title}{{Accelerated expansion from
  structure formation}},{''} \bibinfo{journal}{Journal of Cosmology and
  Astroparticle Physics} \textbf{\bibinfo{volume}{11}}, \bibinfo{eid}{003}
  (pages~\bibinfo{numpages}{40}).

\bibitem[{\citenamefont{Robertson}(1928)}]{Robertson:1928}
\bibinfo{author}{\bibnamefont{Robertson}, \bibfnamefont{H.~P.}},
  \bibinfo{year}{1928}, {``}\bibinfo{title}{On Relativistic Cosmology},{''}
  \bibinfo{journal}{Philosophical Magazine} \textbf{\bibinfo{volume}{5}},
  \bibinfo{pages}{835--848}.

\bibitem[{\citenamefont{{Rosales} and
  {Sanchez-Gomez}}(1998)}]{Rosales.Sanchez-Gomez:1998}
\bibinfo{author}{\bibnamefont{{Rosales}}, \bibfnamefont{J.}}, and
  \bibinfo{author}{\bibfnamefont{J.}~\bibnamefont{{Sanchez-Gomez}}},
  \bibinfo{year}{1998}, {``}\bibinfo{title}{{The ``Pioneer effect'' as a
  manifestation of the cosmic expansion in the solar system}},{''}
  \eprint{arXiv:gr-qc/99810085}.

\bibitem[{\citenamefont{{Rosales}}(2002)}]{Rosales:2002}
\bibinfo{author}{\bibnamefont{{Rosales}}, \bibfnamefont{J.~L.}},
  \bibinfo{year}{2002}, {``}\bibinfo{title}{{The Pioneer's acceleration anomaly
  and Hubble's constant}},{''} \eprint{arXiv:gr-qc/0212019}.

\bibitem[{\citenamefont{{Sch\"ucking}}(1954)}]{Schuecking:1954}
\bibinfo{author}{\bibnamefont{{Sch\"ucking}}, \bibfnamefont{E.}},
  \bibinfo{year}{1954}, {``}\bibinfo{title}{{Das Schwarzschildsche
  Linienelement und die Expansion des Weltalls}},{''}
  \bibinfo{journal}{Zeitschrift {f\"ur} Physik} \textbf{\bibinfo{volume}{137}},
   \bibinfo{pages}{595--603}.

\bibitem[{\citenamefont{{Senovilla} and {Vera}}(1997)}]{Senovilla.Vera:1997}
\bibinfo{author}{\bibnamefont{{Senovilla}}, \bibfnamefont{J.~M.~M.}}, and
  \bibinfo{author}{\bibfnamefont{R.}~\bibnamefont{{Vera}}},
  \bibinfo{year}{1997}, {``}\bibinfo{title}{{Impossibility of the Cylindrically
  Symmetric Einstein-Straus Model}},{''} \bibinfo{journal}{Physical Review
  Letters} \textbf{\bibinfo{volume}{78}},  \bibinfo{pages}{2284--2287}.

\bibitem[{\citenamefont{Sereno and Jetzer}(2007)}]{Sereno.Jetzer:2007}
\bibinfo{author}{\bibnamefont{Sereno}, \bibfnamefont{M.}}, and
  \bibinfo{author}{\bibfnamefont{P.}~\bibnamefont{Jetzer}},
  \bibinfo{year}{2007}, {``}\bibinfo{title}{Evolution of Gravitational Orbits
  in the Expanding Universe},{''} \bibinfo{journal}{Physical Review D}
  \textbf{\bibinfo{volume}{75}}(\bibinfo{number}{6}), \bibinfo{eid}{064031}
  (pages~\bibinfo{numpages}{8}).

\bibitem[{\citenamefont{{Standish}}(2004)}]{Standish:2004}
\bibinfo{author}{\bibnamefont{{Standish}}, \bibfnamefont{E.~M.}},
  \bibinfo{year}{2004}, {``}\bibinfo{title}{{The Astronomical Unit now}},{''}
  in \emph{\bibinfo{booktitle}{{Transits of Venus: New Views of the Solar
  System and Galaxy, Proceedings of the IAU Colloquium No.~196, 2004}}}, edited
  by \bibinfo{editor}{\bibfnamefont{D.~W.} \bibnamefont{{Kurtz}}}
  (\bibinfo{publisher}{Cambridge University Press},
  \bibinfo{address}{Cambridge}),  \bibinfo{pages}{163--179}.

\bibitem[{\citenamefont{Straumann}(2004)}]{Straumann:2004}
\bibinfo{author}{\bibnamefont{Straumann}, \bibfnamefont{N.}},
  \bibinfo{year}{2004}, \emph{\bibinfo{title}{{General Relativity with
  Applications to Astrophysics}}}, Texts and Monographs in Physics
  (\bibinfo{publisher}{Springer-Verlag}, \bibinfo{address}{Berlin}).

\bibitem[{\citenamefont{Sultana and Dyer}(2005)}]{Sultana.Dyer:2005}
\bibinfo{author}{\bibnamefont{Sultana}, \bibfnamefont{J.}}, and
  \bibinfo{author}{\bibfnamefont{C.~C.} \bibnamefont{Dyer}},
  \bibinfo{year}{2005}, {``}\bibinfo{title}{Cosmological black holes: A black
  hole in the {Einstein--de\,Sitter} universe},{''} \bibinfo{journal}{General
  Relativity and Gravitation}
  \textbf{\bibinfo{volume}{37}}(\bibinfo{number}{8}),
  \bibinfo{pages}{1349--1370}.

\bibitem[{\citenamefont{Sussman}(1988)}]{Sussman:1988}
\bibinfo{author}{\bibnamefont{Sussman}, \bibfnamefont{R.~A.}},
  \bibinfo{year}{1988}, {``}\bibinfo{title}{On spherically symmetric shear-free
  perfect fluid configurations (neutral and charged). III. Global view},{''}
  \bibinfo{journal}{Journal of Mathematical Physics}
  \textbf{\bibinfo{volume}{29}}(\bibinfo{number}{5}),
  \bibinfo{pages}{1177--1211}.

\bibitem[{\citenamefont{Szabados}(2004)}]{SzabadosLivingReviews:2004}
\bibinfo{author}{\bibnamefont{Szabados}, \bibfnamefont{L.~B.}},
  \bibinfo{year}{2004}, {``}\bibinfo{title}{Quasi-Local Energy-Momentum and
  Angular Momentum in GR: A Review Article},{''} \bibinfo{journal}{Living
  Reviews in Relativity} \textbf{\bibinfo{volume}{7}}(\bibinfo{number}{4}),
  \bibinfo{pages}{1--135},
  \urlprefix\url{http://www.livingreviews.org/lrr-2004-4}.

\bibitem[{\citenamefont{Turyshev}
  \emph{et~al.}(2005{\natexlab{a}})\citenamefont{Turyshev, Nieto, and
  Anderson}}]{Turyshev.etal:2005a}
\bibinfo{author}{\bibnamefont{Turyshev}, \bibfnamefont{S.~G.}},
  \bibinfo{author}{\bibfnamefont{M.~M.} \bibnamefont{Nieto}}, and
  \bibinfo{author}{\bibfnamefont{J.~D.} \bibnamefont{Anderson}},
  \bibinfo{year}{2005}{\natexlab{a}}, {``}\bibinfo{title}{{A route to
  understanding of the Pioneer anomaly}},{''} \eprint{arXiv:gr-qc/0503021}.

\bibitem[{\citenamefont{Turyshev}
  \emph{et~al.}(2005{\natexlab{b}})\citenamefont{Turyshev, Nieto, and
  Anderson}}]{Turyshev.etal:2005b}
\bibinfo{author}{\bibnamefont{Turyshev}, \bibfnamefont{S.~G.}},
  \bibinfo{author}{\bibfnamefont{M.~M.} \bibnamefont{Nieto}}, and
  \bibinfo{author}{\bibfnamefont{J.~D.} \bibnamefont{Anderson}},
  \bibinfo{year}{2005}{\natexlab{b}}, {``}\bibinfo{title}{{Study of the Pioneer
  anomaly: A problem set}},{''} \bibinfo{journal}{American Journal of Physics}
  \textbf{\bibinfo{volume}{73}}(\bibinfo{number}{11}),
  \bibinfo{pages}{1033--1044}.

\bibitem[{\citenamefont{Vaidya}(1977)}]{Vaidya:1977}
\bibinfo{author}{\bibnamefont{Vaidya}, \bibfnamefont{P.~C.}},
  \bibinfo{year}{1977}, {``}\bibinfo{title}{The {Kerr} metric in cosmological
  background},{''} \bibinfo{journal}{Pramana}
  \textbf{\bibinfo{volume}{8}}(\bibinfo{number}{6}),
  \bibinfo{pages}{512--517}.

\bibitem[{\citenamefont{Vaidya}(1984)}]{Vaidya:1984}
\bibinfo{author}{\bibnamefont{Vaidya}, \bibfnamefont{P.~C.}},
  \bibinfo{year}{1984}, {``}\bibinfo{title}{{Kerr metric in the de\,Sitter
  background.}},{''} \bibinfo{journal}{Pramana} \textbf{\bibinfo{volume}{22}},
  \bibinfo{pages}{151--158}.

\bibitem[{\citenamefont{{van den Bergh} and {Wils}}(1984)}]{vdBergh.Wils:1984}
\bibinfo{author}{\bibnamefont{{van den Bergh}}, \bibfnamefont{N.}}, and
  \bibinfo{author}{\bibfnamefont{P.}~\bibnamefont{{Wils}}},
  \bibinfo{year}{1984}, {``}\bibinfo{title}{{Imbedding a Schwarzschild mass
  into cosmology}},{''} \bibinfo{journal}{Physical Review D}
  \textbf{\bibinfo{volume}{29}},  \bibinfo{pages}{3002--3003}.

\bibitem[{\citenamefont{Vishveshwara}(2000)}]{Vishveshwara:2000}
\bibinfo{author}{\bibnamefont{Vishveshwara}, \bibfnamefont{C.~V.}},
  \bibinfo{year}{2000}, {``}\bibinfo{title}{Black Holes in cosmological
  backgrounds},{''} in \emph{\bibinfo{booktitle}{The Universe: Visions and
  Perspectives}}, edited by \bibinfo{editor}{\bibfnamefont{A.~K.~K.}
  \bibnamefont{Naresh~Dadhich}} (\bibinfo{publisher}{Kluwer Academic
  Publisher}),  \bibinfo{pages}{309--318}.

\bibitem[{\citenamefont{Wiltshire}(2007)}]{Wiltshire:2007}
\bibinfo{author}{\bibnamefont{Wiltshire}, \bibfnamefont{D.~L.}},
  \bibinfo{year}{2007}, {``}\bibinfo{title}{Exact Solution to the Averaging
  Problem in Cosmology},{''} \bibinfo{journal}{Physical Review Letters}
  \textbf{\bibinfo{volume}{99}}(\bibinfo{number}{25}), \bibinfo{eid}{251101}
  (pages~\bibinfo{numpages}{4}).

\bibitem[{\citenamefont{Wiltshire}(2008)}]{Wiltshire:2008}
\bibinfo{author}{\bibnamefont{Wiltshire}, \bibfnamefont{D.~L.}},
  \bibinfo{year}{2008}, {``}\bibinfo{title}{Dark energy without dark
  energy},{''} in \emph{\bibinfo{booktitle}{Dark Matter in Astroparticle and
  Particle Physics: Proceedings of the 6th International Heidelberg
  Conference}} (\bibinfo{publisher}{World Scientific, Singapore}),
  \bibinfo{pages}{565--596}, \eprint{arXiv:0712.3984}.

\bibitem[{\citenamefont{Zannias}(1990)}]{Zannias:1990}
\bibinfo{author}{\bibnamefont{Zannias}, \bibfnamefont{T.}},
  \bibinfo{year}{1990}, {``}\bibinfo{title}{Spacetimes admitting a
  three-parameter group of isometries and quasilocal gravitational mass},{''}
  \bibinfo{journal}{Physical Review D}
  \textbf{\bibinfo{volume}{41}}(\bibinfo{number}{10}),
  \bibinfo{pages}{3252--3254}.

\end{thebibliography}

\end{document}